\documentclass[sigconf, authorversion]{acmart}

\makeatletter
\renewcommand{\@copyrightowner}{Copyright help by the owner/author(s).\\ \includegraphics[width=0.15\linewidth]{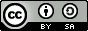}
This work is licensed under a Creative Commons Attribution-Share Alike International 4.0 License.}
\makeatother

\copyrightyear{2024} 
\acmYear{2024} 
\setcopyright{rightsretained} 
\acmBooktitle{Proceedings of the CHI Conference on Human Factors in Computing Systems (CHI '24), May 11--16, 2024, Honolulu, HI, USA}
\acmDOI{10.1145/3613904.3642706}
\acmISBN{979-8-4007-0330-0/24/05}

\usepackage{xcolor} 
\usepackage{colortbl} 
\usepackage{xspace} 
\usepackage{cleveref} 
\usepackage{subcaption} 
\usepackage{listings} 

\AtBeginEnvironment{quote}{\itshape}

\crefname{section}{\S}{\S\S}
\Crefname{section}{\S}{\S\S}
\crefformat{section}{\S#2#1#3}
\Crefname{figure}{Figure}{Figures}
\crefformat{figure}{Figure #2#1#3}
\Crefname{Figure}{Figure}{Figures}

\acmConference[CHI '24]{Proceedings of the CHI Conference on Human Factors in Computing Systems}{May 11--16, 2024}{Honolulu, HI, USA}
\acmBooktitle{Proceedings of the CHI Conference on Human Factors in Computing Systems (CHI '24), May 11--16, 2024, Honolulu, HI, USA}
\acmDOI{10.1145/3613904.3642706}
\acmISBN{979-8-4007-0330-0/24/05}

\begin{document}

\title[How Beginning Programmers and Code LLMs (Mis)read Each Other]{How Beginning Programmers and Code LLMs (Mis)read~Each~Other} 


\author{Sydney Nguyen}
\affiliation{%
\institution{Wellesley College}
\country{USA}}
\author{Hannah McLean Babe}
\affiliation{%
\institution{Oberlin College}
\country{USA}}
\author{Yangtian Zi}
\affiliation{%
\institution{Northeastern University}
\country{USA}}
\author{Arjun Guha}
\affiliation{%
\institution{Northeastern University and Roblox}
\country{USA}}
\email{a.guha@northeastern.edu}
\author{Carolyn Jane Anderson}
\affiliation{%
\institution{Wellesley College}
\country{USA}}
\email{carolyn.anderson@wellesley.edu}
\author{Molly~Q Feldman}
\affiliation{%
\institution{Oberlin College}
\country{USA}}
\email{mfeldman@oberlin.edu}


\newcommand{\pseudonym}[1]{\textsc{\textit{#1}}}

\newcommand{\tool}{Charlie\xspace}
\newcommand{\wellesley}{Wellesley\xspace}
\newcommand{\oberlin}{Oberlin\xspace}
\newcommand{\northeastern}{Northeastern\xspace}
\newcommand{\osf}{\url{https://doi.org/10.17605/OSF.IO/V2C4T}\xspace}

\renewcommand{\shortauthors}{Nguyen et al.}

\begin{abstract} 
Generative AI models, specifically large language models (LLMs), have made strides towards the long-standing goal of text-to-code generation. This progress has invited numerous studies of user interaction. However, less is known about the struggles and strategies of non-experts, for whom each step of the text-to-code problem presents challenges: describing their intent in natural language, evaluating the correctness of generated code, and editing prompts when the generated code is incorrect. This paper presents a large-scale controlled study of how 120 beginning coders across three academic institutions approach writing and editing prompts. A novel experimental design allows us to target specific steps in the text-to-code process and reveals that beginners struggle with writing and editing prompts, even for problems at their skill level and when correctness is automatically determined. Our mixed-methods evaluation provides insight into student processes and perceptions with key implications for non-expert Code LLM use within and outside of education.
\end{abstract}

\maketitle

\begin{figure}[t]
  \centering
  \includegraphics[width=0.75\linewidth]{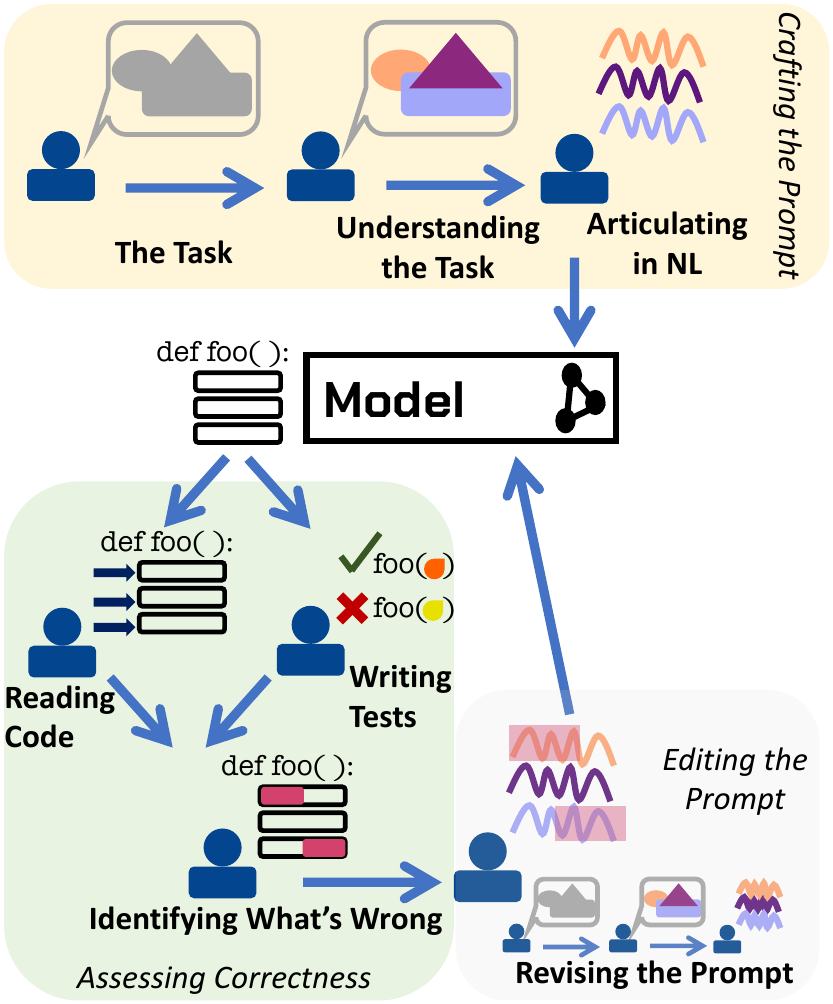}
  \caption{Visualization of the multi-step process of querying a large language mode of code (Code LLM). The user starts with crafting their prompt in natural language (NL). They provide the prompt to the model, which produces code. The user then assesses the correctness of the generated code. If there are errors, they must identify how to resolve them and how to edit the prompt. This continues in an iterative fashion.}
  \label{fig:the-query}
\end{figure}

\section{Introduction}

Computer scientists have been working towards programming in natural language for decades
~\cite{ballard_programming_1979, sammet_use_1966, heidorn_english_1974}, often with the goal of making programming easier for a broader set of users. Recent advances in \emph{generative AI} have brought us nearer to this goal. In programming, along with fields like  digital art~\citep{ramesh_zero-shot_2021,rombach_high-resolution_2022, radford_language_2019}, creative writing~\citep{akoury_storium_2020,ippolito_creative_2022,mirowski_co-writing_2023}, and digital music~\citep{agostinelli_musiclm_2023, liu_generative_2023}, generative AI has reduced the technical skills that users need by allowing them to \emph{prompt} a model with a natural language description of their desired output. 

In many fields, experts have started to use generative AI to accelerate their work, including in software engineering, where \emph{large language models of code} (Code LLMs) have enhanced expert programmer productivity~\cite{peng_impact_2023,ziegler_productivity_2022,murali_codecompose_2023}. However, to fulfill their potential of democratizing these fields, models must be usable without extensive technical training at each stage of creation: 1)~writing prompts for the model, 2)~evaluating model output for quality, and 3)~iteratively refining prompts when generation is unsuccessful. 

Programming presents a particularly challenging domain for non-experts. Like art, computer science has evolved an extensive technical vocabulary; since generative models are trained largely on professional code, they may not work as well if users lack this vocabulary. In visual art, music, and creative writing, a user can quickly determine whether they like the generated output even if they are not an expert (embodying the clich\'e ``I don't know anything about art, but I know what I like''). However, this attitude does not extend to programming. It is very challenging for a non-expert to evaluate the quality of a generated program. Even when a user knows enough to determine a generated program is incorrect, they also need to understand it well enough to know what needs to change and how to update their prompt. 

In order to use a Code LLM, non-experts must grapple with a multi-step process (\Cref{fig:the-query}). First, they must have a clear understanding of what they want the code to do. This may seem trivial, but research on requirements engineering has shown that it can be challenging~\citep{parnas_functional_1995}. Next, the user must clearly articulate the intended behavior of the program in natural language to the model. Once the model generates code, the user must evaluate its correctness by reading it or writing tests. If the code is not correct, they must determine what has gone wrong, and update their prompt accordingly. This requires not only understanding the generated code, but also, understanding the model's generative process. These barriers mirror well-known challenges for non-experts with end-user programming \cite{ko_six_2004} and classical AI systems \cite{lau_why_2009}.

There is a growing body of work studying how non-expert programmers use AI-assisted programming systems in naturalistic settings~\citep{prather_its_2023,kazemitabaar_studying_2023}. However, in open-ended tasks, it is difficult to decouple the steps of the code generation process, since they feed each other: if the user fails to identify incorrect code and moves on, their editing process can't be observed. We present results from a carefully-controlled experiment targeting two steps in the code generation process: prompt creation (\textit{How do users describe the intended program in natural language?}) and prompt modification (\textit{How do users modify their prompts when a generated program is incorrect?}).

One challenge in studying how non-experts use Code LLMs is selecting tasks that make sense to them. For example, replicating \citet{barke_grounded_2023}'s insightful study of experienced programmers would not be appropriate for novices, because the tasks presuppose technical knowledge. Novices have diverse goals, backgrounds, and familiarity with mathematical and computational thinking. Our solution is to target a large population of near-novices with similar experience levels: university students who have completed a single introductory computer science course (CS1). This allows us to select tasks that are conceptually familiar to them. 

\paragraph{Our Approach} We ask whether students who have completed CS1 can effectively prompt a Code LLM to solve tasks from their previous course. In order to isolate students' experiences in writing and editing prompts, our experiment presents tasks as input/output pairs and tests the generated code for correctness. This provides in-depth insight into the processes they develop for describing code in natural language and iteratively refining their prompts. We pose three main research questions: 
\begin{itemize}
    \item \textsc{RQ1}: Can students who have completed a CS1 course effectively prompt a Code LLM to generate code for questions from their previous courses?
    \item \textsc{RQ2}: What is the origin of student challenges with Code LLMs? Do these differ across different groups of students?
    \item \textsc{RQ3}: What are students' mental models of Code LLMs and how do they effect their interactions?
\end{itemize}


We find that students struggle significantly with this task, even though we pose problems tailored to their skill level and test code correctness for them. In essence, \textit{beginning programmers and current Code LLMs tend to misread each other}: the Code LLM fails to generate working code based on student descriptions and students have a hard time adapting their descriptions to the model. Our study has concerning implications for democratizing programming: if these students, who already have basic skills in code explanation and understanding, struggle with this simplified task, the full natural language-to-code task---where the user has to determine correctness themselves---must be very challenging indeed for true novices. This finding also has important implications for education. Code LLMs have sparked an intense debate over the future of computing education, including claims that traditional programming training is no longer necessary~\citep{manjoo_its_2023, welsh_end_2022}. By contrast, our findings highlight the continuing importance of teaching students technical communication and code understanding.

Our work differentiates itself from previous work in three key ways: scale, population, and experimental design. First, we study 120 students solving 48 different programming problems. To our knowledge, no previous work has studied user interactions with Code LLMs at this scale. Second, we focus on a near-novice population with fairly uniform levels of experience, allowing us to carefully tailor tasks to their skill level. Finally, we use an experimental paradigm that allows us to isolate the prompt writing and editing aspects of the task.\footnote{Data collected as part of this work is publicly available at \osf.} 

\section{Related Work}

Our work focuses on how programmers use LLMs to turn natural language into code. Programming with natural language is a decades old proposition~\cite{miller_natural_1981} and has led to several ideas about bringing programming closer to how users communicate~\cite{myers_programmers_2016}.
For instance, \citet{hindle_naturalness_2012} imagined that future language models could be effective at turning natural language to code, a prediction that has been borne out with Code LLMs.

By exploring beginner interactions with Code LLMs, our study contributes to a growing body of work on how non-experts interact with emerging automated technologies~\cite{van_mechelen_emerging_2023}, ranging from automated feedback \cite{feldman_automatic_2018, hsu_attitudes_2021, suzuki_tracediff_2017} to augmented reality \cite{holstein_student_2018, radu_what_2019}. We situate our study within existing work on user interactions with Code LLMs below.

\paragraph{Experienced programmers and LLMs}

We study how beginning programmers interact with a Code LLM, the same foundational technology that powers autocomplete tools such as GitHub Copilot and others~\cite{copilot_github_2023,codewhisperer_ml-powered_2023,tabnine_ai_2023}. These tools are promoted as productivity-boosting technology for experienced programmers. Recent in-the-wild studies and surveys indicate that these tools are popular with expert programmers, improve their self-perception of productivity, and shift their work from writing code to understanding LLM outputs~\cite{bird_taking_2023,murali_codecompose_2023,liang_large-scale_2023}.
In contrast, our study of beginners' interactions with a Code LLM reveals that (1)~they have mixed success with writing natural language prompts, (2)~and they often struggle to understand LLM-generated code.

\citet{vaithilingam_towards_2023} present the earliest academic study of GitHub Copilot with 24 students (undergraduate--PhD) and three tasks. Their main finding is that although participants enjoyed using it, Copilot did not help them code faster or write more correct code. We design our study for less experienced participants. For example, we developed a web interface that is much simpler than a professional IDE.
The same study reports that their participants often struggled to validate LLM-generated code, and we avoid this by testing generated code for our participants automatically.

Since Copilot is a general autocomplete tool, one can use it in several ways: to produce code given code, to generate documentation from code, to turn natural language into code, and so on. Grounded Copilot~\cite{barke_grounded_2023} studies experienced programmers and reports that they prefer using it to turn natural language into code~\cite[Section 4.2.3]{barke_grounded_2023}. Thus our study design focuses on the natural language to code task, but with beginning programmers.

\paragraph{Non-experts and LLMs}

Like us, several researchers have considered the impact of using Code LLMs for the text-to-code task with non-experts, specifically in educational settings. Our work is larger in scale than prior work (120 students from 3 institutions and 48 problems in 8 categories), which allows us to perform statistical analyses that require large sample sizes to be reliable. Moreover, our experiment design allows us to investigate key research questions that prior work has not been able to ask, such as identifying the prompting strategies that beginners use, determining how they modify prompts that do not work, and studying several factors that affect their success.

\citet{prather_its_2023} study 19 students using Copilot for a final project in a CS1 course: building the game Minesweeper. They found that students struggled to use Copilot, even over the course of a week. We reach a similar conclusions with our study, with $48$ problems that are much simpler than building a working video game.

\citet{kazemitabaar_studying_2023} develop CodingSteps, a web-based Python learning environment that allows users to query Codex. The paper compares 33 participants (10--17 years old) with access to Codex to 36 students programming independently, working on the same set of 45 programming problems over several weeks. Their findings indicate that Code LLMs may benefit student learning outcomes. However, because CodingSteps presents students with expert-written problem descriptions, their results do not shed light on whether beginners can write natural language prompts independently. They report that 32\% of student prompts are verbatim copies of the expert-written problem descriptions. In contrast, our study is carefully designed to avoid this problem by showing students input/output examples instead of natural language descriptions. We also investigate the strategies that students use to understand model output and modify their prompts. \citet{kazemitabaar_studying_2023} do not address these kinds of questions, partly because their students received  feedback from instructors throughout the experiment.

Promptly~\cite{denny_promptly_2023} studies 54 students writing prompts for three CS1 problems. Our substantially larger scale (120 students and 48 problems) allows us to explore research questions beyond what they study, such as the how students change their prompting strategies, and demographic factors that influence success rates. Our paper also presents a detailed analysis of LLM output, such as the kinds of errors that appear in LLM-generated code, and the impact of non-determinism on participants' success.

\citet{lau_ban_2023} interviewed 20 CS1/CS2 instructors in early 2023 about their perceptions of ChatGPT and LLM technologies. They report that instructors hold a diverse set of perspectives: some wanted to ``ban it'' and others felt urged to integrate these technologies into curricula to prepare students for future jobs that may require using LLM technology. The students in our study echo many of the concerns and desires raised by instructors in \citet{lau_ban_2023}.

It is also possible to use language models to assist students learning to program, without having the model write code for the student. For example, \citet{geng_novice_2022} use language models to localize type errors in OCaml, but not to correct them. Like our study, this work isolates the interaction mode in which students use Code LLMs; however, we study prompt writing and editing, while they study error detection and explanation.

\paragraph{Alternatives to inline code completion}

Copilot and related tools suggest inline code completions, but there are other ways to interact with AI-assisted programming tools. \citet{vaithilingam_towards_2023} present new interfaces for Visual Studio that present code changes. \citet{liu_what_2023} build a new interaction model, grounded abstraction matching, which targets spreadsheets and data frames,  constraining the generated code to support grounding. These ideas are exciting parallel directions for Code LLM interaction in addition to the natural language prompting approach we study here.

\paragraph{Code LLMs beyond text-to-code}

For a beginning programmer, feedback from an expert teacher or teaching assistant can be invaluable. However, access to expert feedback is limited. There is a long line of research that tries to address this shortage by developing systems that generate actionable feedback for students \cite{singh_automated_2013, rocha_use_2023, hollingsworth_automatic_1960, suzuki_tracediff_2017, head_writing_2017}. \citet{phung_generating_2023} show that LLMs can help build these systems and generate higher quality feedback than prior rule-based approaches. In contrast to our human experiment, they evaluate on benchmark problems. Moreover, their system is intended to help beginners write code directly, whereas our experiment focuses on prompt writing.

Another body of work focuses on automated program repair \cite{goues_automated_2019}, which can be used to fix trivial mistakes that frustrate beginners. Traditional automated program repair systems have required significant engineering for each programming language and problem domain. \Citet{joshi_repair_2023} show that an LLM trained to generate code can be employed to repair simple coding mistakes.

Similarly, \citet{leinonen_comparing_2023} report that Code LLMs are better at explaining code than beginning students, and \citet{leinonen_using_2023} show that an LLMs explanation of a program error can be better than default error messages. This is further evidence that LLM technology may help students learn to write code directly.

Recent additional efforts include \citet{finnie-ansley_robots_2022}, who report that Codex is remarkably good at generating code from natural language prompts from a CS1 class and several variations of the Rainfall Problem; \citet{dakhel_github_2022}, who compare the quality of Codex-generated code to student-written code; and \citet{babe_studenteval_2023}, who use student-written prompts to benchmark Code LLMs.
Finally, Code LLMs have applications that go beyond natural-language-to-code, and
researchers are using them as building blocks for a variety of other tasks~\cite{schafer_adaptive_2023,bareis_code_2022,murali_codecompose_2023,nam_using_2024,ross_programmers_2023,lemieux_codamosa_2023,xia_fuzz4all_2024,joshi_repair_2023,phung_generating_2023,first_baldur_2023,chen_data_2023,ferdowsi_live_2023}.  The aforementioned papers present new tools, benchmarks, and studies of LLM capabilities. But, they do not study users' abilities to prompt models, which is the focus of our work.

\paragraph{Using LLMs for non-programming tasks}

Researchers are currently exploring a wide variety of applications for LLMs beyond computational tasks. While we do not survey the full range of such work, two recent papers are particularly relevant to our task. \citet{zamfirescu-pereira_why_2023} study non-experts prompting an LLM to produce recipes. Their participants actively avoided systemic testing, which we address by automating testing. Like them, we find that participants' mental models of LLMs are very different from how they actually work. 
\citet{singh_where_2022} compare user interactions with a multimedia writing interface with LLM-generated audio, text, and image suggestions. Our post-study interview and survey was inspired by their exploration of participant's perceptions of AI.

\section{Study Design}\label{sec:new_design}

Our work explores whether beginning programmers can effectively prompt Code LLMs. We investigate this question through a multi-institutional \cite{fincher_multi-institutional_2005}, lab-based study, asking 120 students who completed a CS1 course to describe 8 out of 48 possible problems presented via input/output examples. 

In this section, we discuss three major aspects of our study design: 
\begin{enumerate}
\item Why do we use a controlled experiment?
\item How do we successfully present problems to students?
\item How do we select problems that are appropriate for students?
\end{enumerate}
We discuss the logistics of implementing the study in Section~\ref{sec:logistics}.

\subsection{Experimental Environment: In the Lab vs. In the Classroom}
Studies of student interactions with programming tools can be grouped into three main categories: studies within the context of a course during the term, post-hoc analyses of educational data, or controlled, lab-based experiments. Post-hoc analyses are not currently possible, since there is a lack of available educational Code LLM data. We discuss the decision between a course-based study and lab-based study below.

There are many benefits to real-world studies conducted in a course context, including ease of access to participants and normalized educational background \cite{prather_its_2023}. It is easier to study how technology directly impacts learning by using it alongside instruction \cite{kazemitabaar_studying_2023} or as an evaluative method \cite{hsu_attitudes_2021}. At the same time, these studies cannot be as easily controlled: participation may be optional (only around 12\% of students chose to participate in \citet{denny_promptly_2023}); participants may explicitly be learning through the task, making it hard to compare their responses across problems~\cite{kazemitabaar_studying_2023}; and in-depth interviews are challenging to conduct.

Lab-based studies benefit from greater uniformity in observations, which facilitates statistical analysis, and longer experimental sessions. We chose a lab-based experiment because our research questions focus on the \emph{usability} of Code LLMs for beginning programmers and on their \emph{processes}, rather than their educational outcomes. Specifically, the process of working with a Code LLM requires multiple, interdependent steps: (1)~forming an intent, (2)~crafting a prompt to describe the intent, (3)~evaluating the quality of the LLM-generated code, (4)~editing the prompt when the code is wrong, (5)~editing the code manually, or (6)~giving up and writing code manually (\Cref{fig:the-query}). Our goal was to isolate processes (2) and (4). 

Our study limits user interactions in order to isolate prompt writing and editing strategies. One key feature of our paradigm is that we automatically test the generated code. In most observational studies, programmers determine on their own whether the generated code is correct. This is itself an interesting process. However, studying this aspect of Code LLM interaction comes at the cost of studying prompt editing: if a programmer mistakenly accepts incorrect code, they will move on to the next task without editing. \citet{prather_its_2023} report that many of their participants mistakenly accepted incorrect code. Beginning students are particularly likely to err in this way: they may struggle to understand generated code, and their lack of confidence in their own abilities may make them trust the automated system over their own judgment (an example of \textit{automation bias}~\citep{skita_accountability_2000,goddard_automation_2012,gadala_automation_2017,de-arteaga_case_2020}).

Finally, a key contribution of our work is its scale: we study 120 participants across 3 institutions and 48 programming tasks, while previous studies have had fewer participants and problems. We recruit participants from three U.S. institutions: an R1 university (Northeastern University), a small liberal arts college (Oberlin College), and a women’s college (Wellesley College). This selection increases the likelihood that our findings will generalize across institutions. Our scale allows us to explore how diverse factors, such as prior non-curricular programming experience, first-generation status, and mathematics coursework, affect participant success. These kind of statistical analyses require large sample sizes and work best with even observations of participants and problems, which are challenging to obtain in course settings. 

\subsection{How to Describe Problems to Students: Input/Output Examples vs. Written Descriptions}

A key design decision for studies of Code LLM interactions is how to present the task. In classroom environments, students are usually given instructions for what to program via written descriptions. This makes sense, given that the student's goal is to write code. However, natural language presentation poses critical issues for our key research questions. In our study, the goal is to write natural language descriptions of problems, \emph{not} to write code. A core goal is to understand how students approach the natural-language-to-code task. If the task is presented in natural language, students may simply reuse this text rather than putting the task into their own words; our results would no longer measure beginning programmer success, but instead expert description success. Prior work shows that this is a serious concern: in \citet{kazemitabaar_studying_2023}'s study of K-12 students, up to 49\% of submissions for challenging problem categories were copied from the expert-written task description.

Even if participants do not directly copy a description, its wording could influence how participants describe the task. One challenge for beginning programmers is recalling and applying technical vocabulary; presenting them with a natural language description of the task might remind them of terminology that they would not have recalled on their own. This would endanger our goal of assessing beginning programmers' abilities to prompt code generation models, since in many natural settings, they would not have an expert description to rely on.

We therefore rely on a popular alternative for describing program behavior: input/output examples (\Cref{fig:charlie_screenshots}). Students also could reference the function name and parameter names. Our participants had taken CS1 classes where natural language descriptions are frequently accompanied by input/output examples (see Appendix~\ref{app:adaptation}), making this a familiar way of communicating program behavior. Several CS1 courses, and some of the assignments used in our CS1 courses, go beyond this and require students to construct their own examples or even practice test-driven development~\cite{flatt_how_2001,edwards_using_2004}. However, our study does not require students to write their own tests. 


Avoiding natural language presentation is critical in order to study how beginning programmers describe problems in their own words. However, it comes with two risks. First, the input/output paradigm may increase task difficulty, since participants must identify the key pattern on their own. Although understanding natural language descriptions of coding tasks is not always easy for beginning programmers, it is likely easier than our input/output paradigm. Second, input/output examples run the risk of underspecification \cite{shaw_inferring_1975, gulwani_program_2017} -- there may be more than one program that performs the correct input-output mapping. To determine that the provided tests adequately described the problem, we confirmed that our provided test sets had 100\% code coverage for a correct solution and performed mutation testing \cite{jia_analysis_2011}. We also calculated participants' success using only the provided test cases: if the generated code passed the provided tests, it was deemed correct, ensuring that the problem presentation aligned directly with the feedback to the user. 

We feel that these potential issues pose less of a risk to our  key research questions than the copy/paste or word bias risks posed by a natural language presentation. Other researchers have also used an input/output presentation paradigm in studying beginner interactions with Code LLMs~\citep{denny_promptly_2023}.

\subsection{Problem Selection: Previously Seen Tasks vs. New Tasks} \label{sec:problem_choice}

The natural language-to-code task requires participants to describe specific programming problems. Previous work exhibits varied approaches to problem selection, from a single challenging problem in \citet{prather_its_2023} to three simple problems in \citet{denny_promptly_2023} to a set of 45 problems in 5 categories in \citet{kazemitabaar_studying_2023}. 

Our main goal was to select problems at an appropriate level for students who had completed only CS1. Since our research questions focus on student prompting processes, not learning outcomes, we chose problems at a similar level to what participants might be able to code independently. Asking students to solve \emph{new} or more complex problem types increases the likelihood that the Code LLM will generate unfamiliar or difficult to understand code, making the prompt editing process more difficult. We therefore adapted Python problems specifically from CS1 course materials at each institution. We made small changes to facilitate input/output testing or adjust problem difficulty. Appendix~\ref{app:details} contains two examples of how source problems were adapted.

\begin{figure*}[t]

   \includegraphics[width=0.8\textwidth]{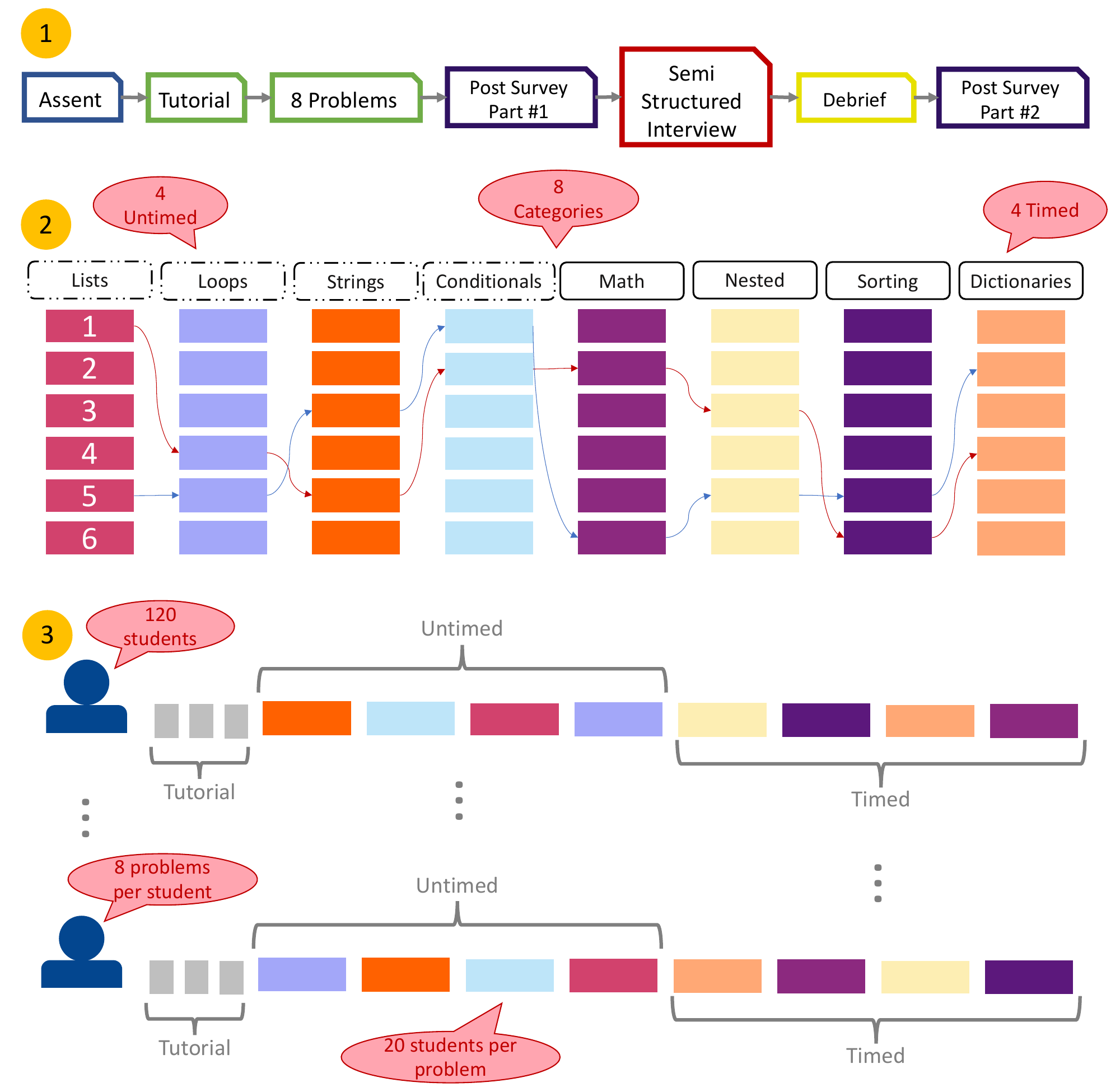}

\caption{Study overview. (1) describes the overall student trajectory through the study. We split the post survey into two sections, divided by the semi-structured interview, to delay collecting demographic information to prevent self-bias. (2) outlines the 8 problem categories (4 timed versus 4 untimed) and the 6 problems per category. Students took individual trajectories through one problem in each category, as shown by the thin arrows. (3) showcases an example trajectory for students through the problems. Students spent, on average, 42.6 minutes (SD=10.6) completing the study, with an average of 26.6 minutes (SD=9.1) on the untimed section and 15.9 minutes (SD=3.3) on the timed section. }
\label{fig:setup}
\end{figure*}

\begin{figure*}
\begin{subfigure}{0.6\textwidth}
\includegraphics[width=\textwidth]{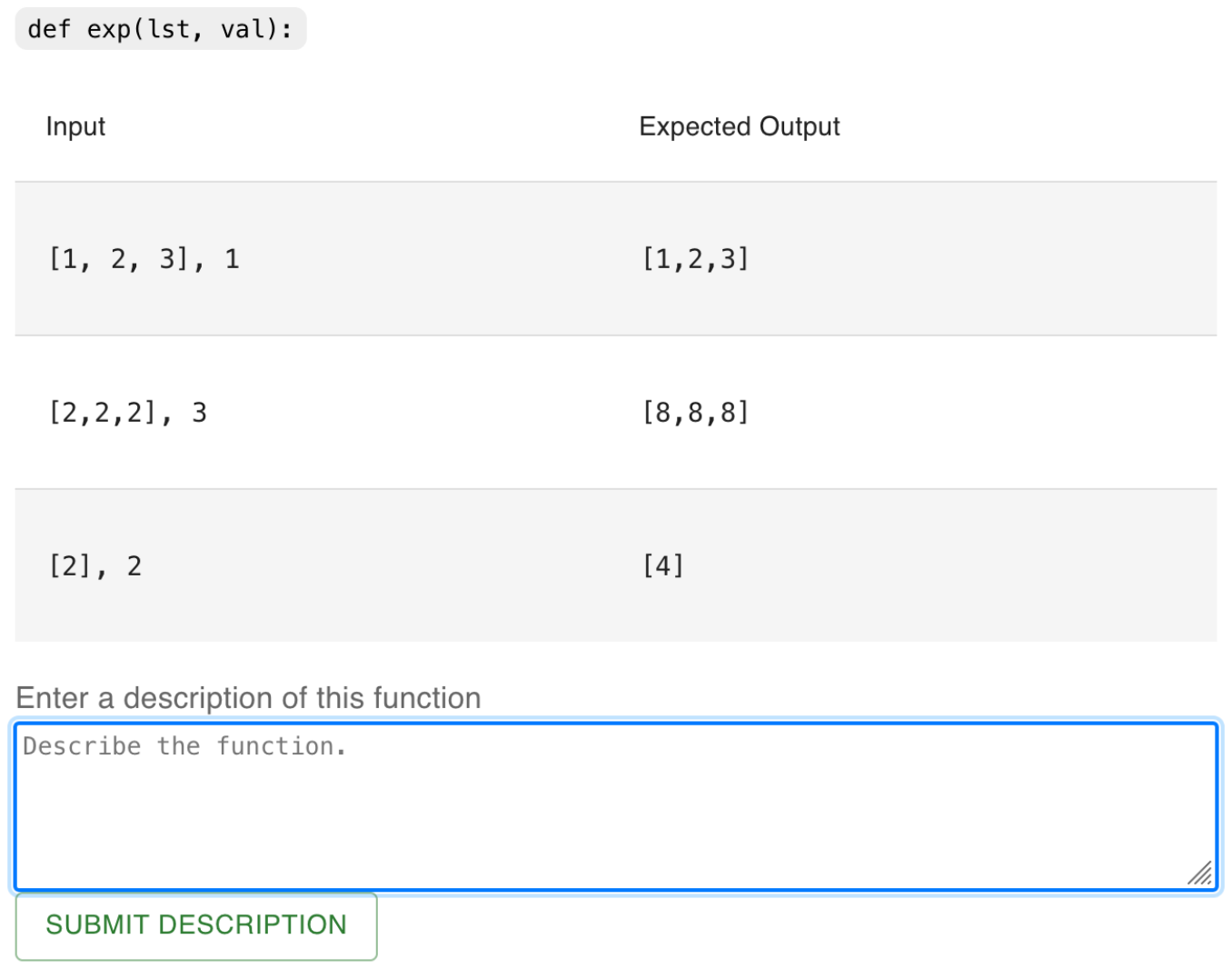}
\vspace{5ex}
\caption{An example task posed to a participant. The interface displays the function name and several input/output examples. Participants write and submit a description in the text box. During our study, 85\% of students who attempted this problem wrote a successful description after a single CS1 course.}
\label{fig:charlie_start}
\end{subfigure}
\hfill
\begin{subfigure}{0.35\textwidth}
\includegraphics[width=\textwidth]{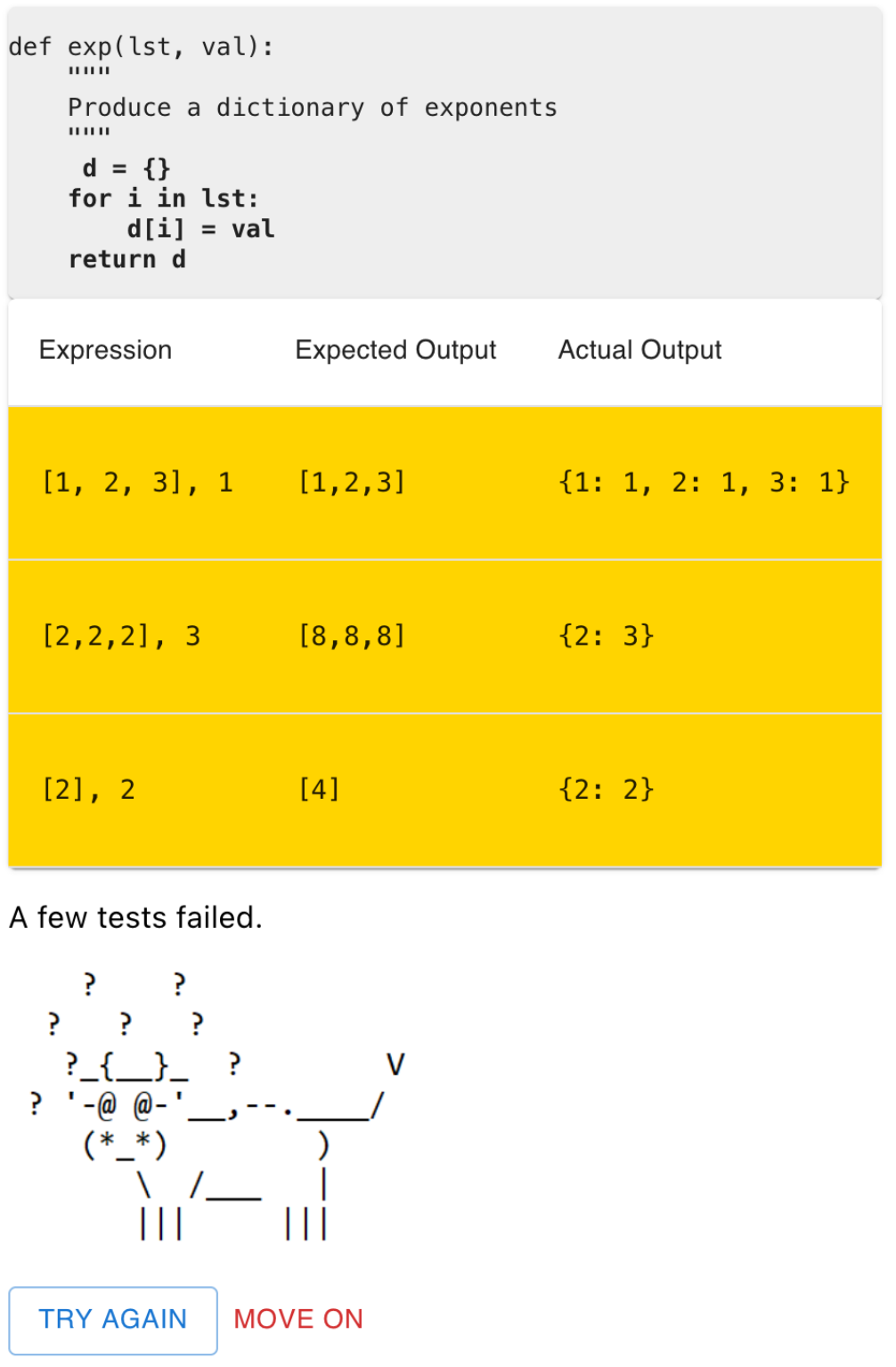}
\caption{We run expert tests automatically and highlight ones that fail. Students are then able to either edit their description by pressing "Try Again" or move on to another problem.}
\label{fig:charlie_fail}
\end{subfigure}
\caption{The \textit{Charlie the Coding Cow} interface.}
\label{fig:charlie_screenshots}
\end{figure*}

We selected 48 problems balanced across eight conceptual categories from CS1 (\Cref{fig:setup}), similar to \citet{kazemitabaar_studying_2023}, but with more categories and problems. Each individual problem was assigned to 20 students; we balanced the experimental lists to control for ordering effects, so that each participant solved one problem in each category, and the average difficulty of each problem list was roughly the same. To facilitate difficulty and category coverage, previous CS1 instructors were asked to provide additional problems as needed. Problems such as \texttt{exp} (Figure~\ref{fig:charlie_screenshots}), for instance, require students to only recognize that numbers in a list are being squared. Other problems ask students to remember complex data structures (e.g. lists, dictionaries), but not the specific Python syntax for them. We further discuss student understanding of the problems in Section ~\ref{sec:hard_problems} and Appendix~\ref{sec:app_difficulty}.

In order to study interactions between Code LLMs and students, it is important to select problems that cannot be trivially solved by a Code LLM without any natural language description. Very common functions (for instance, \texttt{shorten\textunderscore url}) can be solved from a function signature alone, regardless of the accompanying description. To validate our problems, we first checked that the model could not solve problems from their function/parameter names alone and, if they could, edited the names accordingly. We also solved each problem using the Code LLM to ensure that a working natural language description existed. Finally, to address the nondeterminism of Code LLMs, we ran each validation check multiple times to obtain a stable estimate of these results (\cref{subsec:measures}).

\section{Study Logistics}\label{sec:logistics}

The previous section (\Cref{sec:new_design}) described our multi-institutional experimental design. In this section, we discuss the logistics of participant recruitment and executing the study. 

\subsection{Charlie Interface}

\begin{figure}
  \centering
  \includegraphics[width=\linewidth]{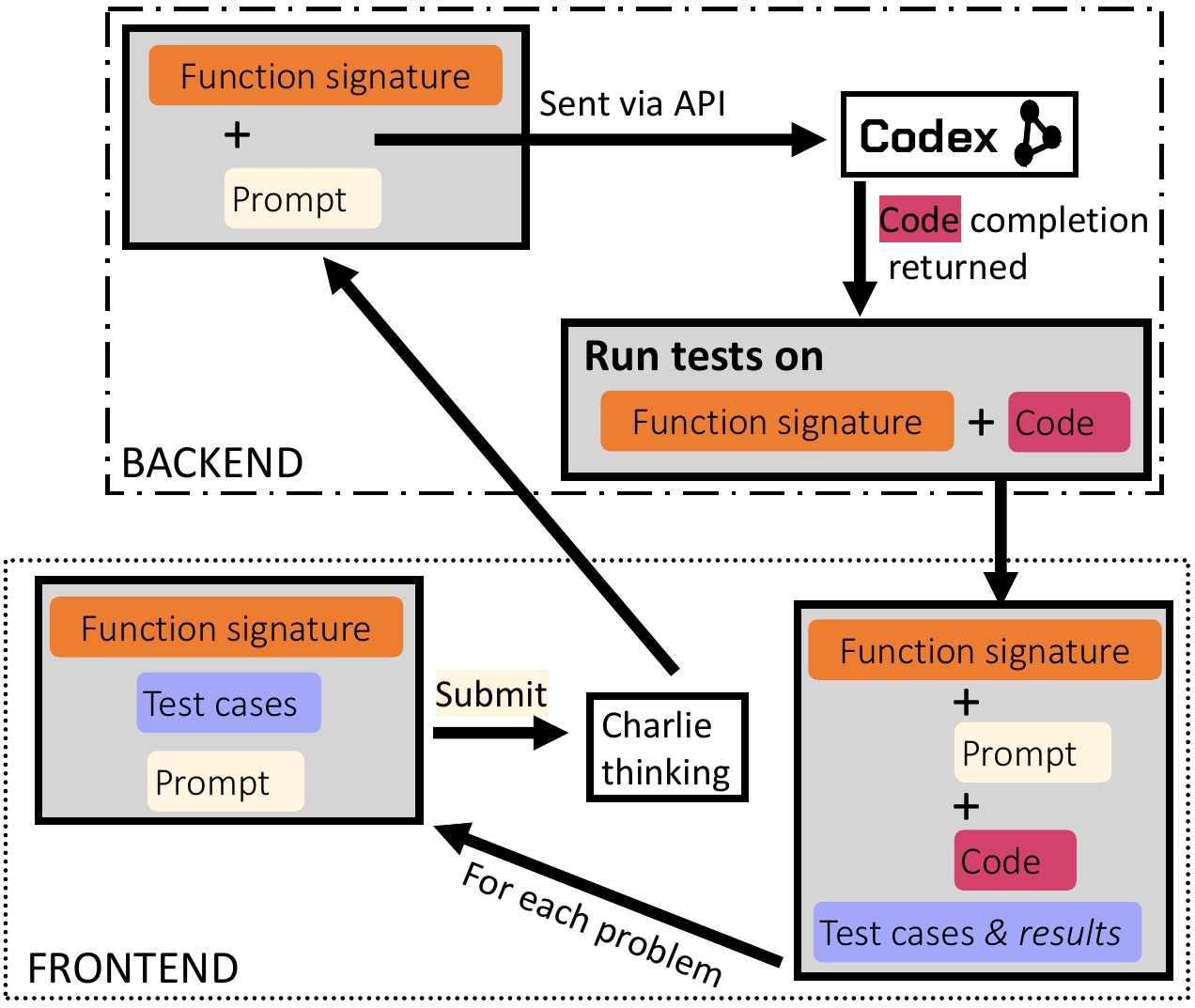}
  \caption{An overview of the experimental platform. For each problem, the frontend provides the participant with the signature and tests and asks them to write a description (prompt). This is then relayed to the backend, where the signature and prompt are sent to Codex via the API. The code completion from Codex is then run on our pre-defined tests. Finally, the results of running the tests and the code completion are presented to the participant in the frontend interface. }
  \Description{A diagram of the Charlie system including interface, backend processing of input and connection to Codex, reprocessing data, and feedback to the user}
  \label{fig:charlie_system}
\end{figure}

We built a web application for the experiment called \emph{Charlie the Coding Cow} or \emph{Charlie}.
Charlie presents one problem per page, displaying the function signature and several input/output examples (\Cref{fig:charlie_start}). Participants write natural language descriptions in a text box. When they submit a description, the Charlie server prompts Codex with the function signature and their description formatted as a docstring (\Cref{fig:charlie_system}). After Codex responds, Charlie shows students the Codex-generated code and displays whether it works on the given input/output examples (\Cref{fig:charlie_fail}).

Charlie does not permit participants to edit the generated code, since we are focused on natural-language-to-code interactions. If the code fails, they can retry the problem or move to the next problem. For retry attempts, we pre-fill the text box with their last prompt to make editing easier. Finally, after every final attempt at a problem, Charlie presents two forced-choice questions with thumbs-up / thumbs-down answers: \emph{Did Charlie generate correct code?} and \emph{Would you have written this code yourself?}. We included these questions to gather information about student perceptions of code style, since the model may produce working code, but in a style that is unfamiliar to students.

Each student worked with Codex to solve 3 tutorial problems and 8 main problems. We used the Charlie character to provide distance from any AI system that students might already know. This suggested a representation that was not human and not robotic. Charlie also provides visual feedback: Charlie animates a ``thinking'' position while Codex generates a completion and appears in different forms when the code does or does not pass all tests. We made this design choice to mitigate frustration with waiting for the model to generate code, a source of annoyance in prior studies of Code LLM interactions~\citep{murali_codecompose_2023}.

\subsection{Model Choice}\label{sec:llm_setup}

When we began piloting in November 2022, the most capable Code LLM was the largest Codex model from OpenAI, \texttt{code-davinci-002}. Although \texttt{code-davinci-002} was first released in 2021, on established Python programming benchmarks, it remains as good as \texttt{gpt-3.5-turbo}, which is the model presently used for GitHub Copilot's inline completions~\cite{zhao_github_2023}, the free version of ChatGPT, and several other commercial products. Specifically, \texttt{gpt-3.5-turbo} and \texttt{code-davinci-002} score 48\% and 46\% respectively on the HumanEval Python programming benchmark~\cite{openai_gpt-4_2023,cassano_multipl-e_2023}, the most commonly used Python benchmark for Code LLMs. Since we started our study, several other LLMs have also appeared, including non-proprietary LLMs that are better for reproducibility (\Cref{sec:open_models}). The best open models perform comparably to \texttt{code-davinci-002}; for instance, CodeLlama (34B) achieves 48\% on HumanEval \citep{roziere_code_2024}. This suggests that the model that we use is as capable at code completion as newer models used in practice.

There are larger models that are more capable, such as GPT-4, which achieves a HumanEval score of 67\%~\cite{openai_gpt-4_2023}. However, GPT-4 is significantly slower and higher latency than the alternatives, and low latency is essential for LLM code completion to be acceptable to users~\cite{murali_codecompose_2023}; if participants have to wait more than a few seconds for the generated code, their frustration might lead them to move on rather than re-attempting the problem.

For consistency, we used the same Codex model throughout the study 
     (\texttt{code-davinci-002}). It is important to note that Code LLMs perform best when their output is sampled; consequently, the model may produce different programs for the same prompt. We generated output using best practices for hyperparameter and sampler settings~\cite{chen_evaluating_2021}.

\subsection{Participants}\label{sec:care_for_participants}

We recruited 40 participants from each institution (n = 120). Eligible participants were at least 18 years old, had taken CS1 at their institution between Fall 2021 and Spring 2023, and had not completed any subsequent CS courses. We recruited participants from March to July 2023 until reaching our sample size of 120. The pilot and main study received IRB approval.



\paragraph{Care for Participants}

Our study design sought to balance obtaining accurate data with addressing potential discomforts and power dynamics. Potential discomforts for participants included frustration regarding their inability to complete a task, which could reinforce negative perceptions of self or CS. In the tutorial, we emphasized that our goal was \emph{not} to evaluate their programming skills, but the collaboration with Charlie. Students were allowed to move on from a problem at any time, resulting in a variable number of attempts per problem.

We took several steps to address potential power dynamics between students and their professors. Recruitment was done through an interest form distributed by other faculty or staff. Scheduling was performed by a researcher at another institution. Finally, research sessions were never run by a professor at the same institution as the participant.

\subsection{Study Execution}\label{sec:study_design}

The study was conducted over Zoom with audio and video recording. Participants signed informed consent material ahead of the experiment and assented at its start. They were compensated with a \$50 gift card for the estimated 75-minute study.

\paragraph{Main Task}

\Cref{fig:setup} (1) outlines the full study design. Students completed 3 tutorial problems to get familiar with the interface and see some possible Codex responses. We supplied participants with a working prompt for the first tutorial problem, then gave them a difficult problem so they could see a failure, and a final easy problem to solve independently. 

The main experiment consisted of 8 problems in two blocks, the first untimed, the second timed. In the second block, students were limited to 5 minutes per problem. We included both timed and untimed blocks in order to balance the need to bound study duration with the desire to observe complete prompt editing cycles. 

Participants were randomly assigned experimental lists, balanced by difficulty, using a modified Latin Square design. Four authors independently assessed the difficulty of writing prompts for each problem; we averaged these scores and developed six roughly equal lists (\Cref{fig:setup}). 

\paragraph{Post-task Interview and Survey}

After the main study, students completed a two-part survey, a semi-structured interview, and an optional debriefing session (\Cref{fig:setup} (1)). The semi-structured interview was interleaved between two survey blocks to mitigate question ordering and priming biases. 

The first part of the survey was designed to study student perceptions of Charlie and of AI more broadly. We adapted validated scales from previous work to understand student perceptions of the usability, trustworthiness, and friendliness of Charlie~\citep{bartneck_measurement_2009,korber_theoretical_2018,wang_towards_2021,druga_how_2021} and the mental workload of the task~\citep{hart_development_1988}.\footnote{In some cases, we removed questions that were not relevant to our study to keep the survey length manageable for participants. Details available via our Supplemental Materials at \osf.} We were also interested in whether students' ability to come up with effective prompting strategies might correlate with fixed versus growth mindsets about computing; we drew on \citet{gorson_why_2020} to measure this. 

The semi-structured interview asked 8 questions covering student editing processes, what they found hard or easy, how they envisioned their interactions with Charlie, and how they imagined Charlie worked. The specific questions were directly inspired by our overarching research questions. Researchers followed a standing script to ask each question - there are a total of 5 missing question responses across the possible 960 interview datapoints, likely due to researcher error or time considerations. In the optional debriefing, we explained the experiment and how Code LLMs work. 

The second part of the survey focused on participants' backgrounds and demographics. These were the last questions of the study to mitigate possible stereotype threat \cite{ncwit_2023}. For questions related to identity (e.g., gender, race, spoken language at home), we followed best practices and solicited responses via open text boxes \cite{spiel_how_2019}. We also asked questions about students' CS1 performance, experience with programming outside of CS1, high school \& educational background, math background, major, and class year. 

\paragraph{Pilot Study}

In late 2022, we ran a pilot study with 19 participants to assess the study design and usability of the interface. Pilot participants were recruited from the same three institutions as in our main study, but were students who had taken more than one CS course. This small pilot allowed us to make sure the web platform was working correctly, identify any problems with specific tasks, refine our time estimates, and assess the quality of the automatic transcriptions of the interview recordings produced by otter.ai.\footnote{\url{https://web.archive.org/web/20231205001012/https://otter.ai/}} During the pilot, we identified one problem with ambiguous test cases, which we changed before the main study. Pilot participants solved an average of 5.5 out of 8 problems (an Eventual Success Rate of 68.8\% using the metric described in \Cref{subsec:measures}).

Because the average pilot participant took 53 minutes, we increased the time estimate and compensation from \$30 for 60 minutes to \$50 for 75 minutes for the main study. We also added a hidden time limit to the first block of questions in case participants spent more than 50 minutes on this portion of the study; this issue never arose in the main study. 

\section{Analysis}

This section presents the analysis framework for \Cref{sec:rq1}, \Cref{sec:rq2}, and \Cref{sec:rq3}. We take a mixed-methods approach to this work.





\subsection{Evaluation Plan}

\paragraph{Qualitative analysis} We collected three types of data which lend themselves to qualitative analysis: (1) information about student experience and demographics, (2) free-response questions about future use of Charlie, and (3) semi-structured interview responses. We employed both inductive and deductive open coding towards consensus. Our aim was to identify common themes present in this specific dataset, rather than to develop a theory. Two researchers with previous qualitative experience conducted the analysis; Section~\ref{app:coding} contains details of the coding methodology. We present selected quotes from the surveys and interviews throughout. Quotations have been lightly edited from the automatically generated transcripts. This includes addressing grammar/punctuation, removing speech errors or filler words, and avoiding the disclosure of any identifiable information. Each participant's quote is accompanied by a pseudonym assigned to them during data collection.

\paragraph{Statistical analysis} We perform statistical testing with a significance level of $\alpha$=0.05 in order to determine whether observed differences in response measures are statistically reliable. For comparisons between two groups, we use Student's $t$-test. For comparisons between multiple groups, we perform ANOVAs; in cases where there is no natural reference group, we use Tukey HSD tests to explore pairwise differences. We report Pearson's $r$ for correlations between continuous variables and Kendall's $\tau$ for correlations between continuous and ordinal variables. Where we are interested in multiple potentially interacting variables, we fit linear mixed-effects models with maximal random effects for participants and problems using the lme4 package in R~\citep{bates_fitting_2015}.

\begin{figure*}[t]

\begin{subfigure}{0.7\textwidth}
\centering
\footnotesize
\begin{tabular}{c|r|r|r}
 & &  \\
     Institution&Mean pass@1& Success Rate & Eventual Success Rate\\ \hline
     \oberlin &0.23 & 26\% & 61\% \\ \hline
     \wellesley &0.23 & 25\% & 57\% \\ \hline
     \northeastern &0.20 & 23\% & 54\% \\ \hline
     Overall       & 0.22 & 24\% & 57\% 
\end{tabular}
\caption{Mean values of different measures of success.}
\label{tab:success_pass_by_institutions}
\end{subfigure}

\begin{subfigure}{0.3\textwidth}
\includegraphics[width=\textwidth]{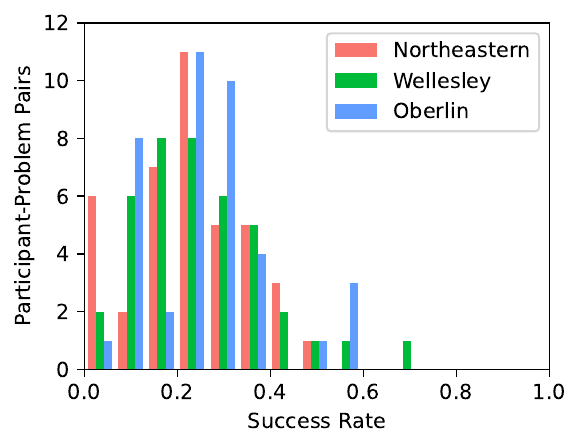}
\caption{Success rates.}
\label{success-rates-hist}
\end{subfigure}
\begin{subfigure}{0.3\textwidth}
\includegraphics[width=\textwidth]{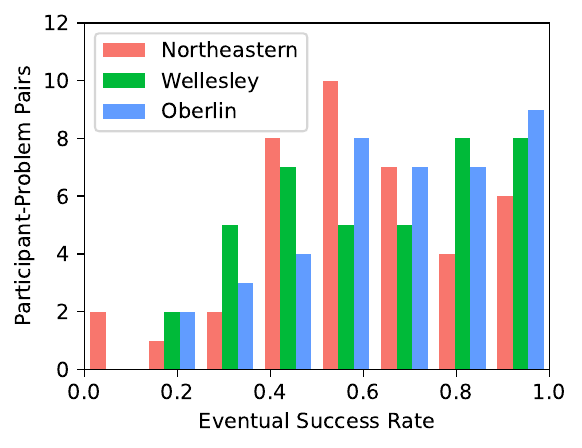}
\caption{Eventual success rates.}
\label{eventual-success-rates-hist}
\end{subfigure}
\begin{subfigure}{0.3\textwidth}
\includegraphics[width=\textwidth]{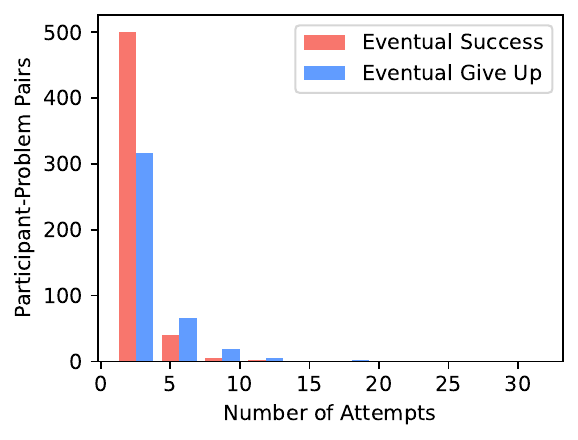}
\caption{Attempts.}
\label{attempts-hist}
\end{subfigure}

\caption{Basic measures of student success at the natural-language-to-code task. \emph{Success rate} is the fraction of all attempts by a participant that succeed. \emph{Eventual success rate} is the fraction of last attempts at a problem by a participant that succeed. Pass@1 resamples the LLM several times to estimate the probability of success. We present these measures by institution. \Cref{tab:success_pass_by_institutions} presents the means. \Cref{success-rates-hist} and \cref{eventual-success-rates-hist} show the distribution of (eventual) success rates. Eventual success rates are higher than success rates, which is to be expected: \Cref{attempts-hist} shows that many students make several attempts at a problem before an eventual success or give up. }
\label{fig:basic_findings}
\end{figure*}

\subsection{Measures of Success}\label{subsec:measures}

There are several ways to measure success when evaluating the natural-language-to-code task. The \emph{success rate} is the fraction of all attempts on which the model generates a working program. Therefore, a participant who takes several attempts to solve a problem will have a lower success rate than another who succeeds in one try. We might also ask whether a participant is ever able to solve a problem; we refer to this as the \emph{eventual success rate}. This metric considers only the participant's final attempt at each assigned problem. The eventual success rate metric is likely specific to this paper, as closely related work \cite{prather_its_2023, denny_promptly_2023, kazemitabaar_studying_2023} studies different notions of success or does not permit controlled, repeated interactions.

Although success rates measure the correctness of the code that students saw during the experiment, LLM generation is non-deterministic.\footnote{Greedy generation is significantly worse for coding tasks than non-deterministic generation~\cite{chen_evaluating_2021}.} Therefore, studying success rates can be misleading: a participant may have just been lucky with a bad prompt or unlucky with a good prompt. For this reason, we also employ an alternative metric called \emph{pass@1}, which accounts for non-deterministic generation~\cite{chen_evaluating_2021}. Since the debut of Codex,
pass@1 has become the standard metric used to evaluate LLMs on the natural-language-to-code task, including GPT-4~\cite{openai_gpt-4_2023}, Code Llama~\cite{roziere_code_2024}, and other models~\cite{li_starcoder_2023,fried_incoder_2023,nijkamp_codegen_2022}.

Given a natural language prompt, \emph{pass@1}~\citep{chen_evaluating_2021} is an estimate of the probability that the LLM will generate working code in one attempt. In the LLM development literature, the accepted best practice for computing pass@1 is to query the LLM 200 times for the same prompt and test every generated program~\citep{chen_evaluating_2021,fried_incoder_2023,xu_systematic_2022,roziere_code_2024}. Sampling 200 generations for all 2,000+ prompts generated as part of this study would be very expensive with the Codex API. Instead, we use a recently released open Code LLM called 
StarCoder~\cite{li_starcoder_2023} that is nearly as capable as the Codex model on Python benchmarks.
Pass@1 with StarCoder will be slightly lower than Codex success rates because of model differences. However, pass@1 is a more stable measure of whether a prompt will succeed than success rate. We use pass@1 for the bulk of our analyses.

\subsection{Positionality}\label{sec:positionality}
All authors were affiliated with the institutions from which participants were recruited (\oberlin, \wellesley, or \northeastern) at the time of the study; we range from undergraduate students to tenured faculty. We developed the problem lists, problem difficulty ratings, and other elements of the study design within a shared educational context. The last three authors are course instructors for CS1. As described in \Cref{sec:care_for_participants}, significant care was taken to address power dynamics between participants and researchers. Some authors also contribute to the development and evaluation of open-source Code LLMs. Overall, the potential incentives for the research team are complex, as we approach this work as both educators and researchers. We aspire to a neutral perspective on Code LLMs, while attempting to center the student experience. 

This research studies students at three selective higher education institutions in the United States. Therefore, while we are able to generalize beyond a single CS curriculum, the educational context is specific: our findings may not generalize to other settings (e.g., community colleges, K-12 education) or cultural contexts.



\section{RQ1: Do Students Succeed at Prompting Code LLMs with Natural Language?}\label{sec:rq1}

In this section, we present how well students do on our Code LLM prompting task and address RQ1: do students succeed at prompting Code LLMs with natural language? We explore differences between students that are linked to their ability to successfully describe problems to Code LLMs.

\subsection{Basic Findings}\label{sec:basic_findings}

\Cref{fig:basic_findings} presents the distribution of participants' success rates and eventual success rates. The average participant solved 4.7 out of 8 assigned problems.  The mean eventual success rate (57\%) is not high, and the mean success rate (24\%) is even lower, since it decreases with every failed attempt. We find no significant institutional difference for either measure of success. 

Participants often submitted a large number of failing attempts (\Cref{attempts-hist}): 153 problems (aggregated across participants) required three or more attempts. In fact, one participant succeeded at a problem only after 32 attempts; another gave up after 26 attempts.
These results suggest that low success rates are not due to a lack of participant effort. Participants struggled to write natural language prompts for the LLM, and often achieved success only after many repeated attempts. The challenging nature of this task is supported by comments from the students themselves (\Cref{sec:student_says_hard}). 

\begin{table}[t]
    \centering
    \begin{tabular}{c|c|c}
         Self-Reported Background&N&Mean pass@1\\\hline
         International&92&0.23\\
         Domestic&27&0.22\\\hline
         First-generation college student&23&0.17\\
         Not first-generation&96&0.23\\\hline
         Attended private high school&38&0.22\\
         Attended public high school&76&0.22\\\hline
         Raised Monolingual in English&49&0.22\\
         Raised Monolingual Not in English&27&0.20\\
         Raised Multilingual Including English&41&0.24\\
         Raised Multilingual Not in English&2&0.22\\
    \end{tabular}
    \caption{Self-reported high school, language, and family background.}
    \label{tab:demo_background}
\end{table}


\subsection{Do Participants Find the Task Challenging?}

In the post-survey, participants completed four items of the NASA TLX \cite{hart_development_1988}. Overall, students found the task mentally demanding (\Cref{tab:nasa}). The questions about mental demand (Q1), time pressure (Q3), and their own performance (Q4) correlate inversely with success rate. Students whose success rates were lower generally rated the task as more demanding (Kendall's $\tau$=-0.16; $p$=0.02); were less likely to say they were successful (Kendall's $\tau$=-0.4; $p$<0.0001); and reported higher levels of stress and insecurity (Kendall's $\tau$=-0.27; $p$<0.0001).

\begin{table*}
    \centering
    \begin{tabular}{l|l|l}
    Abbreviated Question&Scale (1 to 7)&Mean\\\hline
         How mentally demanding was the task?&Very low->Very high&4\\
         How hurried or rushed was the pace of the task?&Very low->Very high& 3.3\\
         How successful were you?&Perfect->Failure& 3.6 \\
         How insecure, stressed, or discouraged were you?&Very low->Very high&3.1 \\
    \end{tabular}
    \caption{Mean NASA-TLX ratings \cite{hart_development_1988}.}
    \label{tab:nasa}
\end{table*}

\subsection{Who Succeeds at the Task?}\label{sec:who_succeeds}

Using data from the post-survey, we analyze the relationship between pass@1 rates and previous knowledge, prior programming experience, and demographics (see \Cref{tab:demo_background} for a summary of demographics). We find only two statistically reliable differences (see Appendix, Table~\ref{tab:dem_ttests} for the full statistical analyses):
\begin{itemize}

\item \textbf{Prior programming experience:}  About 1/3 of participants had no programming experience outside of CS1.
The remaining participants had taken pre-college programming courses (24\%), were in the next CS course (21\%), or had coding experience outside of classes (29\%). There is a statistically reliable difference (t-test; $p$ = 0.02) in pass@1 for students who have only coded in CS1 (0.17) versus those with additional experience (0.24).

\item \textbf{First-generation college students:}  19.1\% of participants identified as first-generation college students. We observe a statistically reliable difference in pass@1 for first-generation participants, who struggle more with the task than others (t-test; $p$=0.04).

\end{itemize}

We examined other factors, but found no significant difference in pass@1 rates:

\begin{itemize}
    \item \textbf{Math courses:} All but one participant had taken at least one college math course and half had taken 2+ courses. Single variable calculus was the most common math course. There is no statistically reliable difference between participants who had or had not taken 2+ math courses  (t-test, $p$=0.42).

    \item \textbf{Computing intensive majors:} 42\% of participants were pursuing computationally intensive majors. We observe identical pass rates for both computing and non-computing majors.

    \item \textbf{International students:} International and U.S. domestic students had similar pass@1 rates.

    \item \textbf{Household language:} Our participants reported growing up in households where a diverse set of languages were spoken: only English (40.8\%), English and other languages (34.2\%), and without English (24.2\%).
    We were surprised to find that pass@1 did not reliably vary by childhood language.
    However, all participants were from selective U.S. institutions that require fluency in English, regardless of childhood language exposure.

    \item \textbf{Public vs private high schools:} 1/3 of participants attended private schools; this had no
    impact on pass rates.

\end{itemize}

\section{RQ2: Where do student difficulties come from?}\label{sec:rq2}

Having shown that students find it hard to prompt a Code LLM in natural language (\cref{sec:rq1}), we explore why. In this section, we present quantitative and qualitative results that address RQ2: when students struggle with the task, where do the struggles come from? What are the most challenging aspects of the natural-language to-code task?

\subsection{What aspects of the task do students say are hard?}\label{sec:student_says_hard}

In the semi-structured interview, we asked participants to reflect on challenges and issues they encountered.  Three common themes emerged: difficulties in getting Charlie to understand them; issues with the generated code; and issues stemming from students' self-reported lack of knowledge or skill (\Cref{tab:issues}).

\begin{table}[t]
    \centering
    \begin{tabular}{l|l}
         Thematic Codes&N\\\hline
        Charlie Doesn't Understand Me&91\\
        Issues With Generated Code&59\\
        Student Struggles&41\\
        No Problems Mentioned&10\\
        Issues with Study Platform&10\\
        Issues With Experimental Design&7\\
        Easier To Write Code Myself&7\\
    \end{tabular}
    \caption{Thematic codes emerging from responses to \textit{What kinds of problems or issues did you run into working with Charlie?}}\
    \label{tab:issues}
\end{table}

\paragraph{Charlie Doesn't Understand Me} The most commonly raised issues related to Charlie's understanding of prompts (n=91); we divided these into subcodes. One of the most common of these was the sentiment that Charlie failed to understand good descriptions (n=23). For instance, \pseudonym{redCoyote} commented, \textit{``It was definitely difficult to have a concept of what you wanted written in your head, and then feel like you're articulating it well, but having it not work properly.''}  Similarly, \pseudonym{aquaLadybug} reports feeling helpless when a good prompt didn't succeed: \textit{``if I was saying it [...] how I thought [...] is the best way to say it, but it still wasn't working, I had no idea where to go from there.''} 


\paragraph{Issues with Generated Code} Another major theme was issues with the generated code. Many commments related to perceived bugs in the generated code or difficulty debugging (26\%). Students also mentioned finding the model's randomness frustrating (8\%). \pseudonym{khakiBee} was alarmed to find that resubmitting the same prompt could generate different programs, commenting \textit{``You feel like you've made progress, and then because it did a different thing the next time, it's like, what do I change? I'm trying to change what I give to the cow. And then that should change what the cow is doing. But if I'm not changing anything, why is that changing?''} Some students also experienced the opposite issue: despite changing their descriptions, the model generated the same incorrect function repeatedly. \pseudonym{purpleCarp} commented, \textit{``Sometimes I changed my [...] description and it just repeated the code the same. And it's just very frustrating''}. This highlights the difficulty of working with stochastic models: students expect the model output to be faithful to their descriptions.


\paragraph{Student Struggles} Participants also reported issues stemming from their own lack of knowledge. 10\% of students reported difficulty understanding a problem, and 8\% reported difficulty in understanding generated code. \pseudonym{yellowChipmunk} said, \textit{``sometimes with the code, just given my knowledge, that's not necessarily the way I would go about coding the code. But I think to even understand it, I would have to know what the code is trying to do, which takes more time than me just trying to reword what I said''}. A handful (n=4) reported that forgetting terminology made it hard to write prompts.

\subsection{Which Problems Do Students Say Are Hard?}\label{sec:hard_problems}

\begin{table*}[t]
    \centering
    \begin{tabular}{c|c|c|c}
         Category&Mean pass@1&Mean Eventual Success Rate& Student Difficulty Ranking\\\hline
         Sorting* &0.09&33\%&1 (Hardest)\\\hline
         Dictionaries* &0.17&43\%&2\\\hline
         Nested* &0.30&68\%&6\\\hline
         Math* &0.16&54\%&4\\\hline
         Loops&0.13&52\%&3\\\hline
         Lists&0.18&61\%&5\\\hline
         Conditionals&0.33&73\%&7\\\hline
         Strings&0.26&74\%&8 (Easiest)\\\hline
    \end{tabular}
    \caption{Pass@1 and success rates by problem category. Each category has six problems, and an equal number of students attempted each problem. The starred (*) problems were timed. Student Difficulty Ranking is done by ordering mean Eventual Success Rate from least to greatest, as that provides as measure of what percentage of students successfully solved a given task.}
    \label{tab:success_pass_by_category}
\end{table*}

Some categories of CS1 problems may be harder to solve with Code LLMs, either because the concepts are difficult or because they are difficult to describe. We examine pass@1 and eventual success rate by category as well as interview responses about which problems were challenging and easy. 

We find that pass@1 and eventual success rates both vary by category (\Cref{tab:success_pass_by_category}). We fit a binomial mixed-effects model to prompt success (1 if the prompt succeeded; 0 otherwise), with fixed effects of category, institution, and their interaction, and random effects of problem and participant (see Appendix, \Cref{tab:mixed_effects}). A statistically reliable difference in success was observed only for Sorting problems, which were the most challenging ($p$=0.045). Participants from \oberlin struggled more in the Nested category compared to other students, but the effect is not statistically reliable ($p$=0.063).

Interviews provide insight into their post-task perspectives. The most commonly mentioned easiest category was Math (n=21), whereas the most common for hardest was Nested (n=19), followed by Dictionaries (n=14). These do not match the ranking in \Cref{tab:success_pass_by_category}, suggesting a disconnect between student performance and perceptions of difficulty.

A common theme that emerged related to the challenge of putting understanding of the problem into English (n=44). \pseudonym{crimsonVole} said, \textit{``the ones that had huge lists of like, strings, and integers, were really hard to solve, because they were really hard to describe for me.''} We differentiated this code both from students' ability to identify patterns (n=35) and their ability to write the code without Charlie (n=8). The opposite code, Easy to Describe, applied to 36 responses from the easiest question: \textit{``I felt like time ones because they're pretty straightforward. They're like [...] exercises that we do in my Intro CS class. And so I guess it will be easier for me to word, the description or my thinking process, like I guess that might be easier.''} (\pseudonym{yelllowWeasel}). 

Three codes that related to student's lack of knowledge emerged, with 27 responses (see \Cref{sec:novel_Python} for more perspectives). 

\subsection{What Role Does the Model Play?}
\label{model-role}

LLMs can fail in surprising ways. We now explore the kinds of model failures that participants encountered.

\subsubsection{Syntax errors}

Contemporary Code LLMs generally produce syntactically well-formed programs. However, 5.5\% of student prompts led to Python syntax errors. We manually examined and categorized them:
\begin{itemize}
  \item 27 generations: Codex produces degenerate, repetitive text~\cite{holtzman_curious_2020} or Python 2 print statements. These are model failures.
  \item 81 generations: Codex could not generate a complete function within the 256 token limit ($\approx$800 characters). Our problems are simple enough to be solvable in far fewer tokens, so increasing the token limit is unlikely to help.
  \item 88 generations: Codex generates incomplete code after a complete function, even with standard stop tokens.
\end{itemize}

The latter two categories arise from a trade-off in system design: the first when the interface does not request enough tokens from the Code LLM; the second when it requests so many that the model generates extraneous additional code. Although these errors are infrequent, they are hard for students to deal with. In 22.4\% of these cases (n=44), students gave up after seeing the syntax error. 

\subsubsection{When the Model Produces Different Programs From the Same Prompt}

Codex is best at coding when its output is sampled  (\cref{sec:llm_setup}), but this stochasticity can frustrate students trying to modify prompts.
In 107 cases (4.2\%), a student submitted a prompt several times, and in most of these cases, Codex generates a new completion. A few of these are trivially different (e.g., different variable names), but most (n=86) are different functions. Some students pointed this out in the interview -- \pseudonym{beigeHalibut} noted that they \textit{``usually would run a couple times, because Charlie is not very consistent with the answers. And sometimes it works. Sometimes it wouldn't work.''}

\subsubsection{When the Model Produces the Same Program Despite Changes to the Prompt}\label{sec:sameprogramdiffprompt}

When the Code LLM produces an incorrect function, and a user edits their prompt,
their intent is to have the LLM produce a different---hopefully
correct---function. Frustratingly, this does not necessarily happen: sometimes the model repeatedly generates the same code despite edits to the prompt.
We observe many instances where this happens (104 submissions, 11\% of total): it occurs in most problems (36 of 48 problems) and is
encountered by a majority of students (72 of 120 students).
This often leads students to give up.
In fact, out of the 340 problems where students gave
up, 70 were cases where the participant
edited the prompt and the LLM repeatedly generated the same code.


\subsection{What Do Students Do When They Encounter Unfamiliar Python?}\label{sec:novel_Python}

\begin{figure}
    \centering
        \begin{tabular}{|c|m{0.3\textwidth}|}
            \hline
            Completion &
\begin{verbatim}
def exp(lst, val):
    return [i ** val for i in lst] 
\end{verbatim}
            \\\hline
            Question & Is this code you would write yourself? \\
            \hline
            Student Responses & \wellesley: Yes, \oberlin: No\\
            \hline
        \end{tabular}
    \caption{An example code completion for the problem \texttt{exp} -- this was generated by multiple different prompts. The completion was rated differently by \oberlin and \wellesley students, likely due to the list comprehension.}
    \label{fig:same-comp-diff-naturalness}
\end{figure}

Code LLMs are trained on online repositories of code and may generate code using language features that students have not seen before.

\paragraph{New Python Constructs.}

In their interviews, some students (n=5) report issues understanding code due to unfamiliar language features. \pseudonym{oliveBear} comments about the lambda construct for anonymous functions: \textit{``I've only ever seen [it] in passing. And so if that hadn't worked, I wouldn't have known what the problem was because I myself don't know how to use that operator.''} Others mentioned \texttt{map}, \texttt{replace}, and \texttt{try}/\texttt{except}. List comprehensions are an interesting case because \wellesley teaches them, but \oberlin does not. When asked about generated code with list comprehensions, 9/24 (37.5\%) \oberlin students indicated that it is similar to code they would write themselves, compared to 20/33 (60.6\%) \wellesley students. Some students responded differently to the same completion (\Cref{fig:same-comp-diff-naturalness}). 

\paragraph{Ratings of Final Completions.}
Students evaluated the correctness and naturalness of the final completion for each problem, producing 960 responses. 
For correctness, 61.8\% of the time students indicated that Charlie's code was correct; the majority (543; 91\%) are cases where all tests passed. However, naturalness responses were more mixed. Students indicated that Charlie's code was like code they would write themselves only 58.3\% of the time. 78.6\% of such responses were made when the code passed all tests. Responses to these questions might diverge when the model generates correct code that is unfamiliar or approaches a problem differently, as well as in cases where the model's code is incorrect, but looks familiar to students.

\section{RQ3: Students' mental models and processes}\label{sec:rq3}

This section addresses RQ3, presenting results related to participants' perceptions of the task, their mental models of Charlie, and their strategies for writing prompts. 

\subsection{How does Charlie work, according to students?}\label{subsec:mental}

In interviews, students were asked how they thought Charlie worked (\Cref{tab:charlie_works}). Comments fell into two broad themes: descriptions of Charlie's knowledge, and descriptions of Charlie's processes. 

\begin{table}[t]
    \centering
    \begin{tabular}{l|l}
    Thematic Codes&N\\\hline
    Knowledge: Keywords - General&30\\
    Knowledge: Keywords - Database/Dictionary&16\\
    Knowledge: ChatGPT&17\\
    Knowledge: Internet Data&12\\
    Knowledge: Intermediate Representation&4\\
    Knowledge: Copilot/Codex&2\\
    Process: Sequential&13\\
    Process: Translation&13\\
    No Guess&13\\
    N/A&24\\
    \end{tabular}
    \caption{Thematic codes emerging from responses to \textit{How did you imagine that Charlie was working?}}
    \label{tab:charlie_works}
\end{table}

\paragraph{Processes} Comments in the Translation theme (n=13) described Charlie in terms of a machine translation process (\pseudonym{fuchsiaBeaver}: \textit{``I thought of him as like a translator, like between English and code}''). Comments in the Sequential theme (n=13) described Charlie as working line-by-line through their prompt. This is plausible but incorrect: Code LLMs condition on the entire prompt at once. This mental model might lead students to focus on individual sentences, rather than how their prompt works as a holistic description. One student actually changed their mental model while answering: \textit{``it looks like he went line by line. Wrote some code for each line that makes sense to him [...] Actually, no, I think he takes in the whole prompt and [...] figures out what to do with the prompt. Because I do remember [...] there were a couple where I give a paragraph and then he returned a line of code, which makes me think that he wasn't going line by line.''} (\pseudonym{khakiClam}).

\paragraph{Charlie's Knowledge.} Most students hypothesized that Charlie relies on keywords (n=46). A large group of students (n=30) had a vague keyword mental model. For instance, \textit{``I guess he probably looks for keywords, “if” and “else” and key coding words, Python words, and he probably has a knowledge of English''} (\pseudonym{wheatOtter}). Another group (n=16) outline a  more specific keyword lookup model, where Charlie uses keywords to retrieve relevant code from a dictionary or database. For instance, \pseudonym{linenBobcat} described Charlie as \textit{``using the code words, and doing it sort of line by line and trying to work from what was given and writing those words with what, like in a directory or some sort of data file, understanding which ones matched up to which functions and which commands.''}

Students with this mental model emphasize the importance of using programming terminology, since they think Charlie may not be able to retrieve code without the right keywords. Some students develop this mental model after observing that their prompts succeed when they use coding words: \textit{``I noticed that if I put in more like, computerized words, I almost had a bit more control. At one point, I forgot to mention that the function returns something. So then when I mentioned that it returned something he put in a return statement. So that felt like very, like logical to me. [...] Charlie's looking for words that kind of line up with different functions, built in functions, and using those.''} (\pseudonym{tanMinnow}). These students correctly observe that sounding like a programmer is important, but explain this with an incorrect mental model.

Some students did correctly identify Charlie as similar to an LLM such as ChatGPT (n=17) or Copilot/Codex (n=2). Success rates for this group were slightly higher (0.27 versus 0.22; $p$=0.03). 


\subsection{What strategies do students develop?}

\begin{table}[t]
    \centering
    \begin{tabular}{l|l}
    Thematic Codes&N\\\hline
Added Detail&48\\
Looked at Tests First&30\\
Looked at Code First&29\\
Added Coding Language&21\\
Comment Not Relevant&16\\
Looked at Code and Tests Together&8\\
Reread the Problem&7\\
Reordered Prompt&5\\
Removed Detail&4\\
Ran Prompt Again&3\\
Fixed Grammar&2\\
    \end{tabular}
    \caption{Thematic codes emerging from responses to \textit{What did you do when you wrote a description, pressed Submit, and it did not work? Describe the steps you took to edit your description.}}
    \label{tab:editing_strategies}
\end{table}

\begin{figure*}[t]
\begin{subfigure}{0.465\textwidth}
\includegraphics[width=\textwidth]{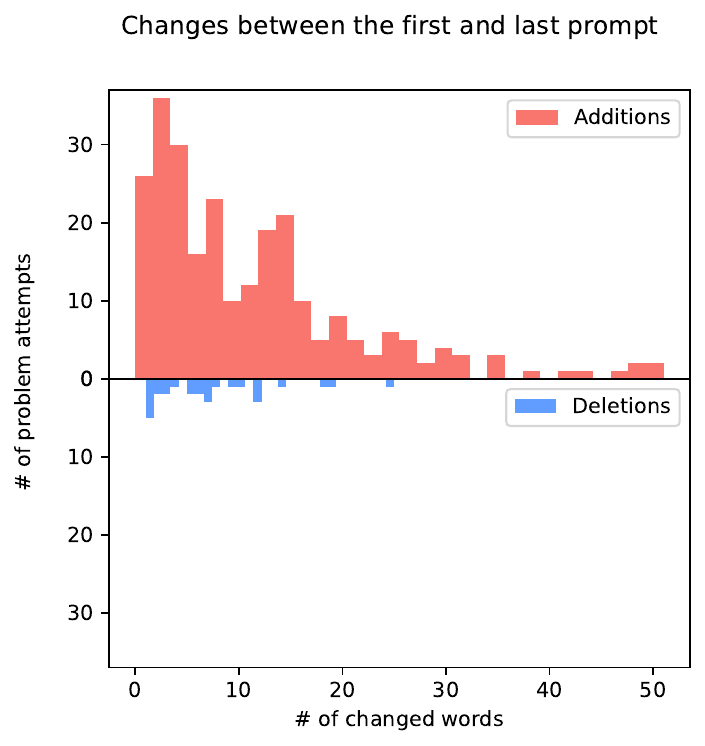}
\label{fig:first_last}
\end{subfigure}
\hfill
\begin{subfigure}{0.49\textwidth}
\includegraphics[width=\textwidth]{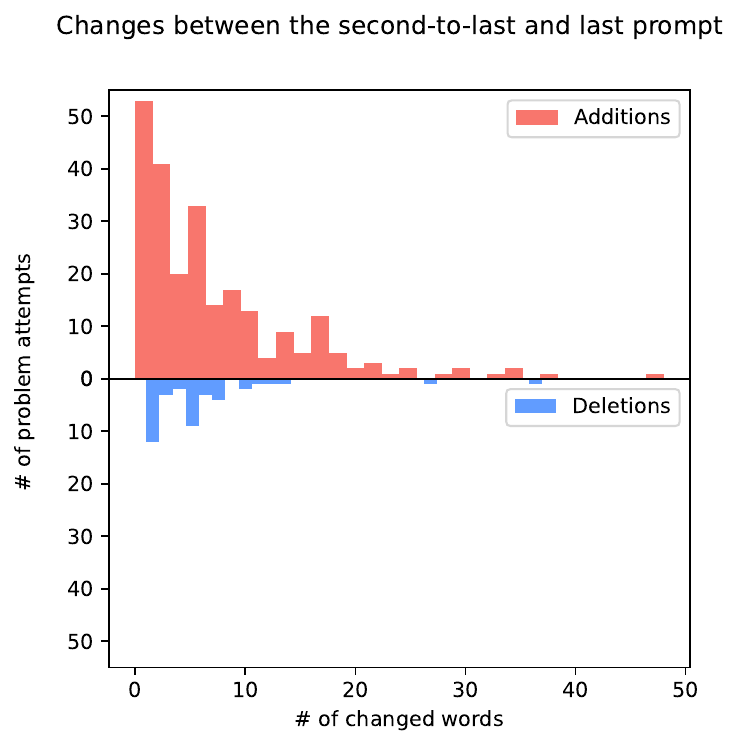}
\label{fig:second_to_last}
\end{subfigure}
\caption{Histograms of the 282 prompts which lead to successes after 2 or more attempts. These represent trends in how students edit prompts. The figure on the (left) shows the number of words changed between a first prompt and last prompt. The figure on the (right) shows the final change that produces a successful final prompt.}
\label{fig:upside_down}
\end{figure*}

The first two semi-structured interview questions asked students about their strategies for writing and editing prompts. We find that students do not have a clear understanding of how models work and that their incorrect mental models appear to affect the strategies they develop for prompting in ways that might be unproductive.

\subsubsection{Editing processes.} Over a third of students (n=48) mentioned adding detail to their descriptions when they did not succeed (\Cref{tab:editing_strategies}). Some students mentioned clarity as a goal in adding detail, like \pseudonym{fuschiaBat}: \textit{``I will go back and try to change the wording to make it more clear, and then try it again. And see if that changes anything. And then just try to repeat that process until it works.''} Others noted that their descriptions needed additional detail because they did not originally fully describe the problem, or as \pseudonym{plumBeetle} puts it, \textit{``I forgot to uppercase Aspen. And that was just my silly mistake. And I will just go back and edit or add changes that I want to add and wish it's gonna work the next time I guess.''} Considering participants' edits quantitatively confirms the popularity of adding detail. When we consider pairs of prompts that ultimately succeed, we find that students, on average, add 9.44 words (SD = 11.34) between their first and last prompt, and 5.36 words (SD = 8.87) between their penultimate and last prompt (\Cref{fig:upside_down}).



While adding details was the most common approach, participants mentioned other strategies, such as reordering (n=5) or removing detail (n=4). There are also eight attempts where rerunning the same prompt resulted in a success; we discuss these cases in \cref{model-role}.

Students looked in different places for insight into how to edit their prompts. Some considered the generated code first (n=29), some the tests (n=30). Others considered both (n=8) or reread the problem (n=7). 

\subsubsection{Strategy changes over time.}

Participants had a range of responses about how their prompting processes changed over time. Some students indicated that they never really developed a process (n=13), while others (n=14) discussed actively testing and adapting to Charlie's capabilities: \textit{``I first [...] was kind of seeing what vocabulary Charlie knew. Like if he knew computer science terms, or if I had to be less computer science-y''} (\pseudonym{beigeBass}).

\begin{figure}[t]
  \centering
  \includegraphics[width=0.95\linewidth]{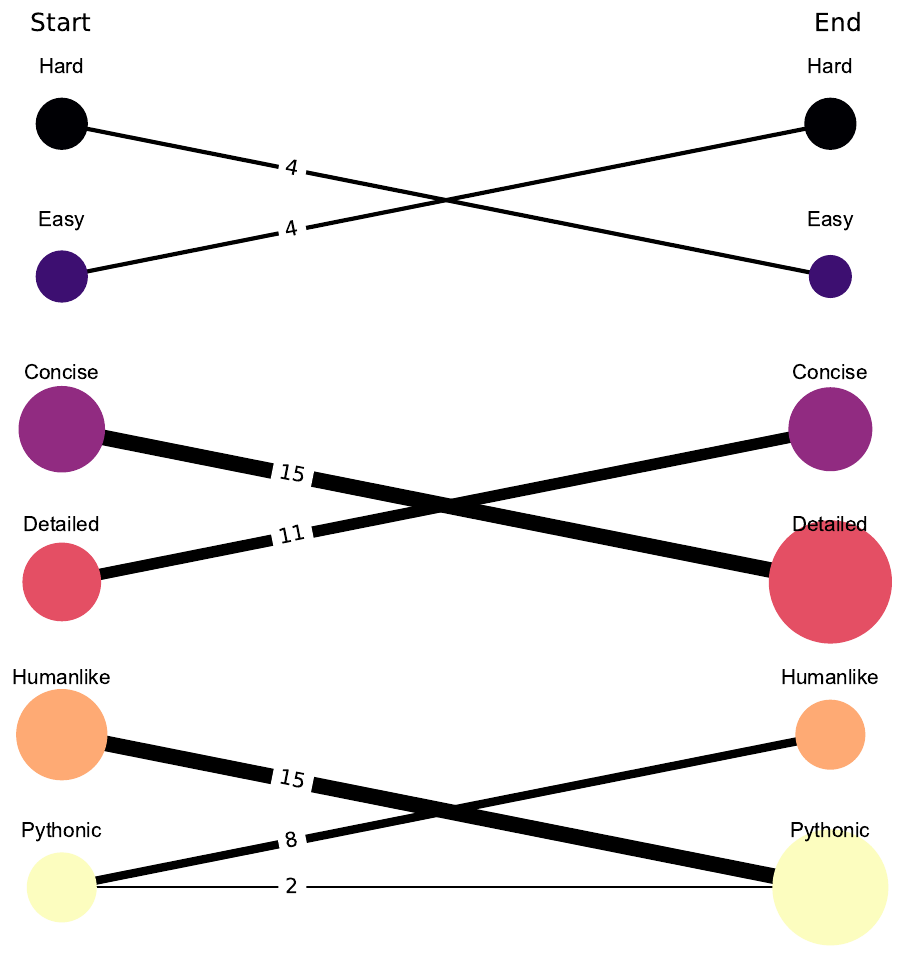}
  \caption{Visualization of how students describe their editing trajectories. The left nodes represent how students described how they began their process. The right nodes represent how students described how they edited prompts at the end of the study. The codes are presented in pairs - Hard versus Easy, Concise versus Detailed, Humanlike versus Pythonic. Only trajectories between pairs are visualized. The size of the nodes is proportional to the total number of students who described their Start or End within that code.}
  \label{fig:trajectories-qual}
\end{figure}

We present key trajectories in \Cref{fig:trajectories-qual}. Overall, we observe a range of reported experiences. Some participants reported starting more human-like and ending more technical (Pythonic), while others said the opposite. For instance, \pseudonym{tomatoBeetle} reported, \textit{``To begin with, I was using less technical terms and then using more computer science terms near the end.  I was thinking that would make Charlie work better, but there wasn’t really any evidence behind that''}, while \pseudonym{grayRabbit} said, \textit{``I kind of treated it like I was just coding but saying things I would like use kind of like if statements and integers and stuff. But towards the end, I tried to focus more on how I could say what was going on at a higher level, so using more plain language versus specific coding language}.''

A large group reported that their prompts became more detailed (n=35) and/or more technical (n=31), mirroring the finding above that students typically add detail when editing. For instance, \pseudonym{tanBat} reports, \textit{``My initial process was just to figure out what the code is doing and then just write generic descriptions, like without any coding language inside of it. But then when I saw that Charlie kept having problems, I started to go to more coding language.''} However, others took the opposite approach, and ended the study writing more human-like (n=11) or concise (n=16) descriptions. 
 
\subsection{Do Students Get Better at Prompting Over Time?}\label{sec:improvement}

It is easy to argue that programming by prompting a Code LLM with prose is more natural than directly writing code and that Code LLM prompting is easy to learn. But how easy is easy? We investigate whether students improve at prompt writing over the course of the study. We explore this by comparing success rates for (1)~students who attempted the problem first with (2)~students who attempted the problem last. Our experiment design ensures that there are exactly 5 students who attempt each problem first and five more who attempt it last. We find no significant difference in success rates between the two groups, indicating that students do not observably improve at prompting within the 75 minute study. 

\subsection{What do students think about Charlie?}

One of the most consistent findings in work on how experts use Code LLMs is that users enjoy using models~\citep{ziegler_productivity_2022,murali_codecompose_2023}, even when no concrete productivity or correctness benefits are observed~\citep{xu_systematic_2022,vaithilingam_expectation_2022}. However, near-novices exhibit different motivations and relationships to technology than expert programmers. This makes it important to investigate how non-experts feel about these systems.

\subsubsection{Charlie's competence and reliability}

The post-task survey asks participants several sets of questions related to their perceptions of Charlie. They completed 5 items from \citet{bartneck_measurement_2009} adapted by \citet{wang_towards_2021} and \citet{druga_how_2021} for non-robotics use. Participants generally give Charlie middling ratings for knowledge and competence. Participants take more extreme positions on Charlie's persona, in opposite directions: they rate Charlie as both friendly and machinelike. Students who experience lower success rates find Charlie somewhat less competent, but do not seem to find Charlie less friendly (\Cref{tab:perceptions}). Students also completed 5 items from \citet{korber_theoretical_2018}'s trust of automation survey. Overall, students see Charlie as somewhat reliable and somewhat interpretable (\Cref{tab:trust}). Students with higher success rates tended to rate Charlie as less error prone, easier to understand, and more reliable.

\begin{table*}[t]
    \centering
    \begin{tabular}{l|l|l}
    Scale&Mean&Correlation with Success Rate ($\tau$)\\\hline
        Ignorant - Knowledgeable &3.68&0.16*  \\
        Machinelike - Humanlike &2.39&0.12* \\
        Responding rigidly - Responding elegantly &3.13&0.09 \\
        Unfriendly - Friendly &4.2&0.008 \\
        Incompetent - Competent &3.58&0.19* \\
    \end{tabular}
    \caption{Mean student responses to Charlie perception questions (1=left endpoint, 5=right endpoint), adapted from \citet{wang_towards_2021}, and correlation with success rate. * indicates statistical significance.}
    \label{tab:perceptions}
\end{table*}

\begin{table*}[t]
    \centering
    \begin{tabular}{l|l|l}
    Question&Mean&Correlation with Success Rate ($\tau$)\\\hline
         Charlie is capable of taking over complicated tasks.&3.24&0.03\\
         Charlie might make sporadic errors.&2.15&0.18*\\
         I was able to understand why things happened.&2.24&-0.34*\\
         I can rely on Charlie.&2.95&-0.17*\\
         Automated systems generally work well.&2.46&-0.14\\
    \end{tabular}
    \caption{Mean student responses to Charlie trust questions (1 = Strongly agree; 5 = Strongly disagree), adapted from \citet{korber_theoretical_2018}, and correlation with success rate. * indicates statistical significance.}
    \label{tab:trust}
    
\end{table*}

\subsubsection{Would they use Charlie?}

The post-survey asked about students' attitudes toward hypothetically using Charlie in (a) the CS1 course they completed and (b) their own future programming practice. We used a thematic analysis approach to analyze this data, as with the interview data (see Appendix \ref{app:coding} for more details). 

Overall, two-thirds (n=83) stated that they would be interested in using Charlie in CS1. Many responses were variants of ``Yes'', but students who responded Maybe (n=13) or No (n=23) typically explained their reasoning. Half (n=19) of these suggested that tools like Charlie would inhibit student learning. For instance, \pseudonym{aquaLadybug} noted, \textit{`If I had questions on how to program a particular thing, using something like Charlie could help me clarify any questions I had by testing out different descriptions. But if I completely relied on something like Charlie as a tool in such a class, I feel like the whole point of me taking the class is overlooked and at some point becomes redundant.''}  Other students, including those who responded Yes, brought up how programmer skill level could play a role. \pseudonym{tealHerring} wrote, \textit{``Yes, but I would want to maybe only try it out towards the end of the course, when I've already learned the process of coding and would like to see how an AI could work to streamline the process.''} Other comments touched on academic integrity (\textit{``I don't think so unless my teacher explicitly endorsed it because I'm terrified of plagiarism!''} - \pseudonym{crimsonWorm}). 

More students supported using tools like Charlie in their own future programming practice (n=95). Maybe (n=20) and No (n=4) respondents again provided more explanation: two common themes included Charlie's limitations and usefulness for different kinds of problems: \textit{``If Charlie improved, then it should be able to generate simple functions for me, in which I don't have to repeat myself''} (\pseudonym{purpleCarp}).

\subsection{AI Attitudes}

Students were asked whether they felt optimistic or pessimistic about AI's future impact on society. About two-thirds of students were optimistic; however, students pursuing a programming major (Computer Science, Data Science, or Media Arts and Science) were notably more optimistic than other students (80\% optimistic compared to 63\% of other majors). There was no difference in task performance between optimists and pessimists (pass@1 rate = 0.22 for both).

\begin{figure}[t]
    \centering
    \includegraphics[width=0.475\textwidth]{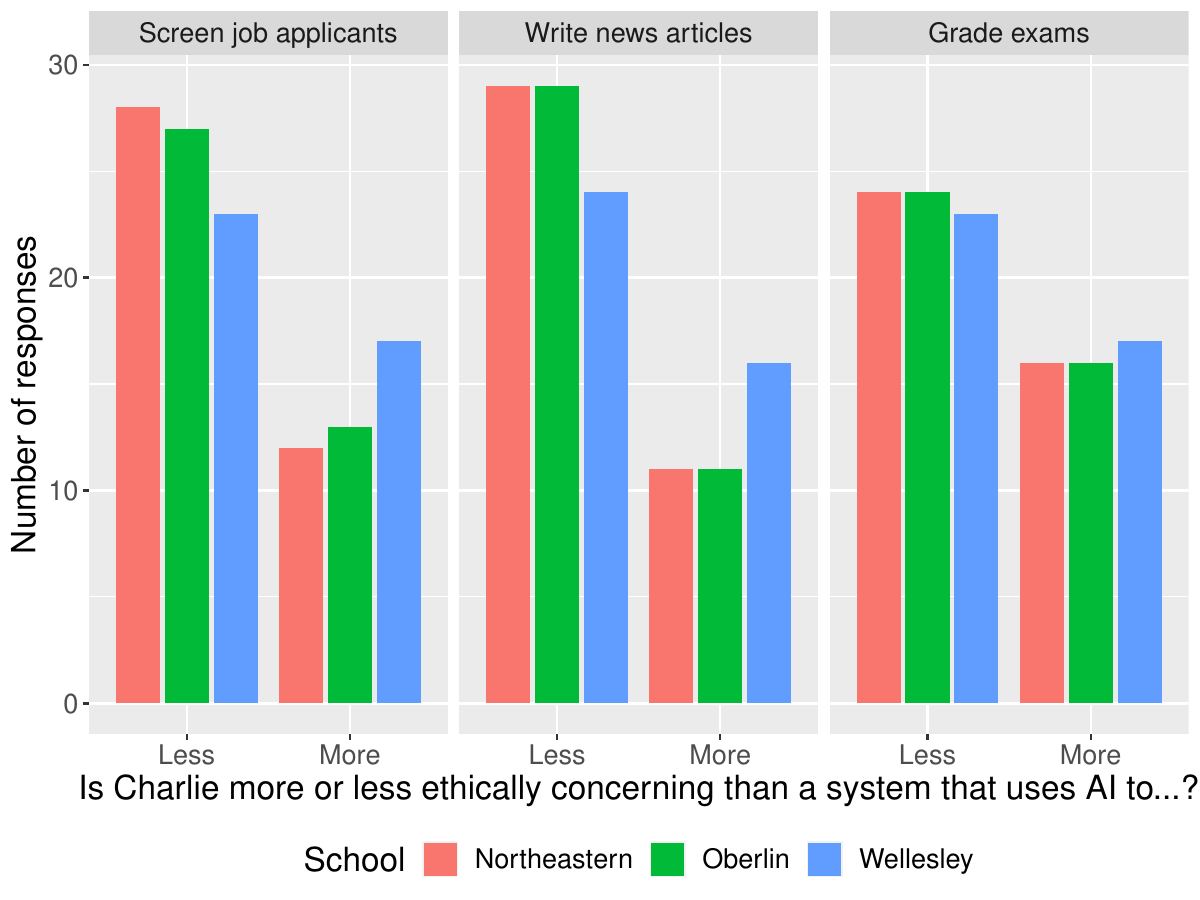}
    \caption{Student perceptions of Charlie's ethicality as compared to other AI scenarios}
    \label{fig:ethics}
\end{figure}

Students were also asked to compare the ethicality of Charlie with three other AI deployment scenarios. Most students found Charlie less ethically concerning in each comparison (\Cref{fig:ethics}). Student responses to these questions did not differ reliably in relation to their success rate or pass rate. 

\section{Discussion}\label{sec:discussion}

In the previous sections we discussed our three main research questions -- we summarize the findings together here:
\begin{itemize}
    \item \textsc{RQ1}: We find that some students can effectively prompt a Code LLM, but it often takes numerous attempts. Students overall found the task mentally demanding. Prior experience and first-generation status are correlated, positively and negatively respectively, with success.
    \item \textsc{RQ2}: The most common issues students report relate to the Code LLM misunderstanding their descriptions and issues with generated code. Both students themselves and our analysis of the data suggest that the stochastic nature of the Code LLM may impact student experiences. We find limited differences between students regarding problem difficulty.  
    \item \textsc{RQ3}: Students' most common mental model for the Code LLM was a data structure with keyword lookup. The most common prompting strategy that students developed was to expand their prompts, making them more detailed and more Pythonic. Students viewed the model as fairly capable and somewhat reliable. However, they expressed a range of opinions about whether Code LLMs would be appropriate for CS1.
\end{itemize}
In this section we draw connections between our findings and related work and discuss their broader implications. 

\subsection{The Natural-Language-to-Code Task is Challenging}

The emergence of LLMs have led some to conclude that this is the ``end of programming''~\cite{welsh_end_2022,manjoo_its_2023}. In contrast, we find that \emph{beginners who can write code nevertheless struggle to write natural language prompts for LLMs}. We carefully select problems that are similar (or identical) to those they completed to pass CS1. The average participant solves 57\% of the assigned problems, but only after several repeated attempts and with automatic feedback on code correctness. Our study contributes to the existing work on beginner interactions with Code LLMs by measuring how well students can use Code LLMs to solve problems at their own programming skill level, rather than in the context of a learning activity, where students may not be expected to able to write the code themselves. Despite the fact that all of our participants had passed CS1, which required writing code to solve problems like those in our study, many of them struggled to write natural language descriptions to lead a Code LLM to solve similar tasks. 


On the whole, our findings reveal a somewhat higher level of difficulty in using Code LLMs than other studies~\citep{prather_its_2023,kazemitabaar_studying_2023,denny_promptly_2023}, though it can be challenging to compare across diverse student populations, study designs, and problem types. Our results align most closely with those from \citet{denny_promptly_2023}'s subsequent study of students with just two weeks of programming instruction. Although their study used only 3 problems and had less experienced programmers, they observed similar challenges: 86\% of students eventually solved their easiest problem, but only 65\% solved their hardest task. This is close to the average eventual success rate that we observe.


\subsection{Not a Panacea for Non-Expert Programming}

Learning an effective process for how to prompt a Code LLM is the key to interacting successfully with it in the long term. Existing work on experts reveals different ``modes'' of interaction~\cite{barke_grounded_2023}. Our findings suggest that unlike experts, near-novices \textit{do not develop well-defined strategies for how to prompt}. Students added more detail to their previous prompts, even when it would have been better to start from scratch. In addition, students' prompting abilities did not observably improve during the study (\Cref{sec:improvement}). Students' failure to develop effective strategies may also be linked to their incorrect mental models of how Code LLMs work (\Cref{subsec:mental}). These results suggest that prompting, like most ways of interacting with code, needs to be explicitly taught to be used effectively. 

\citet{kazemitabaar_studying_2023} present a study of pre-college students that suggests Code LLMs can improve learning outcomes. They compare student performance with and without access to the Code LLM, and provide considerable support to participants, such as instructor feedback and access to expert-written descriptions of the problem. In three of their task categories, both students with and without access to a Code LLM were able to complete 100\% of the tasks, making it difficult to understand the contribution of the Code LLM. In the two more challenging categories, students benefited from the Code LLM, but they also relied heavily on the expert-written description (reusing it around 40\% of the time). Together with our results, we take this to indicate that Code LLMs can be useful to beginners, but that writing prompts remains a barrier. This highlights the importance of understanding why Code LLMs and beginning programmers struggle to understand each other: \citet{kazemitabaar_studying_2023} argue that Code LLMs could positively impact student learning, but our results demonstrate a variety of ways that these interactions currently fail.

Our findings provide fine-grained evidence about student challenges that have implications for complete novices, as well as the beginners we study. The results in \Cref{sec:rq3} highlight how effective prompting requires skills that complete novices do not possess. \Cref{fig:trajectories-qual} visualizes how students described their start and end approaches to editing, showing that many students who started out writing prompts as for a human transition into using more coding terminology by the end of the study. These participants picked up on a key property of Code LLMs: they are trained on expert-written code and documentation and expect natural language prompts to utilize coding terminology. The strategies that were most effective for our beginners would not be available to true novices. 

\subsection{Don't Assume a Mental Model of AI}

Our study suggests that students have \textit{incomplete mental models} of how Code LLMs work. Although participants knew they were interacting with an AI code generation tool and the majority (n=88, 73\% in the post survey) had heard of GPT-3, Github Copilot, or Codex, when asked how they thought our system worked, only 19 students mentioned these models. A notable feature of responses was the number of detailed, but incorrect explanations. The majority of students who gave examples identified a keyword-based lookup strategy, like the dictionaries they had learned about in CS1. 

These mental models fail to explain one aspect of Codex that students find frustrating: its stochastic responses. Students are familiar with errors that persist after editing their code. Code LLMs introduce a related but novel experience: submitting the same prompt and getting a different program (\cref{model-role}). This does not occur in standard CS1 settings and cannot be explained by the database/dictionary mental model of Code LLMs that most participants described. Without a well-developed understanding of why this happens, students have simply added another unknown computational behavior to their coding experience. 

We note that although \citet{prather_its_2023} report that several of their participants described models as having sentience or agency, none of our participants did. This may reflect the growing public awareness of generative AI between their study and ours, resulting in more realistic attitudes about the capabilities of large language models in our population. Our students seem to understand what AI models can do, but not how they do it.

\subsection{Implications for Educators}

Recent work has shown that Code LLMs can solve CS exams or homework assignments \textit{given the educator's description of the problem}~\citep{finnie-ansley_robots_2022,dakhel_github_2022}. Our findings show that although Code LLMs can solve CS1 problems, CS1 students cannot necessarily prompt Code LLMs to solve CS1 problems. Our findings reiterate the importance of key skills taught in CS1: code comprehension, problem decomposition, and the ability to describe computational problems clearly. 

While we do not study learning outcomes explicitly, we find mixed support for Code LLMs as pedagogical tools. The survey portion of our experiment included questions about participants' attitudes towards Code LLMs. About two-thirds of participants expressed interest in using similar technology in CS1. Some participants mentioned that the task helped them remember Python concepts that they had forgotten, or even learn new features (such as list comprehensions for \oberlin students). Others felt that it helped them practice describing technical tasks in natural language; Code LLMs could be used to provide feedback on Explain In Plain English (EiPE) questions~\citep{lopez_relationships_2008, corney_explain_2014}, which many educators see as valuable, but difficult to use without automation \cite{fowler_autograding_2021}. Recent work on students' perceptions of automatically-graded EiPE questions provides guidelines that may serve as a first step towards using Code LLMs as automatic backends \cite{hsu_attitudes_2021}. 

On the other hand, a sizeable number of students did not support using Code LLMs in CS1. Some students expressed ethical concerns. Many questioned whether coming to rely on Code LLMs would diminish their knowledge of programming or their sense of fulfillment. Our survey data also highlights a key challenge of contemporary AI: explainability. Students gave Charlie higher ratings for capability than interpretability. Our findings here complement \citet{sun_investigating_2022}'s exploration of Code LLM explainability needs identified by expert programmers, and \citet{prather_its_2023}'s finding of students' ``slow accept'' mode, where students spent a lot of time reading code generated by Copilot and deciding whether or not to accept it.

By shedding light on how students feel about Code LLMs, our work augments \citet{lau_ban_2023}'s investigation of CS educators' perspectives on Code LLMs. Our studies were conducted at a similar moment when Code LLMs had recently gained public prominence, but few educators or students had much experience with them. Our students and the educators in \citet{lau_ban_2023} raise strikingly similar concerns about ethics and negative impacts on student learning. \citet{denny_promptly_2023}'s subsequent experiment found similar concerns among currently enrolled CS1 students.

The large scale of our study also allows us to contribute data to the debate over equity in \citet{lau_ban_2023}'s study, who show conflicting perspectives among educators: some felt that Code LLMs could strengthen the digital divide between students, while others felt that Code LLMs could improve diversity in CS. On the whole, our findings strengthens concerns. We show that students with extracurricular programming experience have an advantage, echoing \citet{kazemitabaar_studying_2023}'s finding that more experienced programmers benefit more from using Code LLMs. We also show that prompts written by first generation college students have reliably lower pass@1 rates. Educators should weigh the potential benefits of adopting this new technology against the possibility that it might exacerbate existing equity issues \cite{holstein_equity_2021}.

Finally, our students are ambivalent towards AI systems in general. Around two-thirds were optimistic about AI's impact on society in the future, similar to the proportion interested in using Charlie in CS1. This leaves a sizeable number of beginners who are concerned about AI or uninterested in its use in CS1. Our findings capture a nuanced portrait of how young adults perceive generative AI for programming, captured at a moment where generative AI was increasingly prominent in popular media. 

\subsection{Model Selection for Human-AI Interaction Research}\label{sec:open_models}

One issue for studies such as ours is the rapid pace of research and development in machine learning. Running lab experiments with humans takes time. However, current proprietary models are often updated or deprecated with very little warning. This study used OpenAI's Codex, which provides state-of-the-art Code LLM performance but came with significant risks. In the middle of our study, OpenAI announced that Codex would be deprecated within a week, which would have seriously compromised our results; after much public concern, they eventually delayed the deprecation until early 2024.

The mismatch between the timescale of ML development and human-subjects research makes it difficult to complete studies using state-of-the-art models, which are largely proprietary. Based on our experience, we recommend not using proprietary models, although this may come with a trade-off in terms of performance, and imposes significant computational requirements for the research team (since alternatives require access to significant GPU resources). Nonetheless, we strongly suggest the use of open source models \cite{roziere_code_2024, li_starcoder_2023} in future work, and potentially for classroom use, to avoid sudden loss of access. This is an example of an ongoing equity concern for researchers and educators.

\subsection{Timeliness} 

Conducting work with non-experts and Code LLMs in early 2023 captures a specific moment in the evolution of this technology. Our participant pool represents students who mostly completed CS1 before Code LLMs became commonplace. Collecting this data now is paramount to our understanding of baseline interactions with Code LLMs for students without previous exposure. In the future, the controlled background knowledge of this study will become increasingly hard to come by, both at our institutions and farther afield.

We also see our work as timely because of the struggles and strategies, or lack thereof, that we identify. As computing resources become increasingly directed towards Code LLM technology \cite{leswing_chatgpt_2023}, work such as ours has the potential to impact how companies develop their models, tutorials, and interfaces. We find that non-experts struggle to execute the full prompt and edit cycle, even with an interface that identifies output correctness. If this trend generalizes to other non-expert groups, Code LLM technology may strengthen the digital divide between expert and non-expert programmers, adding to the wide ranging list of ethical concerns about generative AI \cite{liao_ai_2023, bender_dangers_2021, khlaaf_hazard_2022}. 

\section{Threats to Validity}

A major challenge of studying human-AI interaction is that AI capabilities and popular awareness of them change quickly. ChatGPT was released between our pilot and main experiment; as a result, students' knowledge and experience with large language models underwent significant growth during our experiment. We observed a statistically significant improvement in task performance for students who took the study in the last month. This may spring from increased familiarity with large language models such as ChatGPT or from more recent exposure to CS1 material. 

Although we recruited participants who had completed CS1 and no subsequent CS courses, their programming backgrounds were not homogeneous. Some participants had taken a prior programming course in high school or in college, and some were concurrently enrolled in a programming course. We study the effects of additional programming experience in \Cref{sec:who_succeeds}. In addition, since we recruited students who had taken CS1 as early as Fall 2021, some participants reported having forgotten programming concepts or terms in the intervening time. 

Several factors may have biased participants towards reporting positive perceptions of our system. While we ensured that the experimenter running the study was not a educator at the participant's institution, participants were aware that the study involved one of their professors and may have responded more positively as a result. 
In addition, students may have answered questions about text-to-code more positively because of the anthropomorphic qualities of our system design; several commented about the appealing affect of the Charlie mascot in post-study questions. Charlie may have also had an effect on students' level of task perseverance~\cite{lee_personifying_2011}. Finally, novelty bias is always a potential concern when evaluating novel interfaces or systems, as-is self-selection bias for stand-alone studies.

\section{Conclusion}
We present results from a large-scale, multi-institution study of how near-novices interact with Code LLMs. Our novel experimental design allows us to isolate the prompt writing and editing tasks, by using a lab-based experiment in which participants write natural language descriptions of tasks and receive automated feedback on the correctness of generated code.

Our results suggest that 
students who have complete a single CS course 
find using Code LLMs challenging, even with tasks at an appropriate skill level. Our findings highlight the various barriers that they face, ranging from distilling their problem understanding into words, using coding terminology, understanding generated code, and grappling with the stochasticity of Code LLM output. We show that certain groups of students, most notably, first-generation college students, face additional difficulties, raising equity issues related to the deployment of Code LLMs in the classroom. We also illustrate how students' incorrect mental models of how Code LLMs operate inhibit their ability to develop effective prompting strategies. Moreover, our qualitative results provide insight into how beginning programmers feel about introducing Code LLMs in the classroom, bringing their voices into an key contemporary debate and complementing existing work on educators' perspectives.

Our findings suggest that Code LLMs do not signal the ``end of programming'': in fact, they highlight the many ways in which Code LLMs remain inaccessible to non-experts. We hope that our findings will motivate renewed effort towards democratizing programming by closing this gap.

\begin{acks}
We thank the anonymous reviewers and Shriram Krishnamurthi for their thoughtful feedback.
We thank our colleagues who helped us recruit participants and who provided CS1 problems that we adapted. 
We would also like to thank Rachelle Hu for her work on the Charlie platform prototype.
This work is partially supported by the National Science Foundation
(SES-2326173, SES-2326174, and SES-2326175). We thank Northeastern Research Computing and the New England Research Cloud for providing computing resources.
\end{acks}

\bibliographystyle{ACM-Reference-Format}

\bibliography{main_camera_ready}


\begin{thebibliography}{105}


\ifx \showCODEN    \undefined \def \showCODEN     #1{\unskip}     \fi
\ifx \showDOI      \undefined \def \showDOI       #1{#1}\fi
\ifx \showISBNx    \undefined \def \showISBNx     #1{\unskip}     \fi
\ifx \showISBNxiii \undefined \def \showISBNxiii  #1{\unskip}     \fi
\ifx \showISSN     \undefined \def \showISSN      #1{\unskip}     \fi
\ifx \showLCCN     \undefined \def \showLCCN      #1{\unskip}     \fi
\ifx \shownote     \undefined \def \shownote      #1{#1}          \fi
\ifx \showarticletitle \undefined \def \showarticletitle #1{#1}   \fi
\ifx \showURL      \undefined \def \showURL       {\relax}        \fi
\providecommand\bibfield[2]{#2}
\providecommand\bibinfo[2]{#2}
\providecommand\natexlab[1]{#1}
\providecommand\showeprint[2][]{arXiv:#2}

\bibitem[Agostinelli et~al\mbox{.}(2023)]%
        {agostinelli_musiclm_2023}
\bibfield{author}{\bibinfo{person}{Andrea Agostinelli},
  \bibinfo{person}{Timo~I. Denk}, \bibinfo{person}{Zalán Borsos},
  \bibinfo{person}{Jesse Engel}, \bibinfo{person}{Mauro Verzetti},
  \bibinfo{person}{Antoine Caillon}, \bibinfo{person}{Qingqing Huang},
  \bibinfo{person}{Aren Jansen}, \bibinfo{person}{Adam Roberts},
  \bibinfo{person}{Marco Tagliasacchi}, \bibinfo{person}{Matt Sharifi},
  \bibinfo{person}{Neil Zeghidour}, {and} \bibinfo{person}{Christian Frank}.}
  \bibinfo{year}{2023}\natexlab{}.
\newblock \bibinfo{title}{{MusicLM}: {Generating} {Music} {From} {Text}}.
\newblock
\newblock
\urldef\tempurl%
\url{http://arxiv.org/abs/2301.11325}
\showURL{%
\tempurl}


\bibitem[Akoury et~al\mbox{.}(2020)]%
        {akoury_storium_2020}
\bibfield{author}{\bibinfo{person}{Nader Akoury}, \bibinfo{person}{Shufan
  Wang}, \bibinfo{person}{Josh Whiting}, \bibinfo{person}{Stephen Hood},
  \bibinfo{person}{Nanyun Peng}, {and} \bibinfo{person}{Mohit Iyyer}.}
  \bibinfo{year}{2020}\natexlab{}.
\newblock \bibinfo{title}{{STORIUM}: {A} {Dataset} and {Evaluation} {Platform}
  for {Machine}-in-the-{Loop} {Story} {Generation}}.
\newblock
\newblock
\urldef\tempurl%
\url{http://arxiv.org/abs/2010.01717}
\showURL{%
\tempurl}
\newblock
\shownote{arXiv:2010.01717 [cs]}.


\bibitem[Babe et~al\mbox{.}(2023)]%
        {babe_studenteval_2023}
\bibfield{author}{\bibinfo{person}{Hannah~McLean Babe}, \bibinfo{person}{Sydney
  Nguyen}, \bibinfo{person}{Yangtian Zi}, \bibinfo{person}{Arjun Guha},
  \bibinfo{person}{Molly~Q. Feldman}, {and} \bibinfo{person}{Carolyn~Jane
  Anderson}.} \bibinfo{year}{2023}\natexlab{}.
\newblock \bibinfo{title}{{StudentEval}: {A} {Benchmark} of {Student}-{Written}
  {Prompts} for {Large} {Language} {Models} of {Code}}.
\newblock
\newblock
\urldef\tempurl%
\url{http://arxiv.org/abs/2306.04556}
\showURL{%
\tempurl}


\bibitem[Ballard and Biermann(1979)]%
        {ballard_programming_1979}
\bibfield{author}{\bibinfo{person}{Bruce~W. Ballard} {and}
  \bibinfo{person}{Alan~W. Biermann}.} \bibinfo{year}{1979}\natexlab{}.
\newblock \showarticletitle{Programming in {Natural} {Language}: “{NLC}” as
  a {Prototype}}. In \bibinfo{booktitle}{\emph{Annual {Conference} of the
  {ACM}}}. \bibinfo{publisher}{Association for Computing Machinery},
  \bibinfo{address}{New York, NY, USA}, \bibinfo{pages}{228--237}.
\newblock
\showISBNx{978-0-89791-008-8}
\urldef\tempurl%
\url{https://doi.org/10.1145/800177.810072}
\showDOI{\tempurl}


\bibitem[Bareiß et~al\mbox{.}(2022)]%
        {bareis_code_2022}
\bibfield{author}{\bibinfo{person}{Patrick Bareiß}, \bibinfo{person}{Beatriz
  Souza}, \bibinfo{person}{Marcelo d'Amorim}, {and} \bibinfo{person}{Michael
  Pradel}.} \bibinfo{year}{2022}\natexlab{}.
\newblock \bibinfo{title}{Code {Generation} {Tools} ({Almost}) for {Free}? {A}
  {Study} of {Few}-{Shot}, {Pre}-{Trained} {Language} {Models} on {Code}}.
\newblock
\newblock
\urldef\tempurl%
\url{http://arxiv.org/abs/2206.01335}
\showURL{%
\tempurl}
\newblock
\shownote{arXiv:2206.01335 [cs]}.


\bibitem[Barke et~al\mbox{.}(2023)]%
        {barke_grounded_2023}
\bibfield{author}{\bibinfo{person}{Shraddha Barke}, \bibinfo{person}{Michael~B.
  James}, {and} \bibinfo{person}{Nadia Polikarpova}.}
  \bibinfo{year}{2023}\natexlab{}.
\newblock \showarticletitle{Grounded {Copilot}: {How} {Programmers} {Interact}
  with {Code}-{Generating} {Models}}.
\newblock \bibinfo{journal}{\emph{Proceedings of the ACM on Programming
  Languages}} \bibinfo{volume}{7}, \bibinfo{number}{OOPSLA1}
  (\bibinfo{date}{April} \bibinfo{year}{2023}), \bibinfo{pages}{85--111}.
\newblock
\showISSN{2475-1421}
\urldef\tempurl%
\url{https://doi.org/10.1145/3586030}
\showDOI{\tempurl}


\bibitem[Bartneck et~al\mbox{.}(2009)]%
        {bartneck_measurement_2009}
\bibfield{author}{\bibinfo{person}{Christoph Bartneck}, \bibinfo{person}{Dana
  Kulić}, \bibinfo{person}{Elizabeth Croft}, {and} \bibinfo{person}{Susana
  Zoghbi}.} \bibinfo{year}{2009}\natexlab{}.
\newblock \showarticletitle{Measurement {Instruments} for the
  {Anthropomorphism}, {Animacy}, {Likeability}, {Perceived} {Intelligence}, and
  {Perceived} {Safety} of {Robots}}.
\newblock \bibinfo{journal}{\emph{International Journal of Social Robotics}}
  \bibinfo{volume}{1}, \bibinfo{number}{1} (\bibinfo{date}{Jan.}
  \bibinfo{year}{2009}), \bibinfo{pages}{71--81}.
\newblock
\showISSN{1875-4791, 1875-4805}
\urldef\tempurl%
\url{https://doi.org/10.1007/s12369-008-0001-3}
\showDOI{\tempurl}


\bibitem[Bates et~al\mbox{.}(2015)]%
        {bates_fitting_2015}
\bibfield{author}{\bibinfo{person}{Douglas Bates}, \bibinfo{person}{Martin
  Mächler}, \bibinfo{person}{Ben Bolker}, {and} \bibinfo{person}{Steve
  Walker}.} \bibinfo{year}{2015}\natexlab{}.
\newblock \showarticletitle{Fitting {Linear} {Mixed}-{Effects} {Models} {Using}
  lme4}.
\newblock \bibinfo{journal}{\emph{Journal of Statistical Software}}
  \bibinfo{volume}{67}, \bibinfo{number}{1} (\bibinfo{year}{2015}),
  \bibinfo{pages}{1--48}.
\newblock
\urldef\tempurl%
\url{https://doi.org/10.18637/jss.v067.i01}
\showDOI{\tempurl}


\bibitem[Bender et~al\mbox{.}(2021)]%
        {bender_dangers_2021}
\bibfield{author}{\bibinfo{person}{Emily~M. Bender}, \bibinfo{person}{Timnit
  Gebru}, \bibinfo{person}{Angelina McMillan-Major}, {and}
  \bibinfo{person}{Shmargaret Shmitchell}.} \bibinfo{year}{2021}\natexlab{}.
\newblock \showarticletitle{On the {Dangers} of {Stochastic} {Parrots}: {Can}
  {Language} {Models} {Be} {Too} {Big}?}. In
  \bibinfo{booktitle}{\emph{Proceedings of the 2021 {ACM} {Conference} on
  {Fairness}, {Accountability}, and {Transparency}}}. \bibinfo{publisher}{ACM},
  \bibinfo{address}{Virtual Event Canada}, \bibinfo{pages}{610--623}.
\newblock
\showISBNx{978-1-4503-8309-7}
\urldef\tempurl%
\url{https://doi.org/10.1145/3442188.3445922}
\showDOI{\tempurl}


\bibitem[Bird et~al\mbox{.}(2023)]%
        {bird_taking_2023}
\bibfield{author}{\bibinfo{person}{Christian Bird}, \bibinfo{person}{Denae
  Ford}, \bibinfo{person}{Thomas Zimmermann}, \bibinfo{person}{Nicole
  Forsgren}, \bibinfo{person}{Eirini Kalliamvakou}, \bibinfo{person}{Travis
  Lowdermilk}, {and} \bibinfo{person}{Idan Gazit}.}
  \bibinfo{year}{2023}\natexlab{}.
\newblock \showarticletitle{Taking {Flight} with {Copilot}: {Early} {Insights}
  and {Opportunities} of {AI}-{Powered} {Pair}-{Programming} {Tools}}.
\newblock \bibinfo{journal}{\emph{Queue}} \bibinfo{volume}{20},
  \bibinfo{number}{6} (\bibinfo{date}{Jan.} \bibinfo{year}{2023}),
  \bibinfo{pages}{35--57}.
\newblock
\showISSN{1542-7730}
\urldef\tempurl%
\url{https://doi.org/10.1145/3582083}
\showDOI{\tempurl}
\newblock
\shownote{Place: New York, NY, USA Publisher: Association for Computing
  Machinery}.


\bibitem[Cassano et~al\mbox{.}(2023)]%
        {cassano_multipl-e_2023}
\bibfield{author}{\bibinfo{person}{Federico Cassano}, \bibinfo{person}{John
  Gouwar}, \bibinfo{person}{Daniel Nguyen}, \bibinfo{person}{Sydney Nguyen},
  \bibinfo{person}{Luna Phipps-Costin}, \bibinfo{person}{Donald Pinckney},
  \bibinfo{person}{Ming-Ho Yee}, \bibinfo{person}{Yangtian Zi},
  \bibinfo{person}{Carolyn~Jane Anderson}, \bibinfo{person}{Molly~Q Feldman},
  \bibinfo{person}{Arjun Guha}, \bibinfo{person}{Michael Greenberg}, {and}
  \bibinfo{person}{Abhinav Jangda}.} \bibinfo{year}{2023}\natexlab{}.
\newblock \showarticletitle{{MultiPL}-{E}: {A} {Scalable} and {Polyglot}
  {Approach} to {Benchmarking} {Neural} {Code} {Generation}}.
\newblock \bibinfo{journal}{\emph{IEEE Transactions on Software Engineering}}
  \bibinfo{volume}{49}, \bibinfo{number}{7} (\bibinfo{date}{July}
  \bibinfo{year}{2023}), \bibinfo{pages}{3675--3691}.
\newblock
\showISSN{0098-5589, 1939-3520, 2326-3881}
\urldef\tempurl%
\url{https://doi.org/10.1109/TSE.2023.3267446}
\showDOI{\tempurl}


\bibitem[Chen et~al\mbox{.}(2023)]%
        {chen_data_2023}
\bibfield{author}{\bibinfo{person}{Le Chen}, \bibinfo{person}{Xianzhong Ding},
  \bibinfo{person}{Murali Emani}, \bibinfo{person}{Tristan Vanderbruggen},
  \bibinfo{person}{Pei-Hung Lin}, {and} \bibinfo{person}{Chunhua Liao}.}
  \bibinfo{year}{2023}\natexlab{}.
\newblock \showarticletitle{Data {Race} {Detection} {Using} {Large} {Language}
  {Models}}. In \bibinfo{booktitle}{\emph{Proceedings of the {SC} '23
  {Workshops} of {The} {International} {Conference} on {High} {Performance}
  {Computing}, {Network}, {Storage}, and {Analysis}}}.
  \bibinfo{publisher}{ACM}, \bibinfo{address}{Denver CO USA},
  \bibinfo{pages}{215--223}.
\newblock
\showISBNx{9798400707858}
\urldef\tempurl%
\url{https://doi.org/10.1145/3624062.3624088}
\showDOI{\tempurl}


\bibitem[Chen et~al\mbox{.}(2021)]%
        {chen_evaluating_2021}
\bibfield{author}{\bibinfo{person}{Mark Chen}, \bibinfo{person}{Jerry Tworek},
  \bibinfo{person}{Heewoo Jun}, \bibinfo{person}{Qiming Yuan},
  \bibinfo{person}{Henrique Ponde de~Oliveira Pinto}, \bibinfo{person}{Jared
  Kaplan}, \bibinfo{person}{Harri Edwards}, \bibinfo{person}{Yuri Burda},
  \bibinfo{person}{Nicholas Joseph}, \bibinfo{person}{Greg Brockman},
  \bibinfo{person}{Alex Ray}, \bibinfo{person}{Raul Puri},
  \bibinfo{person}{Gretchen Krueger}, \bibinfo{person}{Michael Petrov},
  \bibinfo{person}{Heidy Khlaaf}, \bibinfo{person}{Girish Sastry},
  \bibinfo{person}{Pamela Mishkin}, \bibinfo{person}{Brooke Chan},
  \bibinfo{person}{Scott Gray}, \bibinfo{person}{Nick Ryder},
  \bibinfo{person}{Mikhail Pavlov}, \bibinfo{person}{Alethea Power},
  \bibinfo{person}{Lukasz Kaiser}, \bibinfo{person}{Mohammad Bavarian},
  \bibinfo{person}{Clemens Winter}, \bibinfo{person}{Philippe Tillet},
  \bibinfo{person}{Felipe~Petroski Such}, \bibinfo{person}{Dave Cummings},
  \bibinfo{person}{Matthias Plappert}, \bibinfo{person}{Fotios Chantzis},
  \bibinfo{person}{Elizabeth Barnes}, \bibinfo{person}{Ariel Herbert-Voss},
  \bibinfo{person}{William~Hebgen Guss}, \bibinfo{person}{Alex Nichol},
  \bibinfo{person}{Alex Paino}, \bibinfo{person}{Nikolas Tezak},
  \bibinfo{person}{Jie Tang}, \bibinfo{person}{Igor Babuschkin},
  \bibinfo{person}{Suchir Balaji}, \bibinfo{person}{Shantanu Jain},
  \bibinfo{person}{William Saunders}, \bibinfo{person}{Christopher Hesse},
  \bibinfo{person}{Andrew~N. Carr}, \bibinfo{person}{Jan Leike},
  \bibinfo{person}{Josh Achiam}, \bibinfo{person}{Vedant Misra},
  \bibinfo{person}{Evan Morikawa}, \bibinfo{person}{Alec Radford},
  \bibinfo{person}{Matthew Knight}, \bibinfo{person}{Miles Brundage},
  \bibinfo{person}{Mira Murati}, \bibinfo{person}{Katie Mayer},
  \bibinfo{person}{Peter Welinder}, \bibinfo{person}{Bob McGrew},
  \bibinfo{person}{Dario Amodei}, \bibinfo{person}{Sam McCandlish},
  \bibinfo{person}{Ilya Sutskever}, {and} \bibinfo{person}{Wojciech Zaremba}.}
  \bibinfo{year}{2021}\natexlab{}.
\newblock \bibinfo{title}{Evaluating {Large} {Language} {Models} {Trained} on
  {Code}}.
\newblock
\newblock
\urldef\tempurl%
\url{http://arxiv.org/abs/2107.03374}
\showURL{%
\tempurl}
\newblock
\shownote{arXiv:2107.03374 [cs]}.


\bibitem[{CodeWhisperer}(2023)]%
        {codewhisperer_ml-powered_2023}
\bibfield{author}{\bibinfo{person}{{CodeWhisperer}}.}
  \bibinfo{year}{2023}\natexlab{}.
\newblock \bibinfo{title}{{ML}-powered {Coding} {Companion} – {Amazon}
  {CodeWhisperer} – {Amazon} {Web} {Services}}.
\newblock
\newblock
\urldef\tempurl%
\url{https://aws.amazon.com/codewhisperer/}
\showURL{%
\tempurl}


\bibitem[Copilot(2023)]%
        {copilot_github_2023}
\bibfield{author}{\bibinfo{person}{Github Copilot}.}
  \bibinfo{year}{2023}\natexlab{}.
\newblock \bibinfo{title}{Github {Copilot} {Your} {AI} pair programmer}.
\newblock
\newblock
\urldef\tempurl%
\url{https://github.com/features/copilot}
\showURL{%
\tempurl}


\bibitem[Corney et~al\mbox{.}(2014)]%
        {corney_explain_2014}
\bibfield{author}{\bibinfo{person}{Malcolm Corney}, \bibinfo{person}{Sue
  Fitzgerald}, \bibinfo{person}{Brian Hanks}, \bibinfo{person}{Raymond Lister},
  \bibinfo{person}{Renee McCauley}, {and} \bibinfo{person}{Laurie Murphy}.}
  \bibinfo{year}{2014}\natexlab{}.
\newblock \showarticletitle{'{Explain} in {Plain} {English}' {Questions}
  {Revisited}: {Data} {Structures} {Problems}}. In
  \bibinfo{booktitle}{\emph{Proceedings of the 45th {ACM} technical symposium
  on {Computer} science education}}. \bibinfo{publisher}{ACM},
  \bibinfo{address}{Atlanta Georgia USA}, \bibinfo{pages}{591--596}.
\newblock
\showISBNx{978-1-4503-2605-6}
\urldef\tempurl%
\url{https://doi.org/10.1145/2538862.2538911}
\showDOI{\tempurl}


\bibitem[Dakhel et~al\mbox{.}(2022)]%
        {dakhel_github_2022}
\bibfield{author}{\bibinfo{person}{Arghavan~Moradi Dakhel},
  \bibinfo{person}{Vahid Majdinasab}, \bibinfo{person}{Amin Nikanjam},
  \bibinfo{person}{Foutse Khomh}, \bibinfo{person}{Michel~C. Desmarais},
  \bibinfo{person}{Zhen Ming}, {and} \bibinfo{person}{{Jiang}}.}
  \bibinfo{year}{2022}\natexlab{}.
\newblock \bibinfo{title}{{GitHub} {Copilot} {AI} pair programmer: {Asset} or
  {Liability}?}
\newblock
\newblock
\urldef\tempurl%
\url{https://doi.org/10.48550/ARXIV.2206.15331}
\showDOI{\tempurl}


\bibitem[De-Arteaga et~al\mbox{.}(2020)]%
        {de-arteaga_case_2020}
\bibfield{author}{\bibinfo{person}{Maria De-Arteaga}, \bibinfo{person}{Riccardo
  Fogliato}, {and} \bibinfo{person}{Alexandra Chouldechova}.}
  \bibinfo{year}{2020}\natexlab{}.
\newblock \showarticletitle{A {Case} for {Humans}-in-the-{Loop}: {Decisions} in
  the {Presence} of {Erroneous} {Algorithmic} {Scores}}. In
  \bibinfo{booktitle}{\emph{Proceedings of the 2020 {CHI} {Conference} on
  {Human} {Factors} in {Computing} {Systems}}} \emph{(\bibinfo{series}{{CHI}
  '20})}. \bibinfo{publisher}{Association for Computing Machinery},
  \bibinfo{address}{New York, NY, USA}, \bibinfo{pages}{1--12}.
\newblock
\showISBNx{978-1-4503-6708-0}
\urldef\tempurl%
\url{https://doi.org/10.1145/3313831.3376638}
\showDOI{\tempurl}


\bibitem[Denny et~al\mbox{.}(2023)]%
        {denny_promptly_2023}
\bibfield{author}{\bibinfo{person}{Paul Denny}, \bibinfo{person}{Juho
  Leinonen}, \bibinfo{person}{James Prather}, \bibinfo{person}{Andrew
  Luxton-Reilly}, \bibinfo{person}{Thezyrie Amarouche},
  \bibinfo{person}{Brett~A. Becker}, {and} \bibinfo{person}{Brent~N. Reeves}.}
  \bibinfo{year}{2023}\natexlab{}.
\newblock \bibinfo{title}{Promptly: {Using} {Prompt} {Problems} to {Teach}
  {Learners} {How} to {Effectively} {Utilize} {AI} {Code} {Generators}}.
\newblock
\newblock
\urldef\tempurl%
\url{http://arxiv.org/abs/2307.16364}
\showURL{%
\tempurl}
\newblock
\shownote{arXiv:2307.16364 [cs]}.


\bibitem[Druga and Ko(2021)]%
        {druga_how_2021}
\bibfield{author}{\bibinfo{person}{Stefania Druga} {and} \bibinfo{person}{Amy~J
  Ko}.} \bibinfo{year}{2021}\natexlab{}.
\newblock \showarticletitle{How do children’s perceptions of machine
  intelligence change when training and coding smart programs?}. In
  \bibinfo{booktitle}{\emph{Interaction {Design} and {Children}}}.
  \bibinfo{publisher}{ACM}, \bibinfo{address}{Athens Greece},
  \bibinfo{pages}{49--61}.
\newblock
\showISBNx{978-1-4503-8452-0}
\urldef\tempurl%
\url{https://doi.org/10.1145/3459990.3460712}
\showDOI{\tempurl}


\bibitem[Edwards(2004)]%
        {edwards_using_2004}
\bibfield{author}{\bibinfo{person}{Stephen~H. Edwards}.}
  \bibinfo{year}{2004}\natexlab{}.
\newblock \showarticletitle{Using {Software} {Testing} to {Move} {Students}
  from {Trial}-and-{Error} to {Reflection}-in-{Action}}. In
  \bibinfo{booktitle}{\emph{Proceedings of the 35th {SIGCSE} {Technical}
  {Symposium} on {Computer} {Science} {Education}}}. \bibinfo{publisher}{ACM},
  \bibinfo{address}{Norfolk Virginia USA}, \bibinfo{pages}{26--30}.
\newblock
\showISBNx{978-1-58113-798-9}
\urldef\tempurl%
\url{https://doi.org/10.1145/971300.971312}
\showDOI{\tempurl}


\bibitem[Feldman et~al\mbox{.}(2018)]%
        {feldman_automatic_2018}
\bibfield{author}{\bibinfo{person}{Molly~Q Feldman}, \bibinfo{person}{Ji~Yong
  Cho}, \bibinfo{person}{Monica Ong}, \bibinfo{person}{Sumit Gulwani},
  \bibinfo{person}{Zoran Popović}, {and} \bibinfo{person}{Erik Andersen}.}
  \bibinfo{year}{2018}\natexlab{}.
\newblock \showarticletitle{Automatic diagnosis of students' misconceptions in
  k-8 mathematics}. In \bibinfo{booktitle}{\emph{Proceedings of the 2018 {CHI}
  {Conference} on {Human} {Factors} in {Computing} {Systems}}}.
  \bibinfo{pages}{1--12}.
\newblock


\bibitem[Ferdowsi et~al\mbox{.}(2023)]%
        {ferdowsi_live_2023}
\bibfield{author}{\bibinfo{person}{Kasra Ferdowsi},
  \bibinfo{person}{Ruanqianqian Huang}, \bibinfo{person}{Michael~B. James},
  \bibinfo{person}{Nadia Polikarpova}, {and} \bibinfo{person}{Sorin Lerner}.}
  \bibinfo{year}{2023}\natexlab{}.
\newblock \bibinfo{title}{Live {Exploration} of {AI}-{Generated} {Programs}}.
\newblock
\newblock
\urldef\tempurl%
\url{http://arxiv.org/abs/2306.09541}
\showURL{%
\tempurl}
\newblock
\shownote{arXiv:2306.09541 [cs]}.


\bibitem[Fincher et~al\mbox{.}(2005)]%
        {fincher_multi-institutional_2005}
\bibfield{author}{\bibinfo{person}{Sally Fincher}, \bibinfo{person}{Raymond
  Lister}, \bibinfo{person}{Tony Clear}, \bibinfo{person}{Anthony Robins},
  \bibinfo{person}{Josh Tenenberg}, {and} \bibinfo{person}{Marian Petre}.}
  \bibinfo{year}{2005}\natexlab{}.
\newblock \showarticletitle{Multi-institutional, multi-national studies in
  {CSEd} {Research}: some design considerations and trade-offs}. In
  \bibinfo{booktitle}{\emph{Proceedings of the first international workshop on
  {Computing} education research}} \emph{(\bibinfo{series}{{ICER} '05})}.
  \bibinfo{publisher}{Association for Computing Machinery},
  \bibinfo{address}{New York, NY, USA}, \bibinfo{pages}{111--121}.
\newblock
\showISBNx{978-1-59593-043-9}
\urldef\tempurl%
\url{https://doi.org/10.1145/1089786.1089797}
\showDOI{\tempurl}


\bibitem[Finnie-Ansley et~al\mbox{.}(2022)]%
        {finnie-ansley_robots_2022}
\bibfield{author}{\bibinfo{person}{James Finnie-Ansley}, \bibinfo{person}{Paul
  Denny}, \bibinfo{person}{Brett~A. Becker}, \bibinfo{person}{Andrew
  Luxton-Reilly}, {and} \bibinfo{person}{James Prather}.}
  \bibinfo{year}{2022}\natexlab{}.
\newblock \showarticletitle{The {Robots} {Are} {Coming}: {Exploring} the
  {Implications} of {OpenAI} {Codex} on {Introductory} {Programming}}. In
  \bibinfo{booktitle}{\emph{Australasian {Computing} {Education} {Conference}}}
  \emph{(\bibinfo{series}{{ACE} '22})}. \bibinfo{publisher}{Association for
  Computing Machinery}, \bibinfo{address}{New York, NY, USA},
  \bibinfo{pages}{10--19}.
\newblock
\showISBNx{978-1-4503-9643-1}
\urldef\tempurl%
\url{https://doi.org/10.1145/3511861.3511863}
\showDOI{\tempurl}


\bibitem[First et~al\mbox{.}(2023)]%
        {first_baldur_2023}
\bibfield{author}{\bibinfo{person}{Emily First}, \bibinfo{person}{Markus Rabe},
  \bibinfo{person}{Talia Ringer}, {and} \bibinfo{person}{Yuriy Brun}.}
  \bibinfo{year}{2023}\natexlab{}.
\newblock \showarticletitle{Baldur: {Whole}-{Proof} {Generation} and {Repair}
  with {Large} {Language} {Models}}. In \bibinfo{booktitle}{\emph{Proceedings
  of the 31st {ACM} {Joint} {European} {Software} {Engineering} {Conference}
  and {Symposium} on the {Foundations} of {Software} {Engineering}}}.
  \bibinfo{publisher}{ACM}, \bibinfo{address}{San Francisco CA USA},
  \bibinfo{pages}{1229--1241}.
\newblock
\showISBNx{9798400703270}
\urldef\tempurl%
\url{https://doi.org/10.1145/3611643.3616243}
\showDOI{\tempurl}


\bibitem[Flatt et~al\mbox{.}(2001)]%
        {flatt_how_2001}
\bibfield{author}{\bibinfo{person}{Matthew Flatt}, \bibinfo{person}{Matthias
  Felleisen}, \bibinfo{person}{Robert~Bruce Findler}, {and}
  \bibinfo{person}{Shriram Krishnamurthi}.} \bibinfo{year}{2001}\natexlab{}.
\newblock \bibinfo{booktitle}{\emph{How {To} {Design} {Programs}}}.
\newblock \bibinfo{publisher}{MIT Press}.
\newblock


\bibitem[Fowler et~al\mbox{.}(2021)]%
        {fowler_autograding_2021}
\bibfield{author}{\bibinfo{person}{Max Fowler}, \bibinfo{person}{Binglin Chen},
  \bibinfo{person}{Sushmita Azad}, \bibinfo{person}{Matthew West}, {and}
  \bibinfo{person}{Craig Zilles}.} \bibinfo{year}{2021}\natexlab{}.
\newblock \showarticletitle{Autograding "{Explain} in {Plain} {English}"
  questions using {NLP}}. In \bibinfo{booktitle}{\emph{Proceedings of the 52nd
  {ACM} {Technical} {Symposium} on {Computer} {Science} {Education}}}.
  \bibinfo{publisher}{ACM}, \bibinfo{address}{Virtual Event USA},
  \bibinfo{pages}{1163--1169}.
\newblock
\showISBNx{978-1-4503-8062-1}
\urldef\tempurl%
\url{https://doi.org/10.1145/3408877.3432539}
\showDOI{\tempurl}


\bibitem[Fried et~al\mbox{.}(2023)]%
        {fried_incoder_2023}
\bibfield{author}{\bibinfo{person}{Daniel Fried}, \bibinfo{person}{Armen
  Aghajanyan}, \bibinfo{person}{Jessy Lin}, \bibinfo{person}{Sida Wang},
  \bibinfo{person}{Eric Wallace}, \bibinfo{person}{Freda Shi},
  \bibinfo{person}{Ruiqi Zhong}, \bibinfo{person}{Wen-tau Yih},
  \bibinfo{person}{Luke Zettlemoyer}, {and} \bibinfo{person}{Mike Lewis}.}
  \bibinfo{year}{2023}\natexlab{}.
\newblock \bibinfo{title}{{InCoder}: {A} {Generative} {Model} for {Code}
  {Infilling} and {Synthesis}}.
\newblock
\newblock
\urldef\tempurl%
\url{http://arxiv.org/abs/2204.05999}
\showURL{%
\tempurl}
\newblock
\shownote{arXiv:2204.05999 [cs]}.


\bibitem[Gadala(2017)]%
        {gadala_automation_2017}
\bibfield{author}{\bibinfo{person}{Marwa Gadala}.}
  \bibinfo{year}{2017}\natexlab{}.
\newblock \showarticletitle{Automation bias: exploring causal mechanisms and
  potential mitigation strategies}.
\newblock
\urldef\tempurl%
\url{https://api.semanticscholar.org/CorpusID:41123263}
\showURL{%
\tempurl}


\bibitem[Geng et~al\mbox{.}(2022)]%
        {geng_novice_2022}
\bibfield{author}{\bibinfo{person}{Chuqin Geng}, \bibinfo{person}{Haolin Ye},
  \bibinfo{person}{Yixuan Li}, \bibinfo{person}{Tianyu Han},
  \bibinfo{person}{Brigitte Pientka}, {and} \bibinfo{person}{Xujie Si}.}
  \bibinfo{year}{2022}\natexlab{}.
\newblock \bibinfo{title}{Novice {Type} {Error} {Diagnosis} with {Natural}
  {Language} {Models}}.
\newblock
\newblock
\urldef\tempurl%
\url{http://arxiv.org/abs/2210.03682}
\showURL{%
\tempurl}
\newblock
\shownote{arXiv:2210.03682 [cs]}.


\bibitem[Goddard et~al\mbox{.}(2012)]%
        {goddard_automation_2012}
\bibfield{author}{\bibinfo{person}{Kate Goddard}, \bibinfo{person}{Abdul~V.
  Roudsari}, {and} \bibinfo{person}{Jeremy~C. Wyatt}.}
  \bibinfo{year}{2012}\natexlab{}.
\newblock \showarticletitle{Automation bias: a systematic review of frequency,
  effect mediators, and mitigators}.
\newblock \bibinfo{journal}{\emph{Journal of the American Medical Informatics
  Association : JAMIA}}  \bibinfo{volume}{19 1} (\bibinfo{year}{2012}),
  \bibinfo{pages}{121--7}.
\newblock


\bibitem[Gorson and O'Rourke(2020)]%
        {gorson_why_2020}
\bibfield{author}{\bibinfo{person}{Jamie Gorson} {and} \bibinfo{person}{Eleanor
  O'Rourke}.} \bibinfo{year}{2020}\natexlab{}.
\newblock \showarticletitle{Why do {CS1} {Students} {Think} {They}'re {Bad} at
  {Programming}?: {Investigating} {Self}-efficacy and {Self}-assessments at
  {Three} {Universities}}. In \bibinfo{booktitle}{\emph{Proceedings of the 2020
  {ACM} {Conference} on {International} {Computing} {Education} {Research}}}.
  \bibinfo{publisher}{ACM}, \bibinfo{address}{Virtual Event New Zealand},
  \bibinfo{pages}{170--181}.
\newblock
\showISBNx{978-1-4503-7092-9}
\urldef\tempurl%
\url{https://doi.org/10.1145/3372782.3406273}
\showDOI{\tempurl}


\bibitem[Goues et~al\mbox{.}(2019)]%
        {goues_automated_2019}
\bibfield{author}{\bibinfo{person}{Claire~Le Goues}, \bibinfo{person}{Michael
  Pradel}, {and} \bibinfo{person}{Abhik Roychoudhury}.}
  \bibinfo{year}{2019}\natexlab{}.
\newblock \showarticletitle{Automated program repair}.
\newblock \bibinfo{journal}{\emph{Commun. ACM}} \bibinfo{volume}{62},
  \bibinfo{number}{12} (\bibinfo{year}{2019}), \bibinfo{pages}{56--65}.
\newblock
\newblock
\shownote{Publisher: ACM New York, NY, USA}.


\bibitem[Gulwani et~al\mbox{.}(2017)]%
        {gulwani_program_2017}
\bibfield{author}{\bibinfo{person}{Sumit Gulwani}, \bibinfo{person}{Oleksandr
  Polozov}, \bibinfo{person}{Rishabh Singh}, {and} \bibinfo{person}{{others}}.}
  \bibinfo{year}{2017}\natexlab{}.
\newblock \showarticletitle{Program synthesis}.
\newblock \bibinfo{journal}{\emph{Foundations and Trends® in Programming
  Languages}} \bibinfo{volume}{4}, \bibinfo{number}{1-2}
  (\bibinfo{year}{2017}), \bibinfo{pages}{1--119}.
\newblock
\newblock
\shownote{Publisher: Now Publishers, Inc.}.


\bibitem[Hart and Staveland(1988)]%
        {hart_development_1988}
\bibfield{author}{\bibinfo{person}{Sandra~G. Hart} {and}
  \bibinfo{person}{Lowell~E. Staveland}.} \bibinfo{year}{1988}\natexlab{}.
\newblock \showarticletitle{Development of {NASA}-{TLX} ({Task} {Load}
  {Index}): {Results} of {Empirical} and {Theoretical} {Research}}.
\newblock In \bibinfo{booktitle}{\emph{Advances in {Psychology}}},
  \bibfield{editor}{\bibinfo{person}{Peter~A. Hancock} {and}
  \bibinfo{person}{Najmedin Meshkati}} (Eds.). \bibinfo{series}{Human {Mental}
  {Workload}}, Vol.~\bibinfo{volume}{52}. \bibinfo{publisher}{North-Holland},
  \bibinfo{pages}{139--183}.
\newblock
\urldef\tempurl%
\url{https://doi.org/10.1016/S0166-4115(08)62386-9}
\showDOI{\tempurl}


\bibitem[Head et~al\mbox{.}(2017)]%
        {head_writing_2017}
\bibfield{author}{\bibinfo{person}{Andrew Head}, \bibinfo{person}{Elena
  Glassman}, \bibinfo{person}{Gustavo Soares}, \bibinfo{person}{Ryo Suzuki},
  \bibinfo{person}{Lucas Figueredo}, \bibinfo{person}{Loris D'Antoni}, {and}
  \bibinfo{person}{Björn Hartmann}.} \bibinfo{year}{2017}\natexlab{}.
\newblock \showarticletitle{Writing reusable code feedback at scale with
  mixed-initiative program synthesis}. In \bibinfo{booktitle}{\emph{Proceedings
  of the {Fourth} (2017) {ACM} {Conference} on {Learning}@ {Scale}}}.
  \bibinfo{pages}{89--98}.
\newblock


\bibitem[Heidorn(1974)]%
        {heidorn_english_1974}
\bibfield{author}{\bibinfo{person}{George~E. Heidorn}.}
  \bibinfo{year}{1974}\natexlab{}.
\newblock \showarticletitle{English as a {Very} {High} {Level} {Language} for
  {Simulation} {Programming}}. In \bibinfo{booktitle}{\emph{Proceedings of the
  {ACM} {SIGPLAN} {Symposium} on {Very} {High} {Level} {Languages}}}.
  \bibinfo{publisher}{Association for Computing Machinery},
  \bibinfo{address}{New York, NY, USA}, \bibinfo{pages}{91--100}.
\newblock
\showISBNx{978-1-4503-7884-0}
\urldef\tempurl%
\url{https://doi.org/10.1145/800233.807050}
\showDOI{\tempurl}


\bibitem[Hindle et~al\mbox{.}(2012)]%
        {hindle_naturalness_2012}
\bibfield{author}{\bibinfo{person}{Abram Hindle}, \bibinfo{person}{Earl~T.
  Barr}, \bibinfo{person}{Zhendong Su}, \bibinfo{person}{Mark Gabel}, {and}
  \bibinfo{person}{Premkumar Devanbu}.} \bibinfo{year}{2012}\natexlab{}.
\newblock \showarticletitle{On the naturalness of software}. In
  \bibinfo{booktitle}{\emph{Proceedings of the 34th {International}
  {Conference} on {Software} {Engineering}}} \emph{(\bibinfo{series}{{ICSE}
  '12})}. \bibinfo{publisher}{IEEE Press}, \bibinfo{address}{Zurich,
  Switzerland}, \bibinfo{pages}{837--847}.
\newblock
\showISBNx{978-1-4673-1067-3}


\bibitem[Hollingsworth(1960)]%
        {hollingsworth_automatic_1960}
\bibfield{author}{\bibinfo{person}{Jack Hollingsworth}.}
  \bibinfo{year}{1960}\natexlab{}.
\newblock \showarticletitle{Automatic graders for programming classes}.
\newblock \bibinfo{journal}{\emph{Commun. ACM}} \bibinfo{volume}{3},
  \bibinfo{number}{10} (\bibinfo{year}{1960}), \bibinfo{pages}{528--529}.
\newblock
\newblock
\shownote{Publisher: ACM New York, NY, USA}.


\bibitem[Holstein and Doroudi(2021)]%
        {holstein_equity_2021}
\bibfield{author}{\bibinfo{person}{Kenneth Holstein} {and}
  \bibinfo{person}{Shayan Doroudi}.} \bibinfo{year}{2021}\natexlab{}.
\newblock \bibinfo{title}{Equity and {Artificial} {Intelligence} in
  {Education}: {Will} "{AIEd}" {Amplify} or {Alleviate} {Inequities} in
  {Education}?}
\newblock
\newblock
\urldef\tempurl%
\url{http://arxiv.org/abs/2104.12920}
\showURL{%
\tempurl}
\newblock
\shownote{arXiv:2104.12920 [cs]}.


\bibitem[Holstein et~al\mbox{.}(2018)]%
        {holstein_student_2018}
\bibfield{author}{\bibinfo{person}{Kenneth Holstein}, \bibinfo{person}{Bruce~M
  McLaren}, {and} \bibinfo{person}{Vincent Aleven}.}
  \bibinfo{year}{2018}\natexlab{}.
\newblock \showarticletitle{Student learning benefits of a mixed-reality
  teacher awareness tool in {AI}-enhanced classrooms}. In
  \bibinfo{booktitle}{\emph{Artificial {Intelligence} in {Education}: 19th
  {International} {Conference}, {AIED} 2018, {London}, {UK}, {June} 27–30,
  2018, {Proceedings}, {Part} {I} 19}}. \bibinfo{publisher}{Springer},
  \bibinfo{pages}{154--168}.
\newblock


\bibitem[Holtzman et~al\mbox{.}(2020)]%
        {holtzman_curious_2020}
\bibfield{author}{\bibinfo{person}{Ari Holtzman}, \bibinfo{person}{Jan Buys},
  \bibinfo{person}{Li Du}, \bibinfo{person}{Maxwell Forbes}, {and}
  \bibinfo{person}{Yejin Choi}.} \bibinfo{year}{2020}\natexlab{}.
\newblock \showarticletitle{The {Curious} {Case} of {Neural} {Text}
  {Degeneration}}. In \bibinfo{booktitle}{\emph{8th {International}
  {Conference} on {Learning} {Representations}, {ICLR} 2020, {Addis} {Ababa},
  {Ethiopia}, {April} 26-30, 2020}}. \bibinfo{publisher}{OpenReview.net}.
\newblock
\urldef\tempurl%
\url{https://openreview.net/forum?id=rygGQyrFvH}
\showURL{%
\tempurl}


\bibitem[Hsu et~al\mbox{.}(2021)]%
        {hsu_attitudes_2021}
\bibfield{author}{\bibinfo{person}{Silas Hsu}, \bibinfo{person}{Tiffany~Wenting
  Li}, \bibinfo{person}{Zhilin Zhang}, \bibinfo{person}{Max Fowler},
  \bibinfo{person}{Craig Zilles}, {and} \bibinfo{person}{Karrie Karahalios}.}
  \bibinfo{year}{2021}\natexlab{}.
\newblock \showarticletitle{Attitudes {Surrounding} an {Imperfect} {AI}
  {Autograder}}. In \bibinfo{booktitle}{\emph{Proceedings of the 2021 {CHI}
  {Conference} on {Human} {Factors} in {Computing} {Systems}}}.
  \bibinfo{publisher}{ACM}, \bibinfo{address}{Yokohama Japan},
  \bibinfo{pages}{1--15}.
\newblock
\showISBNx{978-1-4503-8096-6}
\urldef\tempurl%
\url{https://doi.org/10.1145/3411764.3445424}
\showDOI{\tempurl}


\bibitem[Ippolito et~al\mbox{.}(2022)]%
        {ippolito_creative_2022}
\bibfield{author}{\bibinfo{person}{Daphne Ippolito}, \bibinfo{person}{Ann
  Yuan}, \bibinfo{person}{Andy Coenen}, {and} \bibinfo{person}{Sehmon Burnam}.}
  \bibinfo{year}{2022}\natexlab{}.
\newblock \bibinfo{title}{Creative {Writing} with an {AI}-{Powered} {Writing}
  {Assistant}: {Perspectives} from {Professional} {Writers}}.
\newblock
\newblock
\urldef\tempurl%
\url{http://arxiv.org/abs/2211.05030}
\showURL{%
\tempurl}
\newblock
\shownote{arXiv:2211.05030 [cs]}.


\bibitem[Jia and Harman(2011)]%
        {jia_analysis_2011}
\bibfield{author}{\bibinfo{person}{Yue Jia} {and} \bibinfo{person}{Mark
  Harman}.} \bibinfo{year}{2011}\natexlab{}.
\newblock \showarticletitle{An {Analysis} and {Survey} of the {Development} of
  {Mutation} {Testing}}.
\newblock \bibinfo{journal}{\emph{IEEE Transactions on Software Engineering}}
  \bibinfo{volume}{37}, \bibinfo{number}{5} (\bibinfo{date}{Sept.}
  \bibinfo{year}{2011}), \bibinfo{pages}{649--678}.
\newblock
\showISSN{0098-5589}
\urldef\tempurl%
\url{https://doi.org/10.1109/TSE.2010.62}
\showDOI{\tempurl}


\bibitem[Joshi et~al\mbox{.}(2023)]%
        {joshi_repair_2023}
\bibfield{author}{\bibinfo{person}{Harshit Joshi},
  \bibinfo{person}{José~Cambronero Sanchez}, \bibinfo{person}{Sumit Gulwani},
  \bibinfo{person}{Vu Le}, \bibinfo{person}{Gust Verbruggen}, {and}
  \bibinfo{person}{Ivan Radiček}.} \bibinfo{year}{2023}\natexlab{}.
\newblock \showarticletitle{Repair {Is} {Nearly} {Generation}: {Multilingual}
  {Program} {Repair} with {LLMs}}.
\newblock \bibinfo{journal}{\emph{Proceedings of the AAAI Conference on
  Artificial Intelligence}} \bibinfo{volume}{37}, \bibinfo{number}{4}
  (\bibinfo{date}{June} \bibinfo{year}{2023}), \bibinfo{pages}{5131--5140}.
\newblock
\showISSN{2374-3468}
\urldef\tempurl%
\url{https://doi.org/10.1609/aaai.v37i4.25642}
\showDOI{\tempurl}


\bibitem[Kazemitabaar et~al\mbox{.}(2023)]%
        {kazemitabaar_studying_2023}
\bibfield{author}{\bibinfo{person}{Majeed Kazemitabaar},
  \bibinfo{person}{Justin Chow}, \bibinfo{person}{Carl Ka~To Ma},
  \bibinfo{person}{Barbara~J. Ericson}, \bibinfo{person}{David Weintrop}, {and}
  \bibinfo{person}{Tovi Grossman}.} \bibinfo{year}{2023}\natexlab{}.
\newblock \showarticletitle{Studying the effect of {AI} {Code} {Generators} on
  {Supporting} {Novice} {Learners} in {Introductory} {Programming}}. In
  \bibinfo{booktitle}{\emph{Proceedings of the 2023 {CHI} {Conference} on
  {Human} {Factors} in {Computing} {Systems}}}. \bibinfo{publisher}{ACM},
  \bibinfo{address}{Hamburg Germany}, \bibinfo{pages}{1--23}.
\newblock
\showISBNx{978-1-4503-9421-5}
\urldef\tempurl%
\url{https://doi.org/10.1145/3544548.3580919}
\showDOI{\tempurl}


\bibitem[Khlaaf et~al\mbox{.}(2022)]%
        {khlaaf_hazard_2022}
\bibfield{author}{\bibinfo{person}{Heidy Khlaaf}, \bibinfo{person}{Pamela
  Mishkin}, \bibinfo{person}{Joshua Achiam}, \bibinfo{person}{Gretchen
  Krueger}, {and} \bibinfo{person}{Miles Brundage}.}
  \bibinfo{year}{2022}\natexlab{}.
\newblock \bibinfo{title}{A {Hazard} {Analysis} {Framework} for {Code}
  {Synthesis} {Large} {Language} {Models}}.
\newblock
\newblock
\urldef\tempurl%
\url{http://arxiv.org/abs/2207.14157}
\showURL{%
\tempurl}
\newblock
\shownote{arXiv:2207.14157 [cs]}.


\bibitem[Ko et~al\mbox{.}(2004)]%
        {ko_six_2004}
\bibfield{author}{\bibinfo{person}{Amy~J. Ko}, \bibinfo{person}{Brad~A. Myers},
  {and} \bibinfo{person}{Htet~Htet Aung}.} \bibinfo{year}{2004}\natexlab{}.
\newblock \showarticletitle{Six {Learning} {Barriers} in {End}-{User}
  {Programming} {Systems}}. In \bibinfo{booktitle}{\emph{2004 {IEEE}
  {Symposium} on {Visual} {Languages} - {Human} {Centric} {Computing}}}.
  \bibinfo{publisher}{IEEE}, \bibinfo{address}{Rome, Italy},
  \bibinfo{pages}{199--206}.
\newblock
\showISBNx{978-0-7803-8696-9}
\urldef\tempurl%
\url{https://doi.org/10.1109/VLHCC.2004.47}
\showDOI{\tempurl}


\bibitem[Körber(2018)]%
        {korber_theoretical_2018}
\bibfield{author}{\bibinfo{person}{Moritz Körber}.}
  \bibinfo{year}{2018}\natexlab{}.
\newblock \showarticletitle{Theoretical considerations and development of a
  questionnaire to measure trust in automation}. In
  \bibinfo{booktitle}{\emph{Congress of the {International} {Ergonomics}
  {Association}}}. \bibinfo{publisher}{Springer}, \bibinfo{pages}{13--30}.
\newblock


\bibitem[Lau and Guo(2023)]%
        {lau_ban_2023}
\bibfield{author}{\bibinfo{person}{Sam Lau} {and} \bibinfo{person}{Philip
  Guo}.} \bibinfo{year}{2023}\natexlab{}.
\newblock \showarticletitle{From "{Ban} {It} {Till} {We} {Understand} {It}" to
  "{Resistance} is {Futile}": {How} {University} {Programming} {Instructors}
  {Plan} to {Adapt} as {More} {Students} {Use} {AI} {Code} {Generation} and
  {Explanation} {Tools} such as {ChatGPT} and {GitHub} {Copilot}}. In
  \bibinfo{booktitle}{\emph{Proceedings of the 2023 {ACM} {Conference} on
  {International} {Computing} {Education} {Research} - {Volume} 1}}
  \emph{(\bibinfo{series}{{ICER} '23})}. \bibinfo{publisher}{Association for
  Computing Machinery}, \bibinfo{address}{New York, NY, USA},
  \bibinfo{pages}{106--121}.
\newblock
\showISBNx{978-1-4503-9976-0}
\urldef\tempurl%
\url{https://doi.org/10.1145/3568813.3600138}
\showDOI{\tempurl}


\bibitem[Lau(2009)]%
        {lau_why_2009}
\bibfield{author}{\bibinfo{person}{Tessa Lau}.}
  \bibinfo{year}{2009}\natexlab{}.
\newblock \showarticletitle{Why {Programming}-{By}-{Demonstration} {Systems}
  {Fail}: {Lessons} {Learned} for {Usable} {AI}}.
\newblock \bibinfo{journal}{\emph{AI Magazine}} \bibinfo{volume}{30},
  \bibinfo{number}{4} (\bibinfo{date}{Oct.} \bibinfo{year}{2009}),
  \bibinfo{pages}{65--65}.
\newblock
\showISSN{2371-9621}
\urldef\tempurl%
\url{https://doi.org/10.1609/aimag.v30i4.2262}
\showDOI{\tempurl}


\bibitem[Lee and Ko(2011)]%
        {lee_personifying_2011}
\bibfield{author}{\bibinfo{person}{Michael~J Lee} {and} \bibinfo{person}{Amy~J
  Ko}.} \bibinfo{year}{2011}\natexlab{}.
\newblock \showarticletitle{Personifying programming tool feedback improves
  novice programmers' learning}. In \bibinfo{booktitle}{\emph{Proceedings of
  the seventh international workshop on {Computing} education research}}.
  \bibinfo{pages}{109--116}.
\newblock


\bibitem[Leinonen et~al\mbox{.}(2023a)]%
        {leinonen_comparing_2023}
\bibfield{author}{\bibinfo{person}{Juho Leinonen}, \bibinfo{person}{Paul
  Denny}, \bibinfo{person}{Stephen MacNeil}, \bibinfo{person}{Sami Sarsa},
  \bibinfo{person}{Seth Bernstein}, \bibinfo{person}{Joanne Kim},
  \bibinfo{person}{Andrew Tran}, {and} \bibinfo{person}{Arto Hellas}.}
  \bibinfo{year}{2023}\natexlab{a}.
\newblock \showarticletitle{Comparing {Code} {Explanations} {Created} by
  {Students} and {Large} {Language} {Models}}. In
  \bibinfo{booktitle}{\emph{Proceedings of the 2023 {Conference} on
  {Innovation} and {Technology} in {Computer} {Science} {Education} {V}. 1}}.
  \bibinfo{publisher}{ACM}, \bibinfo{address}{Turku Finland},
  \bibinfo{pages}{124--130}.
\newblock
\showISBNx{9798400701382}
\urldef\tempurl%
\url{https://doi.org/10.1145/3587102.3588785}
\showDOI{\tempurl}


\bibitem[Leinonen et~al\mbox{.}(2023b)]%
        {leinonen_using_2023}
\bibfield{author}{\bibinfo{person}{Juho Leinonen}, \bibinfo{person}{Arto
  Hellas}, \bibinfo{person}{Sami Sarsa}, \bibinfo{person}{Brent Reeves},
  \bibinfo{person}{Paul Denny}, \bibinfo{person}{James Prather}, {and}
  \bibinfo{person}{Brett~A. Becker}.} \bibinfo{year}{2023}\natexlab{b}.
\newblock \showarticletitle{Using {Large} {Language} {Models} to {Enhance}
  {Programming} {Error} {Messages}}. In \bibinfo{booktitle}{\emph{Proceedings
  of the 54th {ACM} {Technical} {Symposium} on {Computer} {Science} {Education}
  {V}. 1}}. \bibinfo{publisher}{ACM}, \bibinfo{address}{Toronto ON Canada},
  \bibinfo{pages}{563--569}.
\newblock
\showISBNx{978-1-4503-9431-4}
\urldef\tempurl%
\url{https://doi.org/10.1145/3545945.3569770}
\showDOI{\tempurl}


\bibitem[Lemieux et~al\mbox{.}(2023)]%
        {lemieux_codamosa_2023}
\bibfield{author}{\bibinfo{person}{Caroline Lemieux},
  \bibinfo{person}{Jeevana~Priya Inala}, \bibinfo{person}{Shuvendu~K. Lahiri},
  {and} \bibinfo{person}{Siddhartha Sen}.} \bibinfo{year}{2023}\natexlab{}.
\newblock \showarticletitle{{CodaMosa}: {Escaping} {Coverage} {Plateaus} in
  {Test} {Generation} with {Pre}-trained {Large} {Language} {Models}}. In
  \bibinfo{booktitle}{\emph{2023 {IEEE}/{ACM} 45th {International} {Conference}
  on {Software} {Engineering} ({ICSE})}}. \bibinfo{publisher}{IEEE},
  \bibinfo{address}{Melbourne, Australia}, \bibinfo{pages}{919--931}.
\newblock
\showISBNx{978-1-66545-701-9}
\urldef\tempurl%
\url{https://doi.org/10.1109/ICSE48619.2023.00085}
\showDOI{\tempurl}


\bibitem[Leswing(2023)]%
        {leswing_chatgpt_2023}
\bibfield{author}{\bibinfo{person}{Jonathan~Vanian Leswing, Kif}.}
  \bibinfo{year}{2023}\natexlab{}.
\newblock \bibinfo{title}{{ChatGPT} and generative {AI} are booming, but the
  costs can be extraordinary}.
\newblock
\newblock
\urldef\tempurl%
\url{https://www.cnbc.com/2023/03/13/chatgpt-and-generative-ai-are-booming-but-at-a-very-expensive-price.html}
\showURL{%
\tempurl}


\bibitem[Li et~al\mbox{.}(2023)]%
        {li_starcoder_2023}
\bibfield{author}{\bibinfo{person}{Raymond Li}, \bibinfo{person}{Loubna~Ben
  Allal}, \bibinfo{person}{Yangtian Zi}, \bibinfo{person}{Niklas Muennighoff},
  \bibinfo{person}{Denis Kocetkov}, \bibinfo{person}{Chenghao Mou},
  \bibinfo{person}{Marc Marone}, \bibinfo{person}{Christopher Akiki},
  \bibinfo{person}{Jia Li}, \bibinfo{person}{Jenny Chim}, \bibinfo{person}{Qian
  Liu}, \bibinfo{person}{Evgenii Zheltonozhskii}, \bibinfo{person}{Terry~Yue
  Zhuo}, \bibinfo{person}{Thomas Wang}, \bibinfo{person}{Olivier Dehaene},
  \bibinfo{person}{Mishig Davaadorj}, \bibinfo{person}{Joel Lamy-Poirier},
  \bibinfo{person}{João Monteiro}, \bibinfo{person}{Oleh Shliazhko},
  \bibinfo{person}{Nicolas Gontier}, \bibinfo{person}{Nicholas Meade},
  \bibinfo{person}{Armel Zebaze}, \bibinfo{person}{Ming-Ho Yee},
  \bibinfo{person}{Logesh~Kumar Umapathi}, \bibinfo{person}{Jian Zhu},
  \bibinfo{person}{Benjamin Lipkin}, \bibinfo{person}{Muhtasham Oblokulov},
  \bibinfo{person}{Zhiruo Wang}, \bibinfo{person}{Rudra Murthy},
  \bibinfo{person}{Jason Stillerman}, \bibinfo{person}{Siva~Sankalp Patel},
  \bibinfo{person}{Dmitry Abulkhanov}, \bibinfo{person}{Marco Zocca},
  \bibinfo{person}{Manan Dey}, \bibinfo{person}{Zhihan Zhang},
  \bibinfo{person}{Nour Fahmy}, \bibinfo{person}{Urvashi Bhattacharyya},
  \bibinfo{person}{Wenhao Yu}, \bibinfo{person}{Swayam Singh},
  \bibinfo{person}{Sasha Luccioni}, \bibinfo{person}{Paulo Villegas},
  \bibinfo{person}{Maxim Kunakov}, \bibinfo{person}{Fedor Zhdanov},
  \bibinfo{person}{Manuel Romero}, \bibinfo{person}{Tony Lee},
  \bibinfo{person}{Nadav Timor}, \bibinfo{person}{Jennifer Ding},
  \bibinfo{person}{Claire Schlesinger}, \bibinfo{person}{Hailey Schoelkopf},
  \bibinfo{person}{Jan Ebert}, \bibinfo{person}{Tri Dao},
  \bibinfo{person}{Mayank Mishra}, \bibinfo{person}{Alex Gu},
  \bibinfo{person}{Jennifer Robinson}, \bibinfo{person}{Carolyn~Jane Anderson},
  \bibinfo{person}{Brendan Dolan-Gavitt}, \bibinfo{person}{Danish Contractor},
  \bibinfo{person}{Siva Reddy}, \bibinfo{person}{Daniel Fried},
  \bibinfo{person}{Dzmitry Bahdanau}, \bibinfo{person}{Yacine Jernite},
  \bibinfo{person}{Carlos~Muñoz Ferrandis}, \bibinfo{person}{Sean Hughes},
  \bibinfo{person}{Thomas Wolf}, \bibinfo{person}{Arjun Guha},
  \bibinfo{person}{Leandro von Werra}, {and} \bibinfo{person}{Harm de Vries}.}
  \bibinfo{year}{2023}\natexlab{}.
\newblock \bibinfo{title}{{StarCoder}: may the source be with you!}
\newblock
\newblock
\urldef\tempurl%
\url{http://arxiv.org/abs/2305.06161}
\showURL{%
\tempurl}
\newblock
\shownote{arXiv:2305.06161 [cs]}.


\bibitem[Liang et~al\mbox{.}(2023)]%
        {liang_large-scale_2023}
\bibfield{author}{\bibinfo{person}{Jenny~T. Liang}, \bibinfo{person}{Chenyang
  Yang}, {and} \bibinfo{person}{Brad~A. Myers}.}
  \bibinfo{year}{2023}\natexlab{}.
\newblock \bibinfo{title}{A {Large}-{Scale} {Survey} on the {Usability} of {AI}
  {Programming} {Assistants}: {Successes} and {Challenges}}.
\newblock
\newblock
\urldef\tempurl%
\url{http://arxiv.org/abs/2303.17125}
\showURL{%
\tempurl}
\newblock
\shownote{arXiv:2303.17125 [cs]}.


\bibitem[Liao and Vaughan(2023)]%
        {liao_ai_2023}
\bibfield{author}{\bibinfo{person}{Q.~Vera Liao} {and}
  \bibinfo{person}{Jennifer~Wortman Vaughan}.} \bibinfo{year}{2023}\natexlab{}.
\newblock \bibinfo{title}{{AI} {Transparency} in the {Age} of {LLMs}: {A}
  {Human}-{Centered} {Research} {Roadmap}}.
\newblock
\newblock
\urldef\tempurl%
\url{http://arxiv.org/abs/2306.01941}
\showURL{%
\tempurl}
\newblock
\shownote{arXiv:2306.01941 [cs]}.


\bibitem[Liu et~al\mbox{.}(2023b)]%
        {liu_what_2023}
\bibfield{author}{\bibinfo{person}{Michael~Xieyang Liu},
  \bibinfo{person}{Advait Sarkar}, \bibinfo{person}{Carina Negreanu},
  \bibinfo{person}{Benjamin Zorn}, \bibinfo{person}{Jack Williams},
  \bibinfo{person}{Neil Toronto}, {and} \bibinfo{person}{Andrew~D. Gordon}.}
  \bibinfo{year}{2023}\natexlab{b}.
\newblock \showarticletitle{“{What} {It} {Wants} {Me} {To} {Say}”:
  {Bridging} the {Abstraction} {Gap} {Between} {End}-{User} {Programmers} and
  {Code}-{Generating} {Large} {Language} {Models}}. In
  \bibinfo{booktitle}{\emph{Proceedings of the 2023 {CHI} {Conference} on
  {Human} {Factors} in {Computing} {Systems}}}. \bibinfo{publisher}{ACM},
  \bibinfo{address}{Hamburg Germany}, \bibinfo{pages}{1--31}.
\newblock
\showISBNx{978-1-4503-9421-5}
\urldef\tempurl%
\url{https://doi.org/10.1145/3544548.3580817}
\showDOI{\tempurl}


\bibitem[Liu et~al\mbox{.}(2023a)]%
        {liu_generative_2023}
\bibfield{author}{\bibinfo{person}{Vivian Liu}, \bibinfo{person}{Tao Long},
  \bibinfo{person}{Nathan Raw}, {and} \bibinfo{person}{Lydia Chilton}.}
  \bibinfo{year}{2023}\natexlab{a}.
\newblock \bibinfo{title}{Generative {Disco}: {Text}-to-{Video} {Generation}
  for {Music} {Visualization}}.
\newblock
\newblock
\urldef\tempurl%
\url{http://arxiv.org/abs/2304.08551}
\showURL{%
\tempurl}


\bibitem[Lopez et~al\mbox{.}(2008)]%
        {lopez_relationships_2008}
\bibfield{author}{\bibinfo{person}{Mike Lopez}, \bibinfo{person}{Jacqueline
  Whalley}, \bibinfo{person}{Phil Robbins}, {and} \bibinfo{person}{Raymond
  Lister}.} \bibinfo{year}{2008}\natexlab{}.
\newblock \showarticletitle{Relationships between reading, tracing and writing
  skills in introductory programming}. In \bibinfo{booktitle}{\emph{Proceedings
  of the {Fourth} international {Workshop} on {Computing} {Education}
  {Research}}} \emph{(\bibinfo{series}{{ICER} '08})}.
  \bibinfo{publisher}{Association for Computing Machinery},
  \bibinfo{address}{New York, NY, USA}, \bibinfo{pages}{101--112}.
\newblock
\showISBNx{978-1-60558-216-0}
\urldef\tempurl%
\url{https://doi.org/10.1145/1404520.1404531}
\showDOI{\tempurl}


\bibitem[Manjoo(2023)]%
        {manjoo_its_2023}
\bibfield{author}{\bibinfo{person}{Farhad Manjoo}.}
  \bibinfo{year}{2023}\natexlab{}.
\newblock \showarticletitle{It’s the {End} of {Computer} {Programming} as
  {We} {Know} {It}. ({And} {I} {Feel} {Fine}.)}.
\newblock \bibinfo{journal}{\emph{The New York Times}} (\bibinfo{date}{June}
  \bibinfo{year}{2023}).
\newblock
\urldef\tempurl%
\url{https://www.nytimes.com/2023/06/02/opinion/ai-coding.html}
\showURL{%
\tempurl}


\bibitem[McDonald et~al\mbox{.}(2019)]%
        {mcdonald_reliability_2019}
\bibfield{author}{\bibinfo{person}{Nora McDonald}, \bibinfo{person}{Sarita
  Schoenebeck}, {and} \bibinfo{person}{Andrea Forte}.}
  \bibinfo{year}{2019}\natexlab{}.
\newblock \showarticletitle{Reliability and {Inter}-rater {Reliability} in
  {Qualitative} {Research}: {Norms} and {Guidelines} for {CSCW} and {HCI}
  {Practice}}.
\newblock \bibinfo{journal}{\emph{Proceedings of the ACM on Human-Computer
  Interaction}} \bibinfo{volume}{3}, \bibinfo{number}{CSCW}
  (\bibinfo{date}{Nov.} \bibinfo{year}{2019}), \bibinfo{pages}{1--23}.
\newblock
\showISSN{2573-0142}
\urldef\tempurl%
\url{https://doi.org/10.1145/3359174}
\showDOI{\tempurl}


\bibitem[Miller(1981)]%
        {miller_natural_1981}
\bibfield{author}{\bibinfo{person}{L.~A.. Miller}.}
  \bibinfo{year}{1981}\natexlab{}.
\newblock \showarticletitle{Natural language programming: styles, strategies,
  and contrasts}.
\newblock \bibinfo{journal}{\emph{IBM Systems Journal}} \bibinfo{volume}{20},
  \bibinfo{number}{2} (\bibinfo{date}{June} \bibinfo{year}{1981}),
  \bibinfo{pages}{184--215}.
\newblock
\showISSN{0018-8670}
\urldef\tempurl%
\url{https://doi.org/10.1147/sj.202.0184}
\showDOI{\tempurl}


\bibitem[Mirowski et~al\mbox{.}(2023)]%
        {mirowski_co-writing_2023}
\bibfield{author}{\bibinfo{person}{Piotr Mirowski}, \bibinfo{person}{Kory~W.
  Mathewson}, \bibinfo{person}{Jaylen Pittman}, {and} \bibinfo{person}{Richard
  Evans}.} \bibinfo{year}{2023}\natexlab{}.
\newblock \showarticletitle{Co-{Writing} {Screenplays} and {Theatre} {Scripts}
  with {Language} {Models}: {Evaluation} by {Industry} {Professionals}}. In
  \bibinfo{booktitle}{\emph{Proceedings of the 2023 {CHI} {Conference} on
  {Human} {Factors} in {Computing} {Systems}}}. \bibinfo{publisher}{ACM},
  \bibinfo{address}{Hamburg Germany}, \bibinfo{pages}{1--34}.
\newblock
\showISBNx{978-1-4503-9421-5}
\urldef\tempurl%
\url{https://doi.org/10.1145/3544548.3581225}
\showDOI{\tempurl}


\bibitem[Murali et~al\mbox{.}(2023)]%
        {murali_codecompose_2023}
\bibfield{author}{\bibinfo{person}{Vijayaraghavan Murali},
  \bibinfo{person}{Chandra Maddila}, \bibinfo{person}{Imad Ahmad},
  \bibinfo{person}{Michael Bolin}, \bibinfo{person}{Daniel Cheng},
  \bibinfo{person}{Negar Ghorbani}, \bibinfo{person}{Renuka Fernandez}, {and}
  \bibinfo{person}{Nachiappan Nagappan}.} \bibinfo{year}{2023}\natexlab{}.
\newblock \bibinfo{title}{{CodeCompose}: {A} {Large}-{Scale} {Industrial}
  {Deployment} of {AI}-assisted {Code} {Authoring}}.
\newblock
\newblock
\urldef\tempurl%
\url{http://arxiv.org/abs/2305.12050}
\showURL{%
\tempurl}
\newblock
\shownote{arXiv:2305.12050 [cs]}.


\bibitem[Myers et~al\mbox{.}(2016)]%
        {myers_programmers_2016}
\bibfield{author}{\bibinfo{person}{Brad~A. Myers}, \bibinfo{person}{Amy~J. Ko},
  \bibinfo{person}{Thomas~D. LaToza}, {and} \bibinfo{person}{YoungSeok Yoon}.}
  \bibinfo{year}{2016}\natexlab{}.
\newblock \showarticletitle{Programmers {Are} {Users} {Too}: {Human}-{Centered}
  {Methods} for {Improving} {Programming} {Tools}}.
\newblock \bibinfo{journal}{\emph{Computer}} \bibinfo{volume}{49},
  \bibinfo{number}{7} (\bibinfo{date}{July} \bibinfo{year}{2016}),
  \bibinfo{pages}{44--52}.
\newblock
\showISSN{0018-9162, 1558-0814}
\urldef\tempurl%
\url{https://doi.org/10.1109/MC.2016.200}
\showDOI{\tempurl}


\bibitem[Nam et~al\mbox{.}(2024)]%
        {nam_using_2024}
\bibfield{author}{\bibinfo{person}{Daye Nam}, \bibinfo{person}{Andrew Macvean},
  \bibinfo{person}{Vincent Hellendoorn}, \bibinfo{person}{Bogdan Vasilescu},
  {and} \bibinfo{person}{Brad Myers}.} \bibinfo{year}{2024}\natexlab{}.
\newblock \bibinfo{title}{Using an {LLM} to {Help} {With} {Code}
  {Understanding}}.
\newblock
\newblock
\urldef\tempurl%
\url{http://arxiv.org/abs/2307.08177}
\showURL{%
\tempurl}
\newblock
\shownote{arXiv:2307.08177 [cs]}.


\bibitem[{National Center for Women \& Information Technology}(2023)]%
        {ncwit_2023}
\bibfield{author}{\bibinfo{person}{{National Center for Women \& Information
  Technology}}.} \bibinfo{year}{2023}\natexlab{}.
\newblock \bibinfo{title}{{NCWIT} {Guide} to {Demographic} {Survey}
  {Questions}}.
\newblock
\newblock
\urldef\tempurl%
\url{https://docs.google.com/document/d/1E_CSANwOqbKjEG27woNbGZ09JIXUfAf4Cp9j8g5DFak/}
\showURL{%
\tempurl}


\bibitem[Nijkamp et~al\mbox{.}(2022)]%
        {nijkamp_codegen_2022}
\bibfield{author}{\bibinfo{person}{Erik Nijkamp}, \bibinfo{person}{Bo Pang},
  \bibinfo{person}{Hiroaki Hayashi}, \bibinfo{person}{Lifu Tu},
  \bibinfo{person}{Huan Wang}, \bibinfo{person}{Yingbo Zhou},
  \bibinfo{person}{Silvio Savarese}, {and} \bibinfo{person}{Caiming Xiong}.}
  \bibinfo{year}{2022}\natexlab{}.
\newblock \bibinfo{title}{{CodeGen}: {An} {Open} {Large} {Language} {Model} for
  {Code} with {Multi}-{Turn} {Program} {Synthesis}}.
\newblock
\newblock
\urldef\tempurl%
\url{https://doi.org/10.48550/ARXIV.2203.13474}
\showDOI{\tempurl}


\bibitem[OpenAI et~al\mbox{.}(2023)]%
        {openai_gpt-4_2023}
\bibfield{author}{\bibinfo{person}{OpenAI}, \bibinfo{person}{Josh Achiam},
  \bibinfo{person}{Steven Adler}, \bibinfo{person}{Sandhini Agarwal},
  \bibinfo{person}{Lama Ahmad}, \bibinfo{person}{Ilge Akkaya},
  \bibinfo{person}{Florencia~Leoni Aleman}, \bibinfo{person}{Diogo Almeida},
  \bibinfo{person}{Janko Altenschmidt}, \bibinfo{person}{Sam Altman},
  \bibinfo{person}{Shyamal Anadkat}, \bibinfo{person}{Red Avila},
  \bibinfo{person}{Igor Babuschkin}, \bibinfo{person}{Suchir Balaji},
  \bibinfo{person}{Valerie Balcom}, \bibinfo{person}{Paul Baltescu},
  \bibinfo{person}{Haiming Bao}, \bibinfo{person}{Mo Bavarian},
  \bibinfo{person}{Jeff Belgum}, \bibinfo{person}{Irwan Bello},
  \bibinfo{person}{Jake Berdine}, \bibinfo{person}{Gabriel Bernadett-Shapiro},
  \bibinfo{person}{Christopher Berner}, \bibinfo{person}{Lenny Bogdonoff},
  \bibinfo{person}{Oleg Boiko}, \bibinfo{person}{Madelaine Boyd},
  \bibinfo{person}{Anna-Luisa Brakman}, \bibinfo{person}{Greg Brockman},
  \bibinfo{person}{Tim Brooks}, \bibinfo{person}{Miles Brundage},
  \bibinfo{person}{Kevin Button}, \bibinfo{person}{Trevor Cai},
  \bibinfo{person}{Rosie Campbell}, \bibinfo{person}{Andrew Cann},
  \bibinfo{person}{Brittany Carey}, \bibinfo{person}{Chelsea Carlson},
  \bibinfo{person}{Rory Carmichael}, \bibinfo{person}{Brooke Chan},
  \bibinfo{person}{Che Chang}, \bibinfo{person}{Fotis Chantzis},
  \bibinfo{person}{Derek Chen}, \bibinfo{person}{Sully Chen},
  \bibinfo{person}{Ruby Chen}, \bibinfo{person}{Jason Chen},
  \bibinfo{person}{Mark Chen}, \bibinfo{person}{Ben Chess},
  \bibinfo{person}{Chester Cho}, \bibinfo{person}{Casey Chu},
  \bibinfo{person}{Hyung~Won Chung}, \bibinfo{person}{Dave Cummings},
  \bibinfo{person}{Jeremiah Currier}, \bibinfo{person}{Yunxing Dai},
  \bibinfo{person}{Cory Decareaux}, \bibinfo{person}{Thomas Degry},
  \bibinfo{person}{Noah Deutsch}, \bibinfo{person}{Damien Deville},
  \bibinfo{person}{Arka Dhar}, \bibinfo{person}{David Dohan},
  \bibinfo{person}{Steve Dowling}, \bibinfo{person}{Sheila Dunning},
  \bibinfo{person}{Adrien Ecoffet}, \bibinfo{person}{Atty Eleti},
  \bibinfo{person}{Tyna Eloundou}, \bibinfo{person}{David Farhi},
  \bibinfo{person}{Liam Fedus}, \bibinfo{person}{Niko Felix},
  \bibinfo{person}{Simón~Posada Fishman}, \bibinfo{person}{Juston Forte},
  \bibinfo{person}{Isabella Fulford}, \bibinfo{person}{Leo Gao},
  \bibinfo{person}{Elie Georges}, \bibinfo{person}{Christian Gibson},
  \bibinfo{person}{Vik Goel}, \bibinfo{person}{Tarun Gogineni},
  \bibinfo{person}{Gabriel Goh}, \bibinfo{person}{Rapha Gontijo-Lopes},
  \bibinfo{person}{Jonathan Gordon}, \bibinfo{person}{Morgan Grafstein},
  \bibinfo{person}{Scott Gray}, \bibinfo{person}{Ryan Greene},
  \bibinfo{person}{Joshua Gross}, \bibinfo{person}{Shixiang~Shane Gu},
  \bibinfo{person}{Yufei Guo}, \bibinfo{person}{Chris Hallacy},
  \bibinfo{person}{Jesse Han}, \bibinfo{person}{Jeff Harris},
  \bibinfo{person}{Yuchen He}, \bibinfo{person}{Mike Heaton},
  \bibinfo{person}{Johannes Heidecke}, \bibinfo{person}{Chris Hesse},
  \bibinfo{person}{Alan Hickey}, \bibinfo{person}{Wade Hickey},
  \bibinfo{person}{Peter Hoeschele}, \bibinfo{person}{Brandon Houghton},
  \bibinfo{person}{Kenny Hsu}, \bibinfo{person}{Shengli Hu},
  \bibinfo{person}{Xin Hu}, \bibinfo{person}{Joost Huizinga},
  \bibinfo{person}{Shantanu Jain}, \bibinfo{person}{Shawn Jain},
  \bibinfo{person}{Joanne Jang}, \bibinfo{person}{Angela Jiang},
  \bibinfo{person}{Roger Jiang}, \bibinfo{person}{Haozhun Jin},
  \bibinfo{person}{Denny Jin}, \bibinfo{person}{Shino Jomoto},
  \bibinfo{person}{Billie Jonn}, \bibinfo{person}{Heewoo Jun},
  \bibinfo{person}{Tomer Kaftan}, \bibinfo{person}{Łukasz Kaiser},
  \bibinfo{person}{Ali Kamali}, \bibinfo{person}{Ingmar Kanitscheider},
  \bibinfo{person}{Nitish~Shirish Keskar}, \bibinfo{person}{Tabarak Khan},
  \bibinfo{person}{Logan Kilpatrick}, \bibinfo{person}{Jong~Wook Kim},
  \bibinfo{person}{Christina Kim}, \bibinfo{person}{Yongjik Kim},
  \bibinfo{person}{Hendrik Kirchner}, \bibinfo{person}{Jamie Kiros},
  \bibinfo{person}{Matt Knight}, \bibinfo{person}{Daniel Kokotajlo},
  \bibinfo{person}{Łukasz Kondraciuk}, \bibinfo{person}{Andrew Kondrich},
  \bibinfo{person}{Aris Konstantinidis}, \bibinfo{person}{Kyle Kosic},
  \bibinfo{person}{Gretchen Krueger}, \bibinfo{person}{Vishal Kuo},
  \bibinfo{person}{Michael Lampe}, \bibinfo{person}{Ikai Lan},
  \bibinfo{person}{Teddy Lee}, \bibinfo{person}{Jan Leike},
  \bibinfo{person}{Jade Leung}, \bibinfo{person}{Daniel Levy},
  \bibinfo{person}{Chak~Ming Li}, \bibinfo{person}{Rachel Lim},
  \bibinfo{person}{Molly Lin}, \bibinfo{person}{Stephanie Lin},
  \bibinfo{person}{Mateusz Litwin}, \bibinfo{person}{Theresa Lopez},
  \bibinfo{person}{Ryan Lowe}, \bibinfo{person}{Patricia Lue},
  \bibinfo{person}{Anna Makanju}, \bibinfo{person}{Kim Malfacini},
  \bibinfo{person}{Sam Manning}, \bibinfo{person}{Todor Markov},
  \bibinfo{person}{Yaniv Markovski}, \bibinfo{person}{Bianca Martin},
  \bibinfo{person}{Katie Mayer}, \bibinfo{person}{Andrew Mayne},
  \bibinfo{person}{Bob McGrew}, \bibinfo{person}{Scott~Mayer McKinney},
  \bibinfo{person}{Christine McLeavey}, \bibinfo{person}{Paul McMillan},
  \bibinfo{person}{Jake McNeil}, \bibinfo{person}{David Medina},
  \bibinfo{person}{Aalok Mehta}, \bibinfo{person}{Jacob Menick},
  \bibinfo{person}{Luke Metz}, \bibinfo{person}{Andrey Mishchenko},
  \bibinfo{person}{Pamela Mishkin}, \bibinfo{person}{Vinnie Monaco},
  \bibinfo{person}{Evan Morikawa}, \bibinfo{person}{Daniel Mossing},
  \bibinfo{person}{Tong Mu}, \bibinfo{person}{Mira Murati},
  \bibinfo{person}{Oleg Murk}, \bibinfo{person}{David Mély},
  \bibinfo{person}{Ashvin Nair}, \bibinfo{person}{Reiichiro Nakano},
  \bibinfo{person}{Rajeev Nayak}, \bibinfo{person}{Arvind Neelakantan},
  \bibinfo{person}{Richard Ngo}, \bibinfo{person}{Hyeonwoo Noh},
  \bibinfo{person}{Long Ouyang}, \bibinfo{person}{Cullen O'Keefe},
  \bibinfo{person}{Jakub Pachocki}, \bibinfo{person}{Alex Paino},
  \bibinfo{person}{Joe Palermo}, \bibinfo{person}{Ashley Pantuliano},
  \bibinfo{person}{Giambattista Parascandolo}, \bibinfo{person}{Joel Parish},
  \bibinfo{person}{Emy Parparita}, \bibinfo{person}{Alex Passos},
  \bibinfo{person}{Mikhail Pavlov}, \bibinfo{person}{Andrew Peng},
  \bibinfo{person}{Adam Perelman}, \bibinfo{person}{Filipe de Avila~Belbute
  Peres}, \bibinfo{person}{Michael Petrov}, \bibinfo{person}{Henrique Ponde
  de~Oliveira Pinto}, \bibinfo{person}{Michael}, \bibinfo{person}{Pokorny},
  \bibinfo{person}{Michelle Pokrass}, \bibinfo{person}{Vitchyr Pong},
  \bibinfo{person}{Tolly Powell}, \bibinfo{person}{Alethea Power},
  \bibinfo{person}{Boris Power}, \bibinfo{person}{Elizabeth Proehl},
  \bibinfo{person}{Raul Puri}, \bibinfo{person}{Alec Radford},
  \bibinfo{person}{Jack Rae}, \bibinfo{person}{Aditya Ramesh},
  \bibinfo{person}{Cameron Raymond}, \bibinfo{person}{Francis Real},
  \bibinfo{person}{Kendra Rimbach}, \bibinfo{person}{Carl Ross},
  \bibinfo{person}{Bob Rotsted}, \bibinfo{person}{Henri Roussez},
  \bibinfo{person}{Nick Ryder}, \bibinfo{person}{Mario Saltarelli},
  \bibinfo{person}{Ted Sanders}, \bibinfo{person}{Shibani Santurkar},
  \bibinfo{person}{Girish Sastry}, \bibinfo{person}{Heather Schmidt},
  \bibinfo{person}{David Schnurr}, \bibinfo{person}{John Schulman},
  \bibinfo{person}{Daniel Selsam}, \bibinfo{person}{Kyla Sheppard},
  \bibinfo{person}{Toki Sherbakov}, \bibinfo{person}{Jessica Shieh},
  \bibinfo{person}{Sarah Shoker}, \bibinfo{person}{Pranav Shyam},
  \bibinfo{person}{Szymon Sidor}, \bibinfo{person}{Eric Sigler},
  \bibinfo{person}{Maddie Simens}, \bibinfo{person}{Jordan Sitkin},
  \bibinfo{person}{Katarina Slama}, \bibinfo{person}{Ian Sohl},
  \bibinfo{person}{Benjamin Sokolowsky}, \bibinfo{person}{Yang Song},
  \bibinfo{person}{Natalie Staudacher}, \bibinfo{person}{Felipe~Petroski Such},
  \bibinfo{person}{Natalie Summers}, \bibinfo{person}{Ilya Sutskever},
  \bibinfo{person}{Jie Tang}, \bibinfo{person}{Nikolas Tezak},
  \bibinfo{person}{Madeleine Thompson}, \bibinfo{person}{Phil Tillet},
  \bibinfo{person}{Amin Tootoonchian}, \bibinfo{person}{Elizabeth Tseng},
  \bibinfo{person}{Preston Tuggle}, \bibinfo{person}{Nick Turley},
  \bibinfo{person}{Jerry Tworek}, \bibinfo{person}{Juan Felipe~Cerón Uribe},
  \bibinfo{person}{Andrea Vallone}, \bibinfo{person}{Arun Vijayvergiya},
  \bibinfo{person}{Chelsea Voss}, \bibinfo{person}{Carroll Wainwright},
  \bibinfo{person}{Justin~Jay Wang}, \bibinfo{person}{Alvin Wang},
  \bibinfo{person}{Ben Wang}, \bibinfo{person}{Jonathan Ward},
  \bibinfo{person}{Jason Wei}, \bibinfo{person}{C.~J. Weinmann},
  \bibinfo{person}{Akila Welihinda}, \bibinfo{person}{Peter Welinder},
  \bibinfo{person}{Jiayi Weng}, \bibinfo{person}{Lilian Weng},
  \bibinfo{person}{Matt Wiethoff}, \bibinfo{person}{Dave Willner},
  \bibinfo{person}{Clemens Winter}, \bibinfo{person}{Samuel Wolrich},
  \bibinfo{person}{Hannah Wong}, \bibinfo{person}{Lauren Workman},
  \bibinfo{person}{Sherwin Wu}, \bibinfo{person}{Jeff Wu},
  \bibinfo{person}{Michael Wu}, \bibinfo{person}{Kai Xiao},
  \bibinfo{person}{Tao Xu}, \bibinfo{person}{Sarah Yoo}, \bibinfo{person}{Kevin
  Yu}, \bibinfo{person}{Qiming Yuan}, \bibinfo{person}{Wojciech Zaremba},
  \bibinfo{person}{Rowan Zellers}, \bibinfo{person}{Chong Zhang},
  \bibinfo{person}{Marvin Zhang}, \bibinfo{person}{Shengjia Zhao},
  \bibinfo{person}{Tianhao Zheng}, \bibinfo{person}{Juntang Zhuang},
  \bibinfo{person}{William Zhuk}, {and} \bibinfo{person}{Barret Zoph}.}
  \bibinfo{year}{2023}\natexlab{}.
\newblock \bibinfo{title}{{GPT}-4 {Technical} {Report}}.
\newblock
\newblock
\urldef\tempurl%
\url{http://arxiv.org/abs/2303.08774}
\showURL{%
\tempurl}
\newblock
\shownote{arXiv:2303.08774 [cs]}.


\bibitem[Parnas and Madey(1995)]%
        {parnas_functional_1995}
\bibfield{author}{\bibinfo{person}{David~Lorge Parnas} {and}
  \bibinfo{person}{Jan Madey}.} \bibinfo{year}{1995}\natexlab{}.
\newblock \showarticletitle{Functional documents for computer systems}.
\newblock \bibinfo{journal}{\emph{Science of Computer Programming}}
  \bibinfo{volume}{25}, \bibinfo{number}{1} (\bibinfo{date}{Oct.}
  \bibinfo{year}{1995}), \bibinfo{pages}{41--61}.
\newblock
\showISSN{01676423}
\urldef\tempurl%
\url{https://doi.org/10.1016/0167-6423(95)96871-J}
\showDOI{\tempurl}


\bibitem[Peng et~al\mbox{.}(2023)]%
        {peng_impact_2023}
\bibfield{author}{\bibinfo{person}{Sida Peng}, \bibinfo{person}{Eirini
  Kalliamvakou}, \bibinfo{person}{Peter Cihon}, {and} \bibinfo{person}{Mert
  Demirer}.} \bibinfo{year}{2023}\natexlab{}.
\newblock \bibinfo{title}{The {Impact} of {AI} on {Developer} {Productivity}:
  {Evidence} from {GitHub} {Copilot}}.
\newblock
\newblock
\urldef\tempurl%
\url{http://arxiv.org/abs/2302.06590}
\showURL{%
\tempurl}
\newblock
\shownote{arXiv:2302.06590 [cs]}.


\bibitem[Phung et~al\mbox{.}(2023)]%
        {phung_generating_2023}
\bibfield{author}{\bibinfo{person}{Tung Phung}, \bibinfo{person}{José
  Cambronero}, \bibinfo{person}{Sumit Gulwani}, \bibinfo{person}{Tobias Kohn},
  \bibinfo{person}{Rupak Majumdar}, \bibinfo{person}{Adish Singla}, {and}
  \bibinfo{person}{Gustavo Soares}.} \bibinfo{year}{2023}\natexlab{}.
\newblock \bibinfo{title}{Generating {High}-{Precision} {Feedback} for
  {Programming} {Syntax} {Errors} using {Large} {Language} {Models}}.
\newblock
\newblock
\urldef\tempurl%
\url{http://arxiv.org/abs/2302.04662}
\showURL{%
\tempurl}
\newblock
\shownote{arXiv:2302.04662 [cs]}.


\bibitem[Prather et~al\mbox{.}(2023)]%
        {prather_its_2023}
\bibfield{author}{\bibinfo{person}{James Prather}, \bibinfo{person}{Brent~N.
  Reeves}, \bibinfo{person}{Paul Denny}, \bibinfo{person}{Brett~A. Becker},
  \bibinfo{person}{Juho Leinonen}, \bibinfo{person}{Andrew Luxton-Reilly},
  \bibinfo{person}{Garrett Powell}, \bibinfo{person}{James Finnie-Ansley},
  {and} \bibinfo{person}{Eddie~Antonio Santos}.}
  \bibinfo{year}{2023}\natexlab{}.
\newblock \showarticletitle{“{It}’s {Weird} {That} it {Knows} {What} {I}
  {Want}”: {Usability} and {Interactions} with {Copilot} for {Novice}
  {Programmers}}.
\newblock \bibinfo{journal}{\emph{ACM Transactions on Computer-Human
  Interaction}} (\bibinfo{date}{Aug.} \bibinfo{year}{2023}),
  \bibinfo{pages}{3617367}.
\newblock
\showISSN{1073-0516, 1557-7325}
\urldef\tempurl%
\url{https://doi.org/10.1145/3617367}
\showDOI{\tempurl}


\bibitem[Radford et~al\mbox{.}(2019)]%
        {radford_language_2019}
\bibfield{author}{\bibinfo{person}{Alec Radford}, \bibinfo{person}{Jeffrey Wu},
  \bibinfo{person}{Rewon Child}, \bibinfo{person}{David Luan},
  \bibinfo{person}{Dario Amodei}, \bibinfo{person}{Ilya Sutskever}, {and}
  \bibinfo{person}{{others}}.} \bibinfo{year}{2019}\natexlab{}.
\newblock \showarticletitle{Language models are unsupervised multitask
  learners}.
\newblock \bibinfo{journal}{\emph{OpenAI}} \bibinfo{volume}{1},
  \bibinfo{number}{8} (\bibinfo{year}{2019}), \bibinfo{pages}{9}.
\newblock
\urldef\tempurl%
\url{https://cdn.openai.com/better-language-models/language_models_are_unsupervised_multitask_learners.pdf}
\showURL{%
\tempurl}


\bibitem[Radu and Schneider(2019)]%
        {radu_what_2019}
\bibfield{author}{\bibinfo{person}{Iulian Radu} {and} \bibinfo{person}{Bertrand
  Schneider}.} \bibinfo{year}{2019}\natexlab{}.
\newblock \showarticletitle{What can we learn from augmented reality ({AR})?
  {Benefits} and drawbacks of {AR} for inquiry-based learning of physics}. In
  \bibinfo{booktitle}{\emph{Proceedings of the 2019 {CHI} conference on human
  factors in computing systems}}. \bibinfo{pages}{1--12}.
\newblock


\bibitem[Ramesh et~al\mbox{.}(2021)]%
        {ramesh_zero-shot_2021}
\bibfield{author}{\bibinfo{person}{Aditya Ramesh}, \bibinfo{person}{Mikhail
  Pavlov}, \bibinfo{person}{Gabriel Goh}, \bibinfo{person}{Scott Gray},
  \bibinfo{person}{Chelsea Voss}, \bibinfo{person}{Alec Radford},
  \bibinfo{person}{Mark Chen}, {and} \bibinfo{person}{Ilya Sutskever}.}
  \bibinfo{year}{2021}\natexlab{}.
\newblock \showarticletitle{Zero-{Shot} {Text}-to-{Image} {Generation}}. In
  \bibinfo{booktitle}{\emph{Proceedings of the 38th {International}
  {Conference} on {Machine} {Learning}}}. \bibinfo{publisher}{PMLR},
  \bibinfo{pages}{8821--8831}.
\newblock
\urldef\tempurl%
\url{https://proceedings.mlr.press/v139/ramesh21a.html}
\showURL{%
\tempurl}
\newblock
\shownote{ISSN: 2640-3498}.


\bibitem[Rocha et~al\mbox{.}(2023)]%
        {rocha_use_2023}
\bibfield{author}{\bibinfo{person}{Hemilis Joyse~Barbosa Rocha},
  \bibinfo{person}{Patrícia Cabral De Azevedo~Restelli Tedesco}, {and}
  \bibinfo{person}{Evandro De~Barros Costa}.} \bibinfo{year}{2023}\natexlab{}.
\newblock \showarticletitle{On the use of feedback in learning computer
  programming by novices: a systematic literature mapping}.
\newblock \bibinfo{journal}{\emph{Informatics in Education}}
  \bibinfo{volume}{22}, \bibinfo{number}{2} (\bibinfo{year}{2023}),
  \bibinfo{pages}{209}.
\newblock
\newblock
\shownote{Publisher: Institute of Mathematics and Informatics}.


\bibitem[Rombach et~al\mbox{.}(2022)]%
        {rombach_high-resolution_2022}
\bibfield{author}{\bibinfo{person}{Robin Rombach}, \bibinfo{person}{Andreas
  Blattmann}, \bibinfo{person}{Dominik Lorenz}, \bibinfo{person}{Patrick
  Esser}, {and} \bibinfo{person}{Björn Ommer}.}
  \bibinfo{year}{2022}\natexlab{}.
\newblock \showarticletitle{High-{Resolution} {Image} {Synthesis} with {Latent}
  {Diffusion} {Models}}. \bibinfo{publisher}{IEEE Computer Society},
  \bibinfo{pages}{10674--10685}.
\newblock
\showISBNx{978-1-66546-946-3}
\urldef\tempurl%
\url{https://doi.org/10.1109/CVPR52688.2022.01042}
\showDOI{\tempurl}


\bibitem[Ross et~al\mbox{.}(2023)]%
        {ross_programmers_2023}
\bibfield{author}{\bibinfo{person}{Steven~I. Ross}, \bibinfo{person}{Fernando
  Martinez}, \bibinfo{person}{Stephanie Houde}, \bibinfo{person}{Michael
  Muller}, {and} \bibinfo{person}{Justin~D. Weisz}.}
  \bibinfo{year}{2023}\natexlab{}.
\newblock \showarticletitle{The {Programmer}’s {Assistant}: {Conversational}
  {Interaction} with a {Large} {Language} {Model} for {Software}
  {Development}}. In \bibinfo{booktitle}{\emph{Proceedings of the 28th
  {International} {Conference} on {Intelligent} {User} {Interfaces}}}
  \emph{(\bibinfo{series}{{IUI} '23})}. \bibinfo{publisher}{Association for
  Computing Machinery}, \bibinfo{address}{New York, NY, USA},
  \bibinfo{pages}{491--514}.
\newblock
\showISBNx{9798400701061}
\urldef\tempurl%
\url{https://doi.org/10.1145/3581641.3584037}
\showDOI{\tempurl}


\bibitem[Rozière et~al\mbox{.}(2024)]%
        {roziere_code_2024}
\bibfield{author}{\bibinfo{person}{Baptiste Rozière}, \bibinfo{person}{Jonas
  Gehring}, \bibinfo{person}{Fabian Gloeckle}, \bibinfo{person}{Sten Sootla},
  \bibinfo{person}{Itai Gat}, \bibinfo{person}{Xiaoqing~Ellen Tan},
  \bibinfo{person}{Yossi Adi}, \bibinfo{person}{Jingyu Liu},
  \bibinfo{person}{Romain Sauvestre}, \bibinfo{person}{Tal Remez},
  \bibinfo{person}{Jérémy Rapin}, \bibinfo{person}{Artyom Kozhevnikov},
  \bibinfo{person}{Ivan Evtimov}, \bibinfo{person}{Joanna Bitton},
  \bibinfo{person}{Manish Bhatt}, \bibinfo{person}{Cristian~Canton Ferrer},
  \bibinfo{person}{Aaron Grattafiori}, \bibinfo{person}{Wenhan Xiong},
  \bibinfo{person}{Alexandre Défossez}, \bibinfo{person}{Jade Copet},
  \bibinfo{person}{Faisal Azhar}, \bibinfo{person}{Hugo Touvron},
  \bibinfo{person}{Louis Martin}, \bibinfo{person}{Nicolas Usunier},
  \bibinfo{person}{Thomas Scialom}, {and} \bibinfo{person}{Gabriel Synnaeve}.}
  \bibinfo{year}{2024}\natexlab{}.
\newblock \bibinfo{title}{Code {Llama}: {Open} {Foundation} {Models} for
  {Code}}.
\newblock
\newblock
\urldef\tempurl%
\url{http://arxiv.org/abs/2308.12950}
\showURL{%
\tempurl}
\newblock
\shownote{arXiv:2308.12950 [cs]}.


\bibitem[Sammet(1966)]%
        {sammet_use_1966}
\bibfield{author}{\bibinfo{person}{Jean~E. Sammet}.}
  \bibinfo{year}{1966}\natexlab{}.
\newblock \showarticletitle{The {Use} of {English} as a {Programming}
  {Language}}.
\newblock \bibinfo{journal}{\emph{Commun. ACM}} \bibinfo{volume}{9},
  \bibinfo{number}{3} (\bibinfo{date}{March} \bibinfo{year}{1966}),
  \bibinfo{pages}{228--230}.
\newblock
\showISSN{0001-0782}
\urldef\tempurl%
\url{https://doi.org/10.1145/365230.365274}
\showDOI{\tempurl}


\bibitem[Schäfer et~al\mbox{.}(2023)]%
        {schafer_adaptive_2023}
\bibfield{author}{\bibinfo{person}{Max Schäfer}, \bibinfo{person}{Sarah Nadi},
  \bibinfo{person}{Aryaz Eghbali}, {and} \bibinfo{person}{Frank Tip}.}
  \bibinfo{year}{2023}\natexlab{}.
\newblock \bibinfo{title}{Adaptive {Test} {Generation} {Using} a {Large}
  {Language} {Model}}.
\newblock
\newblock
\urldef\tempurl%
\url{https://doi.org/10.48550/arXiv.2302.06527}
\showDOI{\tempurl}
\newblock
\shownote{Issue: arXiv:2302.06527 \_eprint: 2302.06527}.


\bibitem[Shaw et~al\mbox{.}(1975)]%
        {shaw_inferring_1975}
\bibfield{author}{\bibinfo{person}{David~E Shaw}, \bibinfo{person}{William~R
  Swartout}, {and} \bibinfo{person}{C~Cordell Green}.}
  \bibinfo{year}{1975}\natexlab{}.
\newblock \showarticletitle{Inferring {LISP} {Programs} {From} {Examples}}. In
  \bibinfo{booktitle}{\emph{{IJCAI}}}, Vol.~\bibinfo{volume}{75}.
  \bibinfo{pages}{260--267}.
\newblock


\bibitem[Singh et~al\mbox{.}(2022)]%
        {singh_where_2022}
\bibfield{author}{\bibinfo{person}{Nikhil Singh}, \bibinfo{person}{Guillermo
  Bernal}, \bibinfo{person}{Daria Savchenko}, {and} \bibinfo{person}{Elena~L.
  Glassman}.} \bibinfo{year}{2022}\natexlab{}.
\newblock \showarticletitle{Where to {Hide} a {Stolen} {Elephant}: {Leaps} in
  {Creative} {Writing} with {Multimodal} {Machine} {Intelligence}}.
\newblock \bibinfo{journal}{\emph{ACM Transactions on Computer-Human
  Interaction}} (\bibinfo{date}{Feb.} \bibinfo{year}{2022}),
  \bibinfo{pages}{3511599}.
\newblock
\showISSN{1073-0516, 1557-7325}
\urldef\tempurl%
\url{https://doi.org/10.1145/3511599}
\showDOI{\tempurl}


\bibitem[Singh et~al\mbox{.}(2013)]%
        {singh_automated_2013}
\bibfield{author}{\bibinfo{person}{Rishabh Singh}, \bibinfo{person}{Sumit
  Gulwani}, {and} \bibinfo{person}{Armando Solar-Lezama}.}
  \bibinfo{year}{2013}\natexlab{}.
\newblock \showarticletitle{Automated feedback generation for introductory
  programming assignments}. In \bibinfo{booktitle}{\emph{Proceedings of the
  34th {ACM} {SIGPLAN} conference on {Programming} language design and
  implementation}}. \bibinfo{pages}{15--26}.
\newblock


\bibitem[Skita et~al\mbox{.}(2000)]%
        {skita_accountability_2000}
\bibfield{author}{\bibinfo{person}{Linda~J Skita}, \bibinfo{person}{Kathleen
  Mosier}, {and} \bibinfo{person}{Mark~D. Burdick}.}
  \bibinfo{year}{2000}\natexlab{}.
\newblock \showarticletitle{Accountability and automation bias}.
\newblock \bibinfo{journal}{\emph{International Journal of Human-Computer
  Studies}} \bibinfo{volume}{52}, \bibinfo{number}{4} (\bibinfo{year}{2000}),
  \bibinfo{pages}{701--717}.
\newblock
\showISSN{1071-5819}
\urldef\tempurl%
\url{https://doi.org/10.1006/ijhc.1999.0349}
\showDOI{\tempurl}


\bibitem[Spiel et~al\mbox{.}(2019)]%
        {spiel_how_2019}
\bibfield{author}{\bibinfo{person}{Katta Spiel}, \bibinfo{person}{Oliver~L.
  Haimson}, {and} \bibinfo{person}{Danielle Lottridge}.}
  \bibinfo{year}{2019}\natexlab{}.
\newblock \showarticletitle{How to do better with gender on surveys: a guide
  for {HCI} researchers}.
\newblock \bibinfo{journal}{\emph{Interactions}} \bibinfo{volume}{26},
  \bibinfo{number}{4} (\bibinfo{date}{June} \bibinfo{year}{2019}),
  \bibinfo{pages}{62--65}.
\newblock
\showISSN{1072-5520, 1558-3449}
\urldef\tempurl%
\url{https://doi.org/10.1145/3338283}
\showDOI{\tempurl}


\bibitem[Sun et~al\mbox{.}(2022)]%
        {sun_investigating_2022}
\bibfield{author}{\bibinfo{person}{Jiao Sun}, \bibinfo{person}{Q.~Vera Liao},
  \bibinfo{person}{Michael Muller}, \bibinfo{person}{Mayank Agarwal},
  \bibinfo{person}{Stephanie Houde}, \bibinfo{person}{Kartik Talamadupula},
  {and} \bibinfo{person}{Justin~D. Weisz}.} \bibinfo{year}{2022}\natexlab{}.
\newblock \showarticletitle{Investigating {Explainability} of {Generative} {AI}
  for {Code} through {Scenario}-based {Design}}. In
  \bibinfo{booktitle}{\emph{27th {International} {Conference} on {Intelligent}
  {User} {Interfaces}}}. \bibinfo{publisher}{ACM}, \bibinfo{address}{Helsinki
  Finland}, \bibinfo{pages}{212--228}.
\newblock
\showISBNx{978-1-4503-9144-3}
\urldef\tempurl%
\url{https://doi.org/10.1145/3490099.3511119}
\showDOI{\tempurl}


\bibitem[Suzuki et~al\mbox{.}(2017)]%
        {suzuki_tracediff_2017}
\bibfield{author}{\bibinfo{person}{Ryo Suzuki}, \bibinfo{person}{Gustavo
  Soares}, \bibinfo{person}{Andrew Head}, \bibinfo{person}{Elena Glassman},
  \bibinfo{person}{Ruan Reis}, \bibinfo{person}{Melina Mongiovi},
  \bibinfo{person}{Loris D'Antoni}, {and} \bibinfo{person}{Bjoern Hartmann}.}
  \bibinfo{year}{2017}\natexlab{}.
\newblock \showarticletitle{Tracediff: {Debugging} unexpected code behavior
  using trace divergences}. In \bibinfo{booktitle}{\emph{2017 {IEEE}
  {Symposium} on {Visual} {Languages} and {Human}-{Centric} {Computing}
  ({VL}/{HCC})}}. \bibinfo{publisher}{IEEE}, \bibinfo{pages}{107--115}.
\newblock


\bibitem[{TabNine}(2023)]%
        {tabnine_ai_2023}
\bibfield{author}{\bibinfo{person}{{TabNine}}.}
  \bibinfo{year}{2023}\natexlab{}.
\newblock \bibinfo{title}{{AI} {Assistant} for {Software} {Developers}
  {\textbar} {Tabnine}}.
\newblock
\newblock
\urldef\tempurl%
\url{https://www.tabnine.com/}
\showURL{%
\tempurl}


\bibitem[Vaithilingam et~al\mbox{.}(2023)]%
        {vaithilingam_towards_2023}
\bibfield{author}{\bibinfo{person}{Priyan Vaithilingam},
  \bibinfo{person}{Elena~L. Glassman}, \bibinfo{person}{Peter Groenwegen},
  \bibinfo{person}{Sumit Gulwani}, \bibinfo{person}{Austin~Z. Henley},
  \bibinfo{person}{Rohan Malpani}, \bibinfo{person}{David Pugh},
  \bibinfo{person}{Arjun Radhakrishna}, \bibinfo{person}{Gustavo Soares},
  \bibinfo{person}{Joey Wang}, {and} \bibinfo{person}{Aaron Yim}.}
  \bibinfo{year}{2023}\natexlab{}.
\newblock \showarticletitle{Towards {More} {Effective} {AI}-{Assisted}
  {Programming}: {A} {Systematic} {Design} {Exploration} to {Improve} {Visual}
  {Studio} {IntelliCode}’s {User} {Experience}}. In
  \bibinfo{booktitle}{\emph{International {Conference} on {Software}
  {Engineering}: {Software} {Engineering} in {Practice} ({ICSE}-{SEIP})}}.
\newblock


\bibitem[Vaithilingam et~al\mbox{.}(2022)]%
        {vaithilingam_expectation_2022}
\bibfield{author}{\bibinfo{person}{Priyan Vaithilingam},
  \bibinfo{person}{Tianyi Zhang}, {and} \bibinfo{person}{Elena~L. Glassman}.}
  \bibinfo{year}{2022}\natexlab{}.
\newblock \showarticletitle{Expectation vs. {Experience}: {Evaluating} the
  {Usability} of {Code} {Generation} {Tools} {Powered} by {Large} {Language}
  {Models}}. In \bibinfo{booktitle}{\emph{Extended {Abstracts} of the 2022
  {CHI} {Conference} on {Human} {Factors} in {Computing} {Systems}}}
  \emph{(\bibinfo{series}{{CHI} {EA} '22})}. \bibinfo{publisher}{Association
  for Computing Machinery}, \bibinfo{address}{New York, NY, USA}.
\newblock
\showISBNx{978-1-4503-9156-6}
\urldef\tempurl%
\url{https://doi.org/10.1145/3491101.3519665}
\showDOI{\tempurl}


\bibitem[Van~Mechelen et~al\mbox{.}(2023)]%
        {van_mechelen_emerging_2023}
\bibfield{author}{\bibinfo{person}{Maarten Van~Mechelen},
  \bibinfo{person}{Rachel~Charlotte Smith}, \bibinfo{person}{Marie-Monique
  Schaper}, \bibinfo{person}{Mariana Tamashiro}, \bibinfo{person}{Karl-Emil
  Bilstrup}, \bibinfo{person}{Mille Lunding}, \bibinfo{person}{Marianne
  Graves~Petersen}, {and} \bibinfo{person}{Ole Sejer~Iversen}.}
  \bibinfo{year}{2023}\natexlab{}.
\newblock \showarticletitle{Emerging technologies in {K}–12 education: {A}
  future {HCI} research agenda}.
\newblock \bibinfo{journal}{\emph{ACM Transactions on Computer-Human
  Interaction}} \bibinfo{volume}{30}, \bibinfo{number}{3}
  (\bibinfo{year}{2023}), \bibinfo{pages}{1--40}.
\newblock
\newblock
\shownote{Publisher: ACM New York, NY}.


\bibitem[Wang et~al\mbox{.}(2021)]%
        {wang_towards_2021}
\bibfield{author}{\bibinfo{person}{Qiaosi Wang}, \bibinfo{person}{Koustuv
  Saha}, \bibinfo{person}{Eric Gregori}, \bibinfo{person}{David Joyner}, {and}
  \bibinfo{person}{Ashok Goel}.} \bibinfo{year}{2021}\natexlab{}.
\newblock \showarticletitle{Towards {Mutual} {Theory} of {Mind} in {Human}-{AI}
  {Interaction}: {How} {Language} {Reflects} {What} {Students} {Perceive}
  {About} a {Virtual} {Teaching} {Assistant}}. In
  \bibinfo{booktitle}{\emph{Proceedings of the 2021 {CHI} {Conference} on
  {Human} {Factors} in {Computing} {Systems}}}. \bibinfo{publisher}{ACM},
  \bibinfo{address}{Yokohama Japan}, \bibinfo{pages}{1--14}.
\newblock
\showISBNx{978-1-4503-8096-6}
\urldef\tempurl%
\url{https://doi.org/10.1145/3411764.3445645}
\showDOI{\tempurl}


\bibitem[Welsh(2022)]%
        {welsh_end_2022}
\bibfield{author}{\bibinfo{person}{Matt Welsh}.}
  \bibinfo{year}{2022}\natexlab{}.
\newblock \showarticletitle{The {End} of {Programming}}.
\newblock \bibinfo{journal}{\emph{Commun. ACM}} \bibinfo{volume}{66},
  \bibinfo{number}{1} (\bibinfo{date}{Dec.} \bibinfo{year}{2022}),
  \bibinfo{pages}{34--35}.
\newblock
\showISSN{0001-0782}
\urldef\tempurl%
\url{https://doi.org/10.1145/3570220}
\showDOI{\tempurl}
\newblock
\shownote{Place: New York, NY, USA Publisher: Association for Computing
  Machinery}.


\bibitem[Xia et~al\mbox{.}(2024)]%
        {xia_fuzz4all_2024}
\bibfield{author}{\bibinfo{person}{Chunqiu~Steven Xia}, \bibinfo{person}{Matteo
  Paltenghi}, \bibinfo{person}{Jia~Le Tian}, \bibinfo{person}{Michael Pradel},
  {and} \bibinfo{person}{Lingming Zhang}.} \bibinfo{year}{2024}\natexlab{}.
\newblock \bibinfo{title}{{Fuzz4All}: {Universal} {Fuzzing} with {Large}
  {Language} {Models}}.
\newblock
\newblock
\urldef\tempurl%
\url{http://arxiv.org/abs/2308.04748}
\showURL{%
\tempurl}
\newblock
\shownote{arXiv:2308.04748 [cs]}.


\bibitem[Xu et~al\mbox{.}(2022)]%
        {xu_systematic_2022}
\bibfield{author}{\bibinfo{person}{Frank~F. Xu}, \bibinfo{person}{Uri Alon},
  \bibinfo{person}{Graham Neubig}, {and} \bibinfo{person}{Vincent~Josua
  Hellendoorn}.} \bibinfo{year}{2022}\natexlab{}.
\newblock \showarticletitle{A systematic evaluation of large language models of
  code}. In \bibinfo{booktitle}{\emph{Proceedings of the 6th {ACM} {SIGPLAN}
  {International} {Symposium} on {Machine} {Programming}}}.
  \bibinfo{publisher}{ACM}, \bibinfo{address}{San Diego CA USA},
  \bibinfo{pages}{1--10}.
\newblock
\showISBNx{978-1-4503-9273-0}
\urldef\tempurl%
\url{https://doi.org/10.1145/3520312.3534862}
\showDOI{\tempurl}


\bibitem[Zamfirescu-Pereira et~al\mbox{.}(2023)]%
        {zamfirescu-pereira_why_2023}
\bibfield{author}{\bibinfo{person}{J.D. Zamfirescu-Pereira},
  \bibinfo{person}{Richmond~Y. Wong}, \bibinfo{person}{Bjoern Hartmann}, {and}
  \bibinfo{person}{Qian Yang}.} \bibinfo{year}{2023}\natexlab{}.
\newblock \showarticletitle{Why {Johnny} {Can}’t {Prompt}: {How} {Non}-{AI}
  {Experts} {Try} (and {Fail}) to {Design} {LLM} {Prompts}}. In
  \bibinfo{booktitle}{\emph{Proceedings of the 2023 {CHI} {Conference} on
  {Human} {Factors} in {Computing} {Systems}}}. \bibinfo{publisher}{ACM},
  \bibinfo{address}{Hamburg Germany}, \bibinfo{pages}{1--21}.
\newblock
\showISBNx{978-1-4503-9421-5}
\urldef\tempurl%
\url{https://doi.org/10.1145/3544548.3581388}
\showDOI{\tempurl}


\bibitem[Zhao(2023)]%
        {zhao_github_2023}
\bibfield{author}{\bibinfo{person}{Shuyin Zhao}.}
  \bibinfo{year}{2023}\natexlab{}.
\newblock \bibinfo{title}{{GitHub} {Copilot} {Now} {Has} a {Better} {AI}
  {Model} and {New} {Capabilities}}.
\newblock
\newblock
\urldef\tempurl%
\url{https://github.blog/2023-02-14-github-copilot-now-has-a-better-ai-model-and-new-capabilities/}
\showURL{%
\tempurl}
\newblock
\shownote{Publication Title: The GitHub Blog}.


\bibitem[Ziegler et~al\mbox{.}(2022)]%
        {ziegler_productivity_2022}
\bibfield{author}{\bibinfo{person}{Albert Ziegler}, \bibinfo{person}{Eirini
  Kalliamvakou}, \bibinfo{person}{X~Alice Li}, \bibinfo{person}{Andrew Rice},
  \bibinfo{person}{Devon Rifkin}, \bibinfo{person}{Shawn Simister},
  \bibinfo{person}{Ganesh Sittampalam}, {and} \bibinfo{person}{Edward
  Aftandilian}.} \bibinfo{year}{2022}\natexlab{}.
\newblock \showarticletitle{Productivity assessment of neural code completion}.
  In \bibinfo{booktitle}{\emph{Proceedings of the 6th {ACM} {SIGPLAN}
  {International} {Symposium} on {Machine} {Programming}}}.
  \bibinfo{pages}{21--29}.
\newblock


\end{thebibliography}

\newpage

\appendix
\section{Additional Methodological Details}
\subsection{Study Design}\label{app:details}

\subsubsection{Pilot Study}
In late 2022, we ran an IRB-approved pilot study with 19 participants from all three institutions.
These students had completed CS1 and at least one additional course, so they were ineligible for the main study. Overall, we made few changes after the pilot. The most consequential were to add an additional 15 minutes (75 minutes total) to the study window, increase participant compensation, and implement word wrapping in the interface to prevent excessive scrolling.

\subsubsection{Problem Adaptation}\label{app:adaptation}

Our problems were based on CS1 problems used at each of our three institutions. In most cases, we made small adaptations to the problems, both to make it less likely for students to recognize the exact problem, and to fit the constraints of the Code LLM task (i.e, changing printed output to returned output, avoiding library imports).

\begin{figure*}
\begin{subfigure}{0.49\textwidth}
    \includegraphics[width=\textwidth]{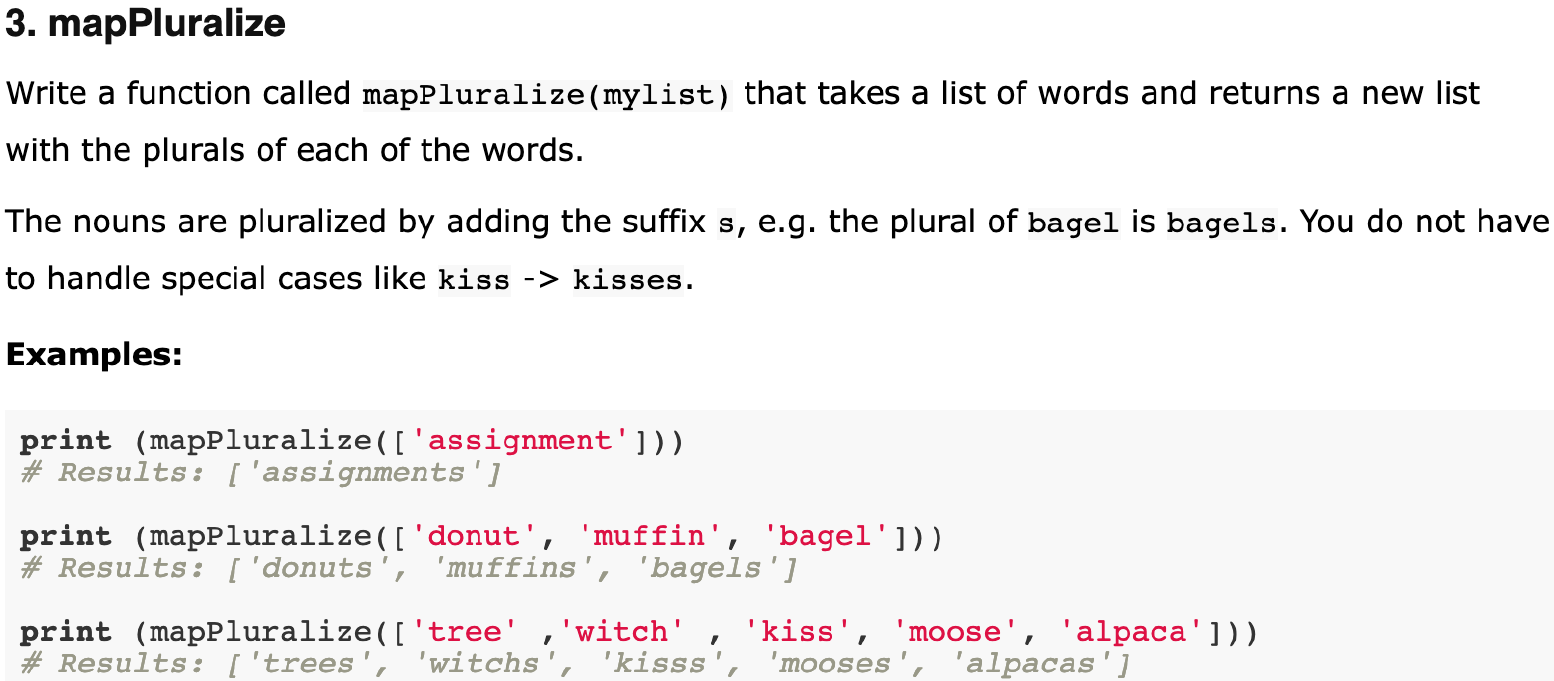}
    \caption{Study problem called \texttt{mod\_end}}
    \label{fig:modEnd}
\end{subfigure}
\hfill
\begin{subfigure}{0.49\textwidth}
    \includegraphics[width=\textwidth]{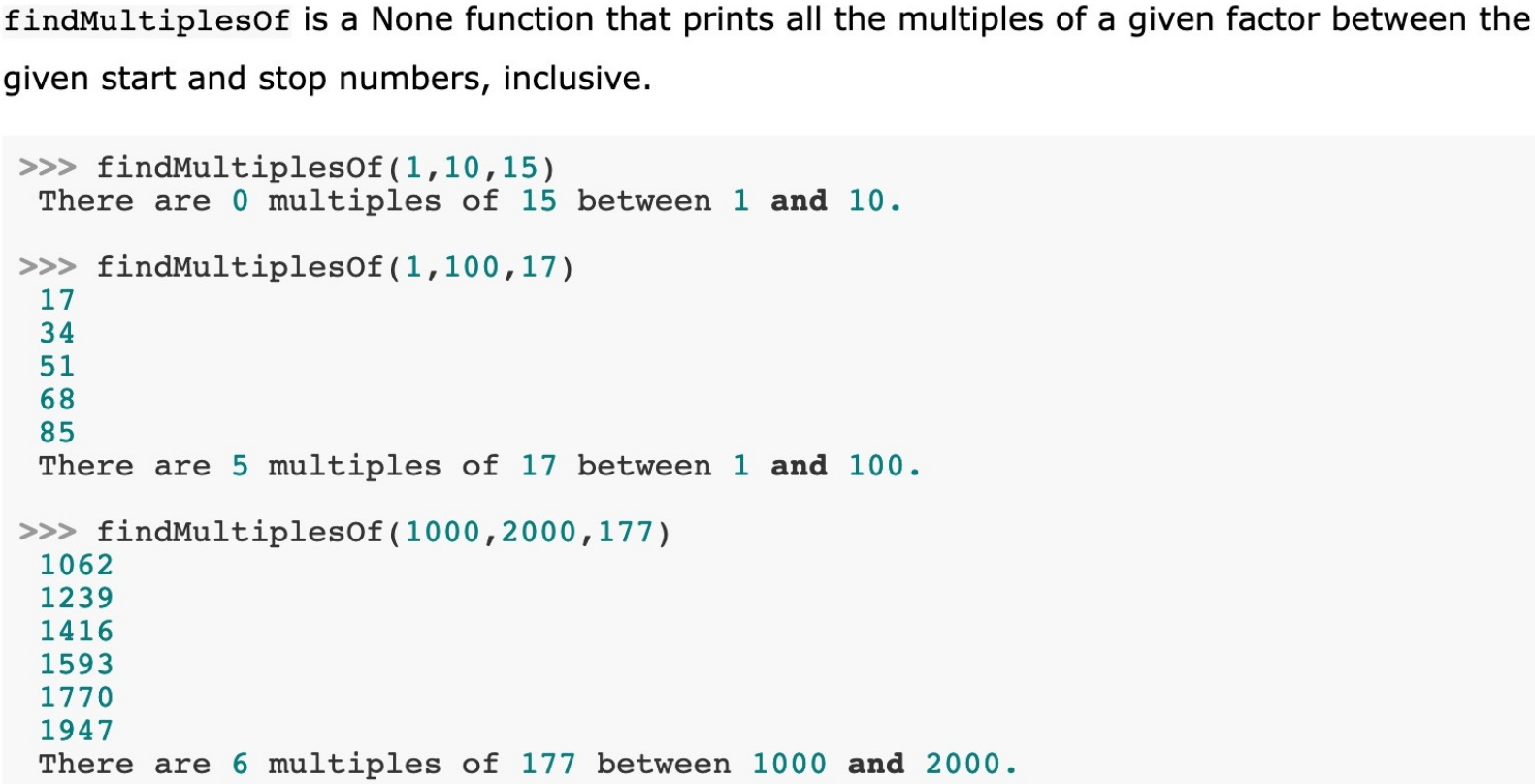}
    \caption{Study problem called \texttt{find\_multiples}}
    \label{fig:findMultiples}
\end{subfigure}
\caption{Original problem presentations}
\label{fig:examples}
\end{figure*}

Figure~\ref{fig:examples} presents two examples of how we adapted problems. Figure \ref{fig:modEnd} shows the original presentation of the problem that was adapted into \textit{mod\_end}. We added an additional parameter so that the function substitutes a given string for the `s' at the end of each string in the list. We also renamed the function. Note that in the original class setting, the problem was presented with three input/output pairs, as in our experimental design.

Figure \ref{fig:findMultiples} shows the original presentation of the problem that was turned into \textit{find\_multiples}. We changed the function to return the list of multiples rather than the number of multiples. As in our experiment, the original problem description contained three input/output pairs.

\subsubsection{Problem Validation}
By selecting from existing problems in the CS1 curricula, we ensured that the problems were at an appropriate difficulty level for our student population. In order to focus specifically on the human-model interaction, we also needed to ensure that the problems were an appropriate difficulty level for the code generation model: the model is capable of generating a solution, but only when it is appropriately prompted. 

Because code generation models memorize common associations between function names and function bodies, it is important to ensure that the model cannot generate a passing implementation from the function name alone. We produced Codex generations from just the function signature for every problem, without any natural language prompt, and measured mean pass@1 rate. We renamed any functions with high pass@1 rates. For our final set of problems, the overall mean pass@1 for function signatures alone is 0.0519. The maximum pass@1 is 0.925, for the problem \texttt{exp}. This means that students generally need to provide a description of the function's intended behavior in order for the model to produce a correct implementation.

We also ensured that there was a prompt that would lead to a correct implementation for every problem. Each problem has an ``expert'' prompt written by one of the authors for which Codex produces a correct implementation. These prompts were not otherwise used as part of the experiment.

\subsubsection{Test Case Validation}
We rely on unit tests to check the correctness of model-generated code. These tests also produce feedback for students about the model's generated code. We built an initial suite of test cases for each problem by taking tests from grading rubrics and other class resources. We used test coverage and mutation testing~\citep{jia_analysis_2011} to identify missing test cases and build more robust test coverage, while keeping the number of test cases per problem to a size that can be easily displayed.


\subsection{Qualitative Analysis}\label{app:coding}

As described in the main body of the paper, the analysis of the qualitative data was done by two researchers with previous qualitative analysis experience. The aim was to identify common themes in the data set, rather than build a generalizable theory. Below we outline the analyses performed on three types of data: (1) data about student experience and demographics, (2) free-response questions about future use of Charlie, and (3) the semi-structured interview responses. We provide the full codebooks, with definitions, for all data types as part of our Supplemental Materials at \osf.

\subsubsection{Student Experience \& Demographics}\label{sec:experience_demo}

We used thematic analysis for the post survey questions, beginning with the Language, Major, and Experience questions. Codes were developed inductively - the two researchers independently developed codes and then iterated on a code set via conversation and consensus. We did not calculate inter-rated reliability for these questions, as their specific use was for quantitative analysis rather than for specific qualitative trends \cite{mcdonald_reliability_2019}. Once the researchers arrived at a tentative codebook they independently coded and iterated until there was complete consensus on all codes for all data points as part of the post survey. This took one round to normalize code application (e.g., Computer Science was not coded as a Natural Science) and then a second round where the codes were complete, but typos were identified. 

\subsubsection{Free Response Questions}

These questions (UseCharlie and Foresee) were coded second out of the three kinds of qualitative data. This process initially followed a similar inductive style to that described above. Due to the open-ended nature of these responses, both researchers then developed independent definitions for each code to provide clearer guidelines for inclusion/exclusion. They then met to merge their definitions and discuss any discrepancies. For instance, normalizing most definitions to start with ``Mentions'' and combining definitions or picking the more detailed. Then the researchers independently coded according to the consensus definitions. Arriving at consensus took two rounds. Two sets of codes were combined (two subcodes of Skill Level and two subcodes of Problem Difficulty) and Documentation/Code Understanding was re-coded due to clarifications in their definitions. The final round of coding identified only typos and unintentional omissions. Again, consensus was reached and inter-rated reliability was not calculated for these codes.

\subsubsection{Semi-Structured Interview Analysis}


We took a different approach to coding the semi-structured interview data than the post-survey data, as the responses varied significantly in length and precision. The details of the codebook development are described below, but the following process was conducted for all 8 questions:
\begin{enumerate}
    \item The two coding researchers independently developed codes for a set of 15 non-overlapping interviews. They met to discuss their codes and general themes.
    \item They then coded 15 shared interviews to test the codes, add additional codes, and finalize definitions. They reached consensus on the codes and their application to the shared 15 interviews.
    \item To confirm their understanding of the codebook, they then coded 20 shared interviews and calculated percent agreement. Any codes with low agreement were discussed, had their definitions changed/edited, or were removed. The researchers then came to consensus on how to apply the codes to these data.
    \item The researchers then divided the remaining 70 interviews and coded independently, making use of a fixed codebook. The researchers did not code interviews they conducted themselves. They also independently recoded their original 15 datapoints.
\end{enumerate}

The two researchers began by coding the last four interview questions, which they deemed more concrete. This process was primarily inductive. Computing percent agreement across the data, 87\% of our 70 codes exhibited 90\% or higher agreement (i.e. disagreement on 2 or less datapoints). 9 codes were less, with the minimum agreement being 75\%. 

The researchers then moved on to the first four interview questions -- the last analysis performed on the qualitative data. This process was more deductive than previous analyses. For example, for Hardest/Easiest, the topic categories of problems (e.g. Loops, Conditionals) were particularly relevant to our analyses and the data suggest those categories as codes. The first two problems had the most variation in student responses -- we attribute this to students' lack of knowledge of their process, as they found this task difficult overall. Therefore, the researchers focused on codes that err on the side of temporal attributes. Percent agreement was again calculated for these codes - 80\% had 90\% agreement or higher. Only two codes had agreement lower than 75\% - the researchers discussed these codes significantly, reaching agreement on the generality of one code (Add Detail) and removing another code entirely.

\section{Additional Quantitative Results}
\subsection{Participant Demographics}\label{app:demo}

In order to protect participant anonymity, we report responses to the open-ended demographics questions only if an identical response was submitted by at least 5 participants. Gender and race responses are shown in \Cref{tab:demo_gender} and \Cref{tab:demo_race}. The majority of responses to the question about ethnicity were unique. Due to the need to protect participant anonymity, we have chosen not to report this data.

\begin{table}[h]
    \centering
    \begin{tabular}{c|c|c}
         Self-Reported Gender&N&Mean pass@1\\\hline
         Female&72&0.22\\
         Male&30&0.21\\
         Nonbinary&5&0.28\\
         All other responses&13&0.24\\
    \end{tabular}
    \caption{Self-reported gender of participants, capitalization normalized to title case.}
    \label{tab:demo_gender}
\end{table}

\begin{table}[h]
    \centering
    \begin{tabular}{c|c|c}
         Self-Reported Race&N&Mean pass@1\\\hline
Asian&38&0.22\\
Black&6&0.11\\
East Asian&6&0.34\\
White&36&0.21\\
All other responses&34&0.23\\
    \end{tabular}
    \caption{Self-reported race of participants, capitalization normalized to title case.}
    \label{tab:demo_race}
\end{table}

\subsubsection{Statistical Analysis}

We used Welch Two Sample t-tests to explore whether there were statistically reliable differences in pass@1 rates for students with different backgrounds. \Cref{tab:dem_ttests} shows the results.

\begin{table*}[ht]
    \centering
    \begin{tabular}{l|l|l|l}
    Group 1&Group 2&$t$&$p$\\\hline
    Domestic student&International student&-0.4&0.68\\
    First generation college student&Not first generation college student&2.1&\textbf{0.04}\\
    English in childhood household&No English in childhood household&1.02&0.31\\
  Public high school&Private high school&0.1&0.92\\
    Coding experience outside of CS1& No other coding experience&2.47&\textbf{0.02}\\
    More than 1 math course&1 college math course&0.8&0.43\\
    \end{tabular}
    \caption{Welch Two Sample t-tests to explore differences in pass@1 rates between demographic groups}
    \label{tab:dem_ttests}
    \vspace{-1mm}
\end{table*}

\subsection{Problem Difficulty}\label{sec:app_difficulty}

\subsubsection{Statistical Analysis of Category Difficulty}

A binomial mixed-effects model (Table~\ref{tab:mixed_effects}) was fitted to prompt success as a binary outcome (1 if the prompt succeeded; 0 otherwise). The model included fixed effects of problem category, institution, and their interaction, and random effects of participant and problem. Treatment coding was used for institution, with \northeastern as the baseline category; deviation coding was used for category, since we were interested in whether any one category differed from the average problem difficulty.

\begin{table}
    \centering
    \begin{tabular}{l|l|l|l}
Fixed effects&$\widehat{\beta}$&$z$&$p$\\\hline
(Intercept)&-0.80 (+/- 0.6)&-1.35&0.18\\  
Dictionaries&-0.70 (+/- 0.8)&-0.8&0.41\\
Lists&-0.55 (+/- 0.8)&-0.7&0.51\\
Loops&-0.92 (+/- 0.8)&-1.1&0.28  \\
Math&-1.10 (+/- 0.8)&-1.3&0.19  \\
Nested&0.83 (+/- 0.8)&1.0&0.32 \\
Sorting&-1.75 (+/- 0.9)&-2.0&\textbf{0.045}\\
Strings&0.14 (+/- 0.8)&0.2&0.87  \\
\wellesley&0.14 (+/- 0.4)&0.4&0.71 \\ 
\oberlin&0.24 (+/- 0.4)&0.6&0.53\\
Dictionaries:\wellesley&-0.04 (+/- 0.6)&-0.1&0.95 \\ 
Lists:\wellesley&0.37 (+/- 0.5)&0.7&0.50  \\
Loops:\wellesley&0.62 (+/- 0.5)&1.2&0.25  \\
Math:\wellesley&0.18 (+/- 0.5)&0.3&0.73  \\
Nested:\wellesley&-0.98 (+/- 0.5)&-1.9&0.062\\
Sorting:\wellesley&-0.16 (+/- 0.6)&-0.3&0.77 \\ 
Strings:\wellesley&-0.17 (+/- 0.5)&-0.4&0.73  \\
Dictionaries:\oberlin&-0.52 (+/- 0.6)&-0.9&0.37 \\ 
Lists:\oberlin&-0.05 (+/- 0.5)&-0.1&0.93  \\
Loops:\oberlin&-0.33 (+/- 0.6)&-0.6&0.57  \\
Math:\oberlin&0.14 (+/- 0.5)&0.3&0.79  \\
Nested:\oberlin&-0.57 (+/- 0.5)&-1.1&0.28\\  
Sorting:\oberlin&-0.10 (+/- 0.6)&-0.2&0.86 \\ 
Strings:\oberlin&0.10 (+/- 0.5)&0.2&0.84 \\
\end{tabular}
    \caption{Full results of binomial mixed-effects model fitted to problem category and institution.}
    \label{tab:mixed_effects}
    \vspace{-10mm}
\end{table}

\subsubsection{Least-Solved Problems}

To understand where struggles arise, we manually examined student responses to two problems: \textit{laugh}, which has one of the lowest number of student successes, and \textit{total\textunderscore bill}, which has a mid-range success rate.

\emph{A challenging problem: \texttt{laugh}.} One of the least-solved problems in our study was \texttt{laugh}. The intended function takes a number $n$ and produces a string of $n$ ``ha''s, where the initial ``ha'' has $n$ ``a''s, and each subsequent laugh has one fewer ``a''. 

Only two students were able to eventually succeed at this task (\pseudonym{orchidWalleye} and \pseudonym{magentaWeasel}). However, a manual inspection of all initial student descriptions reveals only one serious misunderstanding of the task (\pseudonym{tealPossum}) -- see Table~\ref{tab:laugh_submissions} for all students' initial descriptions.

\paragraph{A mid-range problem: \texttt{total\textunderscore bill}} The task in \texttt{total\textunderscore bill} is to compute the total of a grocery bill, using a list of grocery items and a sales tax rate. Each grocery item is itself a list containing the name of the item, a quantity, and a price. One expert description that reliably generates a working program is \textit{Returns the sum of multiplying the second and third indices of each list in grocery\textunderscore list, multiplied by 1 + sales\textunderscore tax. Round to 2 digits}. 

We manually inspect all descriptions for this problem. Of the 20 students who attempted this problem, 12 eventually succeed. All of these students follow a similar path: their first attempt omits the rounding step, leading one of the tests to fail. A handful of students also omit or incorrectly describe the sales tax step initially.

What about the students who never succeed? One student initially misunderstands the task, writing: \textit{This function takes in a list of the item purchased, the price, the tax, and the overall sales tax. All of the prices and tax within the lists are added together. The sales tax is then multiplied by the outcome of the added prices, and then the result of the multiplication is added onto the total price. The total price is then returned as the output.} (\pseudonym{limeSalamander})

The student has misunderstood a key detail in the structure of the lists: the two numbers are the quantity and price, so they should be multiplied, not added. Consequently, this prompt fails. However, their third description is accurate: \textit{This function takes in a list of the item purchased, the amount of the item purchased, the price for each item, and the overall sales tax. The amount purchased is multiplied with the price for each item, creating a total amount. The sales tax is then multiplied by the outcome of the total amount, and then the result of the multiplication is added onto the total price. The total price is then returned as the output.}

Although the student initially misunderstood part of the problem, they are able to reread the input/output pairs and/or code, arriving at the correct interpretation eventually. However, their description still fails. This participant eventually runs out of time. The rest of the participants who never succeed submit accurate descriptions that omit key details, such as how to calculate the sales tax (6 participants) or the list positions of the price and quantity (5 participants). 

Overall, the student prompts for \texttt{total\textunderscore bill} demonstrate more issues in describing the problem than in understanding it. Although one participant misunderstands the task initially, they were able to quickly self-correct.

\begin{table*}[htbp]
\centering
   \begin{tabular}{p{0.14\textwidth}|p{0.65\textwidth}|p{0.11\textwidth}}
   Participant&Initial Description&N submissions\\\hline
         \pseudonym{aquaLadybug}&If n is the input value, returns a combination of n strings, where each of the n strings consists of "h" followed by n occurrences of "a", and there is " " before each "h" except the first "h"&18\\
         \pseudonym{greenMoth}&a function have initial input as `ha' when  input of size(int) is 1, when size+= 1 from 1, `ha' will gain one more `a'&2\\
         \pseudonym{orchidBeetle}&Based on the inputted number, will return a laugh size where the number of "a"'s starts with the initial size, then decreases by one for each additional laugh.&4\\
         \pseudonym{tealPossum}&return the number of words in a string&2\\
         \pseudonym{pinkFisher}&the function laugh will take the input of an int and should output a string with the ha as many times as the input but also the number of a's is based on the number it is currently working with&4 \\
         \pseudonym{magentaWeasel}&Write a function which takes an integer size as an input, and uses a for loop to print an h followed by size a's and then a space, and then an h followed by size-1 a's and then a space, etc. until it prints a h followed by one a&7\\
         \pseudonym{aquamarineShrew}&This function prints an `h' and adds the corresponding amount of a's as the value provided. It then adds a space to the output. It subtracts 1 from the value and prints another h with less a's and repeats until the value of the number is 0&26\\
          \pseudonym{orchidWalleye}&function adds `a' to every `h' based on input and will lower amount of `a' until it reaches only 1 `a' after the `h'&3\\
         \pseudonym{khakiBee}&take in a number and write the word `ha' but with as many `a's as the number&7\\
          \pseudonym{pinkPerch}&Produce a string, with each word starting with h and then however many a's the input says. Decrease the count of a's by one following the h for each word after.&5\\
         \pseudonym{orchidFlounder}&the input generates a string where the number corresponds to how many items are in the string. each item in the string also starts with the letter `h' and the letter `a' is added to the letter `h' based on the number of the input. However, only the first item in the string has the number of `a' equal to the input, the following `a' are added to `h' by subtracting 1 from the input.&1\\
          \pseudonym{beigeBass}&the code increases the number of the letters in "ha," depending on the input in an increasing factorial way&1\\
          \pseudonym{tomatoFisher}&This function takes an integer and an input produces the word "ha" that number of times but the number of times "a" appears in each "ha" decreases by one until "ha"&2\\
          \pseudonym{crimsonVole}&Takes in an integer `n' input and outputs a string with `n' words, `h' as the first letter for each word, and `n' number of `a's after it, followed by `h' as the first letter of the next word and `n-1' number of `a's after it and so on until we reach n = 1&1\\
         \pseudonym{ lavenderPossum}&Given an integer, return a string in the form `ha' where the integer determines the number of a's and repeat the same pattern until there is one a&5\\
          \pseudonym{lavenderBat}&The input takes in a number, say n, and produces a string that has n words. the first word is formed of one "h" and n number of "a". The number of "a" decreases by one for each next word&8\\
          \pseudonym{magentaDolphin}&This function returns the number of laughs in a string, where a laugh is the character `h' followed by any number of the character `a'&9\\
          \pseudonym{linenBobcat}&Counts the number of laughs, beginning with the given number of "a"s within it and descending by each laugh, totaling the given number of laughs.&2\\
          \pseudonym{grayVole}&Takes size and uses recursion to produce that number of "ha" laughs with one less "a" with each "ha" until there is only one "a" left&8\\
          \pseudonym{thistleTrout}&Using the given number, add that number of "a"s after an "h". Count down the number by 1, and add that number of "a"s after another "h" and repeat.&5\\
   \end{tabular}
   \caption{Initial descriptions of the \texttt{laugh} problem from all 20 students who encountered it. \emph{N submissions} describes how many times the specific student attempted \texttt{laugh} before succeeding or giving up.}
   \label{tab:laugh_submissions}
\end{table*}

\end{document}